\documentclass[twocolumn, tighten]{aastex631}
\pdfoutput=1
\usepackage{mathtools}
\usepackage{appendix}
\usepackage{bm}
\usepackage{soul}
\usepackage{natbib}

\defcitealias{Bastian_2015MNRAS}{BN15}
\defcitealias{Brandt_2015ApJ_58}{BH15}
\defcitealias{Breimann_2021ApJ}{B21}
\defcitealias{Cargile_2020ApJ}{C20}
\defcitealias{Gossage_2019ApJ}{G19}
\defcitealias{Lipatov_2020ApJ}{LB20}
\defcitealias{Kamann_2020MNRAS}{K20}
\defcitealias{Zucker_2020A&A}{Z20}

\shorttitle{Rotation and Age in NGC 1846}
\shortauthors{Lipatov, Brandt, and Gossage}
\submitjournal{The Astrophysical Journal}
\received{November 4, 2021}
\accepted{June 13, 2022}

\begin{document}

\title{Rotational Variation Allows for Narrow Age Spread \\ in the Extended Main Sequence Turnoff of Massive Cluster NGC 1846}

\correspondingauthor{Mikhail Lipatov}
\email{mikhail@physics.ucsb.edu}

\author[0000-0001-9939-1758]{Mikhail Lipatov}
\affiliation{Department of Physics, University of California, Santa Barbara, CA 93106, USA}

\author[0000-0003-2630-8073]{Timothy D. Brandt}
\affiliation{Department of Physics, University of California, Santa Barbara, CA 93106, USA}

\author[0000-0001-6692-6410]{Seth Gossage}
\affiliation{Center for Interdisciplinary Exploration and Research in Astrophysics, Northwestern University, Evanston, IL 60201, USA}

\begin{abstract}

The color-magnitude diagrams (CMDs) of intermediate-age star clusters ($\lesssim$2\,Gyr) are much more complex than those predicted by coeval, non-rotating stellar evolution models. Their observed extended main sequence turnoffs (eMSTOs) could result from variations in stellar age, stellar rotation, or both.  The physical interpretation of eMSTOs is largely based on the complex mapping between stellar models---themselves functions of mass, rotation, orientation, and binarity---and the CMD. In this paper, we compute continuous probability densities in three-dimensional color, magnitude, and $v_{\rm e}\sin{i}$ space for individual stars in a cluster's eMSTO, based on a rotating stellar evolution model. 
These densities enable the rigorous inference of cluster properties from a stellar model, or, alternatively, constraints on the stellar model from the cluster's CMD.  We use the \texttt{MIST} stellar evolution models
to jointly infer the age dispersion, the rotational distribution, and the binary fraction of the Large Magellanic Cloud cluster NGC 1846. We derive an  age dispersion of $\sim70-80\,{\rm Myr}$, approximately half the earlier estimates due to non-rotating models. 
This finding agrees with the conjecture that rotational variation is largely responsible for eMSTOs. 
However, the \texttt{MIST} models do not provide a satisfactory fit to all stars in the cluster and achieve their best agreement at an unrealistically high binary fraction. The lack of agreement near the main-sequence turnoff suggests specific physical changes to the stellar evolution models, including a lower mass for the Kraft break and potentially enhanced main sequence lifespans for rapidly rotating stars. 
\\
\end{abstract}


\section{Introduction} \label{introduction}

\subsection{Evolution of Rotating Stars}

According to modern physical science, fundamental principles can explain the diversity of observed stars via stellar structure and evolution \citep{Arny_1990_via,Christensen_2021LRSP}. An early manifestation of this idea is the Vogt-Russell theorem, a proposition that a star's chemical composition structure and its initial mass (or, simply, mass) fully determine the course of its life \citep[][p. 333]{Kaehler_1978IAUS,Carroll_2007}. The addition of rotation to the list of life-determining parameters constitutes an important amendment to this proposition. 

Generally speaking, stars rotate. This phenomenon has been observed in the movement of the Sun's spots \citep{Howard_1984ApJ}, the centrifugally deformed shapes of nearby B- and A-type stars \citep{Monnier_2007Sci,deSouza_2014A&A}, and spectroscopic rotational velocities of unresolved stars \citep{Royer_2002A&A_a,Royer_2002A&A_b,Healy_2020ApJ}. Rotation has important consequences for the evolution and observed properties of stars. It mixes extra hydrogen into the core of a main-sequence star, increasing both its luminosity and lifetime \citep{Brott_2011A&A,Eggenberger_2013EPJWC}. In addition, the equatorial regions of a rotating star are cooler and dimmer than its polar regions due to an effect called gravity darkening \citep{vonZeipel_1924,EspinosaLara_2011A&A}. This makes the star's magnitudes and colors depend on the inclination of its axis with respect to the observer \citep[e.g.,][]{Lipatov_2020ApJ}. 

Stars inherit their angular momenta from ancestral clouds of gas and dust \citep{Prentice_1971MNRAS,Tomisaka_2000ApJ,Larson_2010RPPh}. Subsequently, their rotational speeds evolve to the present day \citep{Maeder_2000ARA&A}. The speeds extend up to appreciable fractions of the centrifugal breakup limit for stars with mass $\gtrsim$1.5$\,M_\odot$ \citep{Zorec_2012A&A,Kamann_2020MNRAS}. Lower-mass stars, on the other hand, spin down rapidly \citep[e.g., see Figure 11 in][]{GodoyRivera_2021arXiv}. This pattern likely results from the emergence of an outer convective zone that supports magnetic field lines that, in turn, rotate with the star and extend away from it. Stellar wind particles move along these lines, depriving the star of angular momentum. This process, termed magnetic braking, results in the Kraft break -- a sharp reduction in observed rotation rates as stellar mass decreases below $\sim1.3\,M_\odot$ \citep{Kraft_1967,Noyes+Hartmann+Baliunas+etal_1984}. Recent analyses tune models of magnetic braking to clusters, i.e., gravitationally bound collections of stars \citep{Matt_2015ApJ,Breimann_2021ApJ,Gossage_2021ApJ}.

\subsection{Stellar Distributions in Massive Clusters} \label{intro_distributions}

In this work, we focus on  NGC 1846, which belongs to the category of massive ($\gtrsim\!10^4\,M_\odot$), intermediate-age ($\sim 1-2 \, {\rm Gyr}$) clusters that reside in the Magellanic Clouds \citep[][hereafter BN15]{Bastian_2015MNRAS}. Like other clusters, it offers an opportunity to tune a model of stellar structure and evolution simultaneously to all of its stars, since their shared cluster membership implies that they share some of their life-determining parameters. 

For example, if the stars in a cluster are all formed from the collapse and fragmentation of the same giant molecular cloud \citep{Klessen_2001ApJ,Bate_2003MNRAS}, they should all have the same chemical composition. This picture is not entirely true for massive clusters, which can contain multiple populations (MPs) with different chemical compositions \citep{Bastian_2018ARA&A,Gratton_2012A&ARv,Piotto_2009IAUS}. On the other hand, massive clusters in the Magellanic Clouds generally show insignificant within-cluster departures from uniform iron abundances [Fe/H] \citep{Piatti_2019AJ,Piatti_2020A&A,Mucciarelli_2008AJ}. This suggests that there is not enough variation in chemical composition to produce appreciable variation in stellar evolution within such clusters. Similarly to [Fe/H] distributions, initial mass distributions in clusters are relatively well-known, with consequently predictable effects on magnitudes and colors. There is evidence that these mass distributions do not differ significantly from the Salpeter initial mass function (IMF) above $\sim$1\,$M_\odot$ \citep{Salpeter_1955ApJ,Kroupa_2001MNRAS,Chabrier_2003PASP,Villaume_2017ApJ}. 

Unlike [Fe/H] and mass distributions, rotational and age distributions of stars within clusters are not established.  Variations in both rotation and age have been invoked to explain the color spreads of the main sequence turnoff (MSTO), termed extended MSTOs (eMSTOs). One of the first eMSTOs was discovered in NGC 1846 \citep{Mackey_2007MNRAS}. Initial photometry-based analysis led to the hypothesis that this pattern results from a wide stellar age distribution, i.e., an extended star formation (eSF) period \citep{Goudfrooij_2009AJ,Rubele_2013MNRAS,Goudfrooij_2011ApJ_3,Goudfrooij_2011ApJ_4}. Subsequently, as eMSTOs were discovered in other clusters, it became apparent that age and rotation spreads could both contribute to this phenomenon, making it difficult to distinguish between the two factors from MSTO photometry alone \citep{Bastian_2009MNRAS,Bastian_2015MNRAS,Brandt_2015ApJ_25,DAntona_2017NatAs}. At the same time, eSF ought to have similar effects on different portions of the CMD -- e.g., the MSTO, the sub-giant branch (SGB), and the red clump (RC). \citetalias{Bastian_2015MNRAS} show that, even under the assumption of zero rotational variation, the SGB and RC morphologies in NGC 1846 are consistent with zero age spread and are significantly narrower than expected if eSF causes the cluster's eMSTO. \citetalias{Bastian_2015MNRAS} go on to suggest that their results can be explained by a rotational distribution that widens the MSTO, but does not necessarily widen the SGB or the RC. 

A variety of additional evidence conflicts with the hypothesis that eSF causes eMSTOs in NGC 1846 and other massive clusters. For example, \citet{Niederhofer_2015MNRAS} show that, under the assumption of zero rotation, age spreads inferred from eMSTOs correlate with cluster age, an observation that is inconsistent with the idea that the age spread of a cluster is set for the duration of its life. Instead, as the authors demonstrate, the observed correlation is in good agreement with the hypothesis that rotation spreads cause eMSTOs. Furthermore, \citet{Bastian_CabreraZiri_2013MNRAS} examine a number of clusters at one to several tens of Myr (Young Massive Clusters, or YMCs); at these ages, one expects significant star formation under the eSF hypothesis. The authors do not find evidence of such formation in spectral emission lines and constrain the maximum mass of the material that could be undergoing star formation to no more than 1-2 \% of the existing stellar mass content. Along the same line of inquiry, \citet{CabreraZiri_2015MNRAS} show that YMCs do not possess the interstellar gas and dust that can form into stars in the course of eSF. 

\subsection{Analysis of Star Clusters}

The morphology of the CMD results from the theory of stellar evolution and the properties---mass, age, composition, rotation, and orientation---of individual stars.  In order to infer cluster parameters from the CMD, or to tune models of stellar evolution, we need to compare theoretical and observed CMDs either qualitatively or quantitatively.  Recent work, which we review here, has advanced toward ever-more rigorous statistical comparisons between theoretical and observed CMDs.

Some statistical approaches infer the parameters of individual stars.  For example, \citet[][henceforth BH15]{Brandt_2015ApJ_58} infer the ages and other present-day parameters of stars from color, magnitude, and projected rotational velocity, under the assumption of the \texttt{SYCLIST} evolutionary model library \citep{Ekstrom_2012A&A,Georgy_2013A&A}. More recently, \citet[][henceforth C20]{Cargile_2020ApJ} accomplish this task under the assumption of the \texttt{MIST} library \citep[\texttt{MESA} Isochrones and Stellar Tracks;][]{Dotter_2016ApJS, Choi_2016ApJ, Gossage_2018ApJ, Gossage_2019ApJ} . Both of these star-by-star approaches are Bayesian, with the goal of computing 
the stellar parameters' joint posterior distribution.  
Both \citetalias{Brandt_2015ApJ_58} and \citetalias{Cargile_2020ApJ}  
write down the likelihood of stellar parameters in terms of instrumental uncertainty and multiply the likelihood by the parameters' prior. \citetalias{Cargile_2020ApJ} approximate the resulting posterior by way of a Monte Carlo methodology called nested sampling \citep{Speagle_2020MNRAS}, while \citetalias{Brandt_2015ApJ_58} calculate it on a deterministic grid. Both methods can estimate multi-modal and/or highly covariant posteriors more efficiently than conventional Monte Carlo (MC) methodologies, although the deterministic method is only viable when the dimensionality of the posterior is small.

One can also simultaneously infer the parameters of many stars under the assumption that they share the values for some of these parameters (e.g.,~age and composition).
For example, \citetalias{Brandt_2015ApJ_58} marginalize the posteriors of many stars over mass, rotation, and orientation 
to infer shared parameters in a star cluster. Building on earlier work \citep{Zucker_2019ApJ,Schlafly_2014ApJ,Green_2014ApJ}, \citet[][henceforth Z20]{Zucker_2020A&A} follow a similar procedure to infer shared parameters for a different sort of object -- a molecular cloud that lies between the stars along lines of sight. 

Intuitively, when the posterior is viewed as a probability density in stellar observable space at constant cloud/cluster parameters, parameter likelihood is the product of density values at the observable-space locations of stars. \citetalias{Brandt_2015ApJ_58}, \citetalias{Zucker_2020A&A}, and \citet{Green_2014ApJ} state this result without proof, but \citet{Walmswell_2013MNRAS} and \citet[][henceforth B21]{Breimann_2021ApJ} prove it as a consequence of data generation via a Poissonian process that is inhomogeneous in observable space. The idea of thus multiplying probability density values at observable-space locations of stars to obtain the likelihood of cluster parameters was introduced earlier \citep{Naylor_2006MNRAS,vanDyk_2009AnApS}. \citetalias{Breimann_2021ApJ}  evaluate the density values and, consequently, the likelihoods, over a range of cluster parameters. In \citetalias{Breimann_2021ApJ}'s case, the latter are synonymous with stellar evolution parameters. These authors find that theoretical probability density values for some of the observed stars are very low, even at maximum-likelihood evolutionary parameters: these stars cannot be explained by the theoretical model.  \citetalias{Breimann_2021ApJ} conclude that the evolutionary model approximations should be modified. Unlike other authors mentioned so far in this section, \citetalias{Breimann_2021ApJ} never evaluate or marginalize single-star posteriors over stellar parameter ranges to calculate the probability densities. Instead, they estimate the densities directly by binning stellar models in observable space.

\citet[][henceforth G19]{Gossage_2019ApJ} also take a binning approach and estimate cluster parameters via comparisons of theoretical densities in color-magnitude space, a.k.a.~Hess diagrams, with their observed counterparts \citep{Dolphin_2002MNRAS}. In \citetalias{Gossage_2019ApJ}'s work, the estimated parameters are the cluster's age, the Gaussian age spread, and the rotation rate distribution. The authors' evolutionary models are from \texttt{MIST}, like those in \citetalias{Cargile_2020ApJ}. Like \citetalias{Breimann_2021ApJ}, \citetalias{Gossage_2019ApJ} do not evaluate single-star posteriors, directly comparing likelihoods of different cluster models. These authors state that their analysis does not conclusively distinguish between age and rotation in causing eMSTOs. However, they suggest that the distinction could be made via the inclusion of rotational data such as projected equatorial velocities. Furthermore, \citetalias{Gossage_2019ApJ}'s detailed analysis allows them to identify the evolutionary processes that one can tune to improve the model's fit to the data and to independent knowledge of cluster structure and formation history. Specifically, the authors propose that the match to the data could improve with the tuning of the model's rotation-related processes, such as magnetic braking. Earlier work in the same vein indicates that other processes, such as rotationally induced mixing, also greatly affect the joint inference of age and rotational distributions \citep{Gossage_2018ApJ}.

\subsection{Our Analysis of NGC 1846}

 In the present work, we follow \citetalias{Gossage_2019ApJ}'s example and compare NGC 1846 data with the \texttt{MIST} rotating stellar model to jointly infer the rotational and age distributions of the cluster's MSTO stars. In line with \citetalias{Gossage_2019ApJ}'s suggestion and similarly to \citetalias{Brandt_2015ApJ_58}, our analysis integrates projected equatorial velocity measurements with multi-band photometry of the stars. Furthermore, much like \citetalias{Gossage_2019ApJ}, we identify evolutionary processes that one can tune to improve the fit between the model and the data. With this work, we intend to provide a generally applicable and statistically quantifiable numerical framework for the derivation of the properties of star clusters based on known aspects of stellar evolution and the derivation of constraints on stellar evolution based on known properties of star clusters.  

The rest of this article is structured as follows. In Section \ref{data}, we present the data that form the basis of our inference. In Section \ref{stellar}, we describe our stellar model, which maps age and other parameters to observables.  In Section \ref{probabilities}, we detail the calculation of theoretical probability densities in observable space, based on the stellar model and measurement error. Section \ref{statistical} describes the statistical model that allows us to combine multiple observed stars in the inference of cluster parameters. We present the resulting parameter estimates in Section \ref{results}. Section \ref{discussion} suggests specific physical changes to evolution models in view of the disagreement between our cluster parameter estimates and independently known values, as well as between our probability densities and individual data points. We summarize this work and suggest additional future directions in Section \ref{summary}.

\section{Data} \label{data}

We base our analysis on recent spectroscopic $v_{\rm e}\sin{i}$ measurements of individual stars in the central 1 arcmin $\times$ 1 arcmin of NGC 1846, collected by \citet[][henceforth K20]{Kamann_2020MNRAS} with the Multi Unit Spectroscopic Explorer \citep[\texttt{MUSE},][]{Bacon_2010SPIE} on the Very Large Telescope. Here, $v_{\rm e}$ is the equatorial velocity of a star and $i$ is the inclination of its rotational axis with respect to the plane of the sky, so that $v \equiv v_{\rm e}\sin{i}$ is the projected equatorial velocity. \citetalias{Kamann_2020MNRAS} estimate $v_{\rm e}\sin{i}$ from transition line broadening via full-spectrum fitting and augment these measurements with previously collected multi-band \texttt{HST} (Hubble Space Telescope) photometry of the same stars \citep{Martocchia_2018MNRAS}. The photometric magnitudes correspond to three filters on the Wide Field Channel of \texttt{HST}'s Advanced Camera for Surveys: $m_{435} \equiv m_{\rm F435W}$, $m_{555} \equiv m_{\rm F555W}$, and $m_{814} \equiv m_{\rm F814W}$. The \texttt{MUSE} data show significant variation in $v_{\rm e}\sin{i}$ across the MSTO, indicating that the stars in this area of the CMD have significantly variable rotation speeds and/or inclinations.

Inference of rotational and age distributions in clusters is sensitive to the modeling of processes relevant to the evolution of stars, in ways that potentially differ between the stages of a star's life and between stars of different masses. Thus, in order to better understand the meaning of our results, we restrict ourselves to a particular portion of the NGC 1846 data set and a particular range of stellar evolution models. Specifically, we work only with the stars observed in the MSTO area of the CMD (see Figure 4 in \citetalias{Kamann_2020MNRAS}) and interpret them solely in terms of 1 to 2 $M_\odot$ main-sequence stellar models. Even when the inference is subject to these restrictions, the data set remains large, while the evolutionary models produce predictions that are sufficiently intricate to warrant taking into account the exact uncertainty on each measurement. To accomplish the latter for the entire data set, we find it advantageous to establish minimum possible errors on measurements, compute corresponding minimum-error theoretical probability distributions, then broaden these distributions as necessary for each individual measurement. Our data selection and error assignment, further described in the rest of this section, are designed in view of the above-mentioned considerations.

We make use of the $n = 2353$ stars in \citetalias{Kamann_2020MNRAS}'s data set that fall in our region of interest (ROI), which satisfies $m \equiv m_{555} \in [19.5, 22.0]$, $c \equiv m_{435} - m_{814} \in [0.4, 1.0]$, and $v \in [0, \infty]$. We refer to a point in the 3-dimensional observable space as ${\bm x} \equiv (m, c, v)$. Since neither of the two filters that produce $c$ is the filter that produces $m$, we assume zero correlation between the errors in these two observables for a given star. We also assume that errors in broadband filter magnitudes do not correlate with errors in the broadening of individual spectral lines, so that the errors in $m$ and $c$ do not correlate with the error in $v$. The rotational measurement $v$ is positive, zero and missing for $n_{\rm p} = 1237$, $n_{\rm z} = 74$ and $n_{\rm m} = 1042$ of these stars, respectively. Every $m$ and $c$ measurement in the data set is associated with its own error value, which we interpret as the standard deviation of the corresponding error distribution. Furthermore, every positive $v$ measurement in the data set is associated with an upper and a lower error value. The average of these latter two values becomes the standard deviation of the corresponding $v$ measurement error distribution. This averaging procedure does not affect inference at $v > 100\,{\rm km/s}$, where the upper and lower errors are equal. We choose to retain the averaging procedure at lower $v$, for the sake of computational speed and simplicity. We further assume that the error distributions are Gaussian and impose a lower limit of $\sigma_m = 0.01$ on the standard deviations of these distributions for magnitudes. This makes the error distributions for color measurements Gaussian as well, with a lower limit on standard deviations $\sigma_c = \sigma_m \sqrt{2} = 0.014$. Our approximation of non-Gaussian error distributions for low $v$ measurements as Gaussian may introduce offsets to our cluster parameter estimates. On the other hand, we expect these offsets to be significantly lower than the offsets due to uncertainties in evolutionary models. The true error distributions are, in any case, likely to be far more complicated than two half-Gaussians with a discontinuity where they meet.  

Our lower limit on the uncertainty of $v$ measurements is $\sigma_v = 10\, {\rm km\,s^{-1}}$, which is on the order of the uncertainty in $v_{\rm e}\sin{i}$ at a given line broadening $\sigma_{\rm LOS}$ in Figure A1 of \citetalias{Kamann_2020MNRAS}. Although we are not explicitly given an error on the $v = 0$ measurements, we set it equal to $50\,{\rm km\,s^{-1}}$, based on an approximate extrapolation of the $v$ measurement standard deviations down to $v = 0$. We collectively refer to the minimum errors on the observables as $\bm{\sigma_{\bm x}} \equiv (\sigma_m, \sigma_c, \sigma_v)$. Each data point is composed of an observed star's ${\bm x}$ and $\bm{\sigma_{\bm x}}$: ${\bm x_{\bm p}} \equiv (m_p, c_p, v_p)$ and $\bm{\sigma_{{\bm x}{\bm p}}} \equiv (\sigma_{mp}, \sigma_{cp}, \sigma_{vp})$, where subscript $p \in [1, n]$ is data point index. A missing rotational measurement corresponds to $\sigma_{vp} = \infty$.

\section{Stellar model} \label{stellar}

In this section, we describe the procedure that yields magnitude, color and $v_{\rm e}\sin{i}$ for a stellar model, given its independent parameters -- initial mass, initial rotation rate, inclination of the rotational axis, age, and initial mass of a binary companion (if present).

\subsection{Evolution} \label{evolution}

We model the evolution of stars according to version 1.0 of the \texttt{MIST} library , which is based on version r7503 of the \texttt{MESA} computer code \citep[Modules for Experiments in Stellar Astrophysics:][]{Paxton_2011ApJS,Paxton_2013ApJS, Paxton_2015ApJS,Paxton_2019ApJS}. \texttt{MESA} models rotation by using pressure as the radial coordinate.  It does not assume spherical symmetry, but rather that certain physical quantities are constant along isobars and that energy transport is perpendicular to the local effective gravity.

At a given age, \texttt{MIST} provides Equivalent Evolutionary Phase (${\rm EEP}$), mass $M$, surface angular speed $\Omega$, dimensionless angular speed $\omega_{{\rm M}}$, luminosity $L$, Eddington ratio $L / L_{\rm Edd}$, and radius $R_{\rm M}$.  Here, $R_{\rm M}$ is the radius of a sphere that encloses the star's volume $V$ \citep{Endal+Sofia_1976ApJ, Paxton_2019ApJS}, so that
\begin{linenomath*}\begin{equation}
    V = \frac{4 \pi}{3} R_{\rm M}^3.
    \label{eq:Vmist}
\end{equation}\end{linenomath*}
Furthermore, $\omega_{\rm M} \equiv \Omega / \Omega_{\rm crit}$, where $\Omega_{\rm crit}$ is defined soon after Equation (26) in \citet{Paxton_2013ApJS}:
\begin{linenomath*}\begin{equation}
    \Omega_{\rm crit} \equiv \sqrt{\left(1 - \frac{L}{L_{\rm Edd}}\right)\frac{G M}{R_{\rm M}^3}}.
    \label{eq:oMcrit}
\end{equation}\end{linenomath*}

We only consider main sequence \texttt{MIST} models, with ${\rm EEP} \in [202, 454]$. We model stars at other EEPs as part of a background distribution, while choosing our region of the CMD to exclude most of these post-main sequence stars. We estimate that only about 1\% of the stars that remain in our ROI on the CMD (i.e., about 20 stars) are post-main sequence stars, given the amount of time that \texttt{MIST} models spend on the subgiant branch before crossing the red edge of our ROI at $m_{435}-m_{814}=1$. Our background distribution also subsumes other types of stars, such as blue stragglers, that are not modeled by the \texttt{MIST} library. Upon visual inspection of the turn-off, we estimate that $\lesssim1\%$ of our observed stars are likely to be blue stragglers. These would have to be modeled via binary evolution, which is beyond the scope of this paper.

Along with age $t$, the models' independent parameters are initial mass, initial angular speed $\omega_{\rm Mi}$, and metallicity ${\rm [M/H]_M}$. Initial mass is designated by $M_{\rm i}$ for a primary in a star system and by $M_{\rm Ci}$ for a secondary companion. Here and elsewhere in the article, subscripts ${\rm M}$ and ${\rm i}$ stand for "\texttt{MIST}"  and "initial", respectively. Furthermore,
\begin{linenomath*}\begin{equation}
    {\rm [M/H]_M} = \log{\frac{Z}{X}} - \log{\frac{Z_{\rm \odot, M}}{X_\odot}},
\label{eq:metM}
\end{equation}\end{linenomath*} 
where $Z$ and $X$ are the respective metal and hydrogen mass fractions of the star, $X_\odot$ is the protosolar hydrogen mass fraction, and $Z_{\rm \odot, M} = 0.0142$ is an estimate of the protosolar metal mass fraction \citep[pp. 2-3 in][]{Choi_2016ApJ,Asplund_2009ARA&A}. In Equation \eqref{eq:metM} and the rest of this work, $\log$ designates logarithm with base ten.

\texttt{MIST} has solar-scaled abundance ratios, so that its metallicity is equivalent to relative iron abundance, i.e., ${\rm [M/H]_M} \equiv {\rm [Fe/H]_M}$. There is some evidence that the LMC and Milky Way (i.e., solar) abundance patterns differ. In particular, the LMC may have relatively low ${\rm Mg}$ to ${\rm Fe}$ and ${\rm O}$ to ${\rm Fe}$ ratios \citep{Pompeia_2008A&A,VanDerSwaelmen2013A&A,Rolleston_2002A&A}. Future work may provide model libraries with LMC-scaled abundances.  We keep metallicity ${\rm [M/H]_M}$ constant at $-0.45$, a value that is based on isochrone fits in \citetalias{Kamann_2020MNRAS}, so that the models start off parametrized by $\{M_{\rm i}, \omega_{\rm Mi}, t\}$. 

Traditionally, an isochrone is a line on the CMD that corresponds to a set of models at constant chemical composition and age, parameterized by initial mass. Here, we define a generalized isochrone as the cloud of points in observable space that corresponds to the full range of our independent model parameters---mass, rotation, and orientation---restricted to a particular age $t$ and composition ${\rm [M/H]_M}$. In this context, equivalent evolutionary phase (${\rm EEP}$) can parametrize isochrones instead of initial mass \citep{Dotter_2016ApJS}. For a point on an isochrone with some initial mass $M_{\rm i}$ and initial rotation rate $\omega_{\rm Mi}$ that translate to some EEP, the point closest in observable space on a neighboring isochrone is approximately the one with the same EEP, not the one with the same mass. Accordingly, when we intepolate between isochrones, i.e.,~in age $t$, we fix EEP and $\omega_{\rm Mi}$ instead of $M_{\rm i}$ and $\omega_{\rm Mi}$.  This recipe could have been complicated by the fact that several values of $M_{\rm i}$ can correspond to the same ${\rm EEP}$ at a given combination of $\omega_{\rm Mi}$ and $t$. However, none of the models we utilize exhibit this behavior.

Here and elsewhere in this article, interpolation is linear unless stated otherwise. Furthermore, all interpolation and integration that involves $t$, luminosity $L$, and Eddington ratio $L/L_{\rm Edd}$ uses the logarithms of these variables. 

\subsection{Rotational Speed Conversion} \label{rot_conv}

The radius used to compute the dimensionless rotation speed $\omega_{\rm M}$ in \texttt{MIST} is a volume-averaged quantity. Because a rapidly rotating star expands in its equatorial regions,  $\omega_{\rm M}$ of unity does not correspond to the critical angular speed where the stellar equator becomes unbound.

Accordingly, in addition to $\omega_{\rm M}$ and average radius $R_{\rm M}$, we consider dimensionless rotational speed $\omega \equiv \Omega / \Omega_{\rm K} \in [0, 1]$ and equatorial radius $R_{\rm e}$. Here, $\Omega_{\rm K}$ is the Keplerian limit on $\Omega$, i.e., the rotational speed at which a star with mass $M$ and equatorial radius $R_{\rm e}$ would start to break up due to the centrifugal effect. Under the assumption that all mass is at the star's center -- i.e., the Roche model of mass distribution,
\begin{linenomath*}\begin{equation}
    \Omega_{\rm K} = \sqrt{\frac{G M}{R_{\rm e}^3}}.
    \label{eq:oK}
\end{equation}\end{linenomath*}

The Roche model admits an analytic expression for a normalized radial cylindrical coordinate of the stellar surface $\tilde{r}$ in terms of a normalized vertical cylindrical coordinate $\tilde{z}$ \citep[][henceforth LB20]{Lipatov_2020ApJ}. We now use that expression to derive a conversion between $\omega_{\rm M}$ and $\omega$ under this model of mass distribution. To start, we define a star's dimensionless volume as
\begin{linenomath*}\begin{equation}
    \tilde{V} \equiv \frac{3}{4 \pi}\frac{V}{R_{\rm e}^3}
\end{equation}\end{linenomath*}
and express it in terms of an integral in dimensionless cylindrical coordinates:
\begin{linenomath*}\begin{equation}
    \tilde{V}(\omega) = \frac{3}{4 \pi} \frac{2}{f} \int_0^1 \pi\, \tilde{r}(\tilde{z})^2\,d\tilde{z},
    \label{eq:Vtilde_int}
\end{equation}\end{linenomath*}
where $\tilde{r} \equiv r / R_{\rm e}$, $\tilde{z} \equiv z / R_{\rm p}$, $R_{\rm p}$ is the polar radius, and $f \equiv R_{\rm e} / R_{\rm p}$, as defined in \citetalias{Lipatov_2020ApJ}. Here, $f$ and $\tilde{r}(\tilde{z})$, and therefore $\tilde{V}$, are functions of $\omega$. We compute $\tilde{V}$ on a fine grid of $\omega$ values using Equation \eqref{eq:Vtilde_int}, the expression for $\tilde{r}(\tilde{z})$ in \citetalias{Lipatov_2020ApJ}, and the composite trapezoidal rule. We then perform cubic interpolation to obtain $\tilde{V}(\omega)$. Dividing Equation \eqref{eq:Vmist} by $R_{\rm e}^3$, we also have 
\begin{linenomath*}\begin{equation}
    \tilde{V} = \left(\frac{R_{\rm M}}{R_{\rm e}}\right)^3,
    \label{eq:Vtilde_rat}
\end{equation}\end{linenomath*}
so that
\begin{linenomath*}\begin{equation}
    R_{\rm e} = \frac{R_{\rm M}}{\sqrt[3]{\tilde{V}}}.
\end{equation}\end{linenomath*}

The definitions of $\omega$ and $\omega_{\rm M}$, in addition to Equations \eqref{eq:oMcrit}, \eqref{eq:oK}, and \eqref{eq:Vtilde_rat}, yield
\begin{linenomath*}\begin{equation}
    \tilde{V}(\omega) \times \omega^2 = \omega_{\rm M}^2 \left(1 - \frac{L}{L_{\rm Edd}}\right) \equiv \omega'^{2}_{\rm M}.
    \label{eq:omega_eq}
\end{equation}\end{linenomath*}

\begin{figure}[ht]
\includegraphics[width=\linewidth]{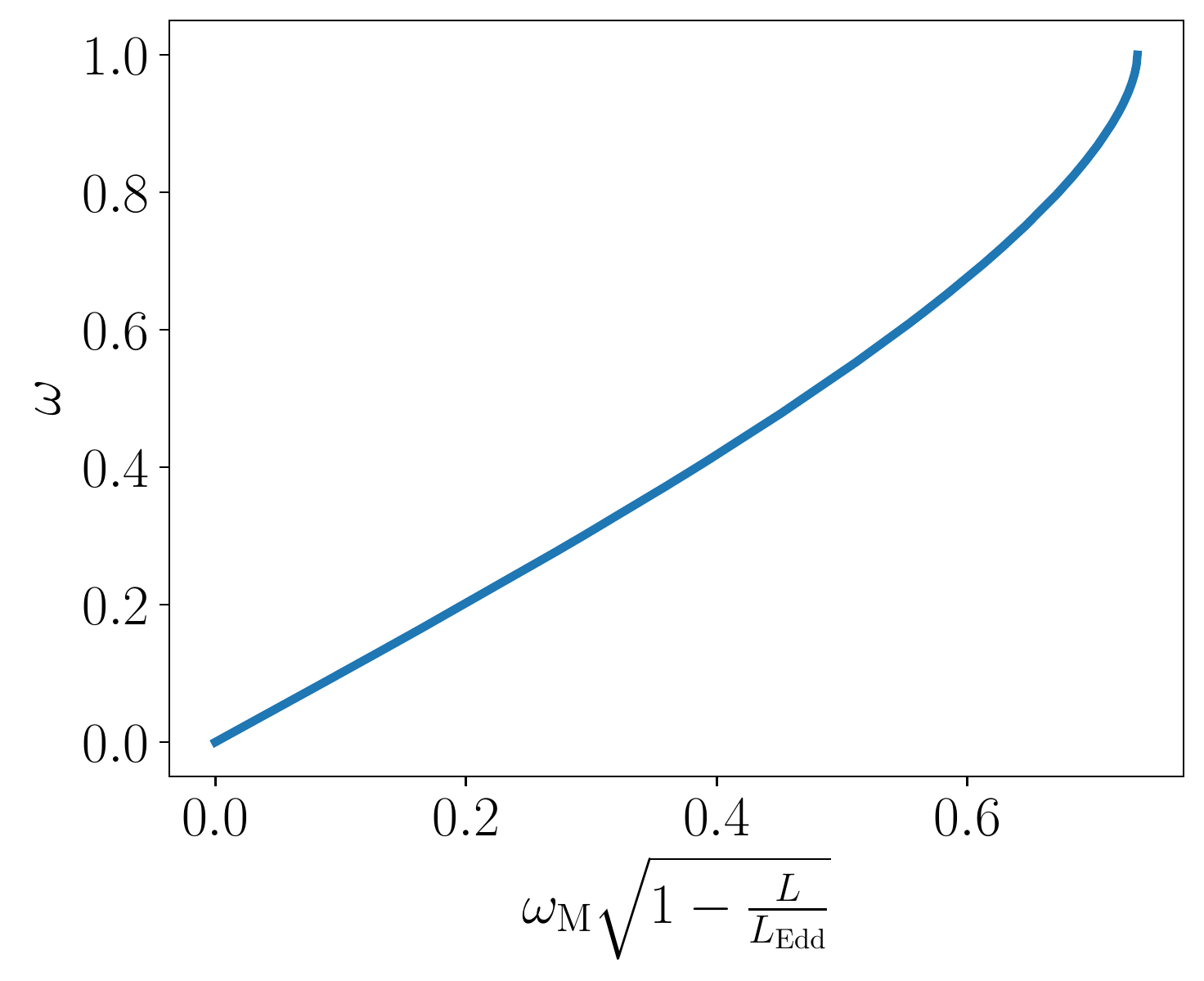}
\caption{Relationship between the proportion of Keplerian rotational limit $\omega$ and a quantity related to the \texttt{MIST} dimensionless rotational velocity $\omega_{\rm M}$.}
\label{fig:omega}
\end{figure}

At $\omega = 0$, rotation doesn't deform the star, so that $R_{\rm M}$, defined in Equation \eqref{eq:Vmist}, is equal to the equatorial radius $R_{\rm e}$. As $\omega$ increases, rotational deformation causes $R_{\rm M} / R_{\rm e}$ to decrease. Thus, according to Equations \eqref{eq:Vtilde_rat} and \eqref{eq:omega_eq}, $\omega'_{\rm M} = 0$ when $\omega = 0$ and $\omega'_{\rm M} / \omega$ decreases from one as $\omega$ increases from zero. When $\omega$ reaches one, so that $\Omega$ is at the Keplerian limit, $\omega'_{\rm M} = 0.7356$, which implies a shape so non-spherical that $R_{\rm M} / R_{\rm e} = (0.7356)^{2/3} = 0.8149$. We solve Equation \eqref{eq:omega_eq} numerically to obtain $\omega(\omega'_{\rm M})$. Figure \ref{fig:omega} presents the result, a monotonically increasing function. We use it to calculate $\omega$ from $\omega_{\rm M}$ and $L / L_{\rm Edd}$.

\texttt{MIST} does not provide the initial value of Eddington ratio $L / L_{\rm Edd}$ for all of its models. However, it does inform that a non-rotating model with initial mass $M_{\rm i} = 2 M_\odot$ has $L / L_{\rm Edd} = 0.017$ at zero age main sequence (ZAMS). All \texttt{MIST} models we use are less massive and most have significantly smaller $L / L_{\rm Edd}$. Therefore, in the case of initial rotational speeds, we set $L / L_{\rm Edd} = 0$ and solve
\begin{linenomath*}\begin{equation}
    \tilde{V}(\omega_{\rm i}) \times \omega_{\rm i}^2 = \omega_{\rm Mi}^2
    \label{eq:omega0_eq}
\end{equation}\end{linenomath*}
to find $\omega_{\rm i}$, the initial value of $\omega$. This corresponds to the dependence in Figure \ref{fig:omega}, with $\omega_{\rm Mi}$ and $\omega_{\rm i}$ re-labelling the horizontal and vertical axes, respectively. The largest $\omega_{\rm Mi}$ below the above-mentioned Keplerian limit of 0.7356 is 0.7 in the \texttt{MIST} models. Setting $\omega_{\rm Mi}$ to 0.7 in Equation \eqref{eq:omega0_eq} yields $\omega_{\rm i} = 0.8590$. This is the maximum $\omega_{\rm i}$ in the present analysis, since we do not extrapolate to higher values of this parameter. Some of the stars in our data set may possess $\omega_{\rm i} > 0.8590$. 
In Section \ref{discussion}, we discuss the implications of the rotational speed limit in our analysis on the quantitative results and suggest the limit's increase as an important future modification to the \texttt{MIST} library. 

\subsection{Synthetic Magnitudes} \label{synth_mag}

In Sections \ref{evolution} and \ref{rot_conv}, we mention present-day parameters that determine a star's magnitude -- its mass $M$, luminosity $L$, average radius $R_{\rm M}$, and \texttt{MIST}'s dimensionless rotational velocity $\omega_{\rm M}$. We also describe a procedure that converts $R_{\rm M}$ and $\omega_{\rm M}$ to equatorial radius $R_{\rm e}$ and another kind of dimensionless rotational velocity $\omega$. In this section, we describe a procedure that yields stellar magnitudes from $M$, $L$, $R_{\rm e}$, $\omega$, and rotational axis inclination $i$.

\subsubsection{Model Parameters for Magnitude Calculation} \label{param_mag}

\citetalias{Lipatov_2020ApJ} introduced \texttt{PARS} - Paint the Atmospheres of Rotating Stars, a program that computes theoretical magnitudes of a rotating star in a given telescope filter. The program is based on a model of internal energy flux due to \citet{EspinosaLara_2011A&A} and \texttt{ATLAS9}, a library of stellar atmosphere models due to \citet{castelli_2004}. 
Our present work necessitates the computation of magnitudes for many stellar models, so that separately employing \texttt{PARS} to compute the magnitude of each would be too slow. Accordingly we aim to interpolate magnitude on a grid of \texttt{PARS} models instead.

\texttt{PARS}'s input stellar parameters are $L$, $M$, $R_{\rm e}$, $\omega$,  $i$, and metallicity ${\rm [M/H]_P}$. Here,
\begin{linenomath*}\begin{equation}
    {\rm [M/H]_P} = \log{\frac{Z}{X}} - \log{\frac{Z_{\rm \odot, P}}{X_\odot}},
\label{eq:metP}
\end{equation}\end{linenomath*}
where $Z$, $X$ and $X_\odot$ are defined as in Equation \eqref{eq:metM}, $Z_{\rm \odot, P} = 0.01886$ is an estimate of the protosolar metal mass fraction \citep{Anders_1989GeCoA}, and the available ${\rm [M/H]_P}$ values are the same as in \texttt{ATLAS9} \citep{castelli_2004}. Subtracting Equation \eqref{eq:metM} from Equation \eqref{eq:metP}, we get
\begin{linenomath*}\begin{equation}
\begin{split}
    {\rm [M/H]_P} - {\rm [M/H]_M} & = \log{Z_{\rm \odot, M}} - \log{Z_{\rm \odot, P}} \\ & = -0.1233.
\end{split}
\label{eq:met_diff}
\end{equation}\end{linenomath*}
Mapping between \texttt{MIST} and \texttt{PARS} models according to Equation \eqref{eq:met_diff} ensures that metal mass fraction $Z$ remains the same, despite the differences in the protosolar mass fraction between the two model libraries.

In order to speed up interpolation on the \texttt{PARS} grid, we wish to decrease its dimensionality. Towards this end, we derive parameters $\gamma$ and $\tau$, which are similar to surface gravity and effective temperature, respectively. We will show that one can interpolate in $\gamma$ and $\tau$ at fixed equatorial radius $R_{\rm e} = R_\odot$ instead of interpolating in $M$, $L$, and $R_{\rm e}$, then add a function of $R_{\rm e} / R_\odot$ to the resulting magnitudes. Parameter $\gamma$ is the logarithm of the gravitational acceleration at the equator in cgs units,
\begin{linenomath*}\begin{equation}
    \gamma \equiv \log{\left(\frac{G M}{R_{\rm e}^2}\middle/ {\rm cm\,s^{-2}}\right)}.
    \label{eq:gamma}
\end{equation}\end{linenomath*}
Parameter $\tau$ is the effective temperature of a spherically symmetric star with luminosity $L$ and radius $R_{\rm e}$,
\begin{linenomath*}\begin{equation}
    \tau \equiv \left(\frac{L}{4 \pi \sigma_{\rm SB} R_{\rm e}^2}\right)^{1/4}.
    \label{eq:tau}
\end{equation}\end{linenomath*}
Quantities $G$ and $\sigma_{\rm SB}$ are the gravitational and Stefan-Boltzmann constants, respectively. Here and in the rest of this work, logarithms of physical quantities are base-ten, while those of likelihood and probability functions are natural, unless stated otherwise. 

\texttt{PARS} adds up luminous power over the set of infinitesimal patches that make up the visible stellar surface, taking into account the viewing angle of each patch. 
Stars of different size but constant $\omega$ and orientation look the same apart from an overall scale--their angular extent on the sky.
This allows us to define a normalized $\omega$-surface, which has unit equatorial radius and depends solely on dimensionless rotational velocity $\omega$. Consequently, we can write down the power emanating from a star at a given wavelength as a product of $R_{\rm e}^2$ and an integral of the star's intensity over the patches on such a normalized surface (see Equation 18 in \citetalias{Lipatov_2020ApJ}).

In addition to $\omega$ and $i$, $\gamma$ and $\tau$ determine the above-mentioned flux from a normalized surface, as follows. The intensity of each surface patch depends on its viewing angle, its temperature $T$, and its value of $g$ -- the combined gravitational and centrifugal acceleration. Equation (36) in \citetalias{Lipatov_2020ApJ} writes $g$ as a product of $10^\gamma$ and a function of the patch's location on the $\omega$-surface. On the other hand, Equation (31) in \citet{EspinosaLara_2011A&A} expresses $T$ as a product of $\tau$ and another function of the $\omega$-surface location. Thus, luminous power can be computed from $\gamma$, $\tau$, $\omega$ and $i$, up to a factor of $R_{\rm e}^2$.

Consequently, to compute the magnitude of a stellar model from \texttt{PARS}'s input parameters, we calculate $\gamma$ and $\tau$ from Equations \eqref{eq:gamma} and \eqref{eq:tau}, interpolate magnitude on the \texttt{PARS} grid, and subtract $5\times\log{R_{\rm e} / R_\odot}$, which is equivalent to multiplying luminous power by $\left(R_{\rm e} / R_\odot\right)^2$.

\subsubsection{Model Grid for Magnitude Calculation} \label{pars_grid}

We compute the \texttt{PARS} grid---multi-band synthetic photometry on a grid of $\tau$, $\omega$, $i$ and $\gamma$---for ${\rm [M/H]_M} = -0.45$, extinction parameter $A_{\rm V} = 0.263$, distance modulus $\mu = 18.45$, and $R_{\rm e} = R_\odot$. Our value for $\mu$ is the same as in \citetalias{Kamann_2020MNRAS} and the value for $A_{\rm V}$ is based on isochrone fits in \citetalias{Kamann_2020MNRAS}. We do not include the uncertainties for ${\rm [M/H]_M}$, $A_{\rm V}$, or $\mu$ in our analysis, since the influence of these uncertainties on our results should be significantly less than that of the uncertainties in the stellar evolution model (see Section \ref{discussion}). The extinction curve is from \citet{Fitzpatrick_1999PASP}, with $R_V = 3.1$. The magnitudes we calculate are $m_{435}$, $m_{555}$, and $m_{814}$. Here, we first convert ${\rm [M/H]_M}$ to ${\rm [M/H]_P}$ via Equation \eqref{eq:met_diff}, then interpolate between the available values of ${\rm [M/H]_P}$.  

The range of the $\tau$ and $\gamma$ portion of the grid is the same as that of temperature and gravity in \texttt{ATLAS9} plane-parallel atmosphere models \citep[Table 1 of][]{castelli_2004}: 3,$500\,{\rm K}\le\tau\le50$,$000\,{\rm K}$ and $0.0\le\gamma\le5.0$. The two grids have similar model coverage since, for example, the surface of a star with parameter $\tau$ has temperatures in the neighborhood of $\tau$. The spacing between adjacent $\tau$ values is $16\,{\rm K}$ below $4500\,{\rm K}$, $31\,{\rm K}$ between $4500\,{\rm K}$ and $6200\,{\rm K}$, and $63\,{\rm K}$ above $6200\,{\rm K}$. The spacing between adjacent $\gamma$ values is 0.5. The $\omega$ grid extends from 0 to 0.95, with a spacing of 0.05. The $i$ grid has 20 values, equally spaced between 0 and $\pi / 2$. Overall, there are close to 1 million models on the $\tau$, $\omega$, $i$ and $\gamma$ grid.

\begin{figure}[ht]
\includegraphics[width=\linewidth]{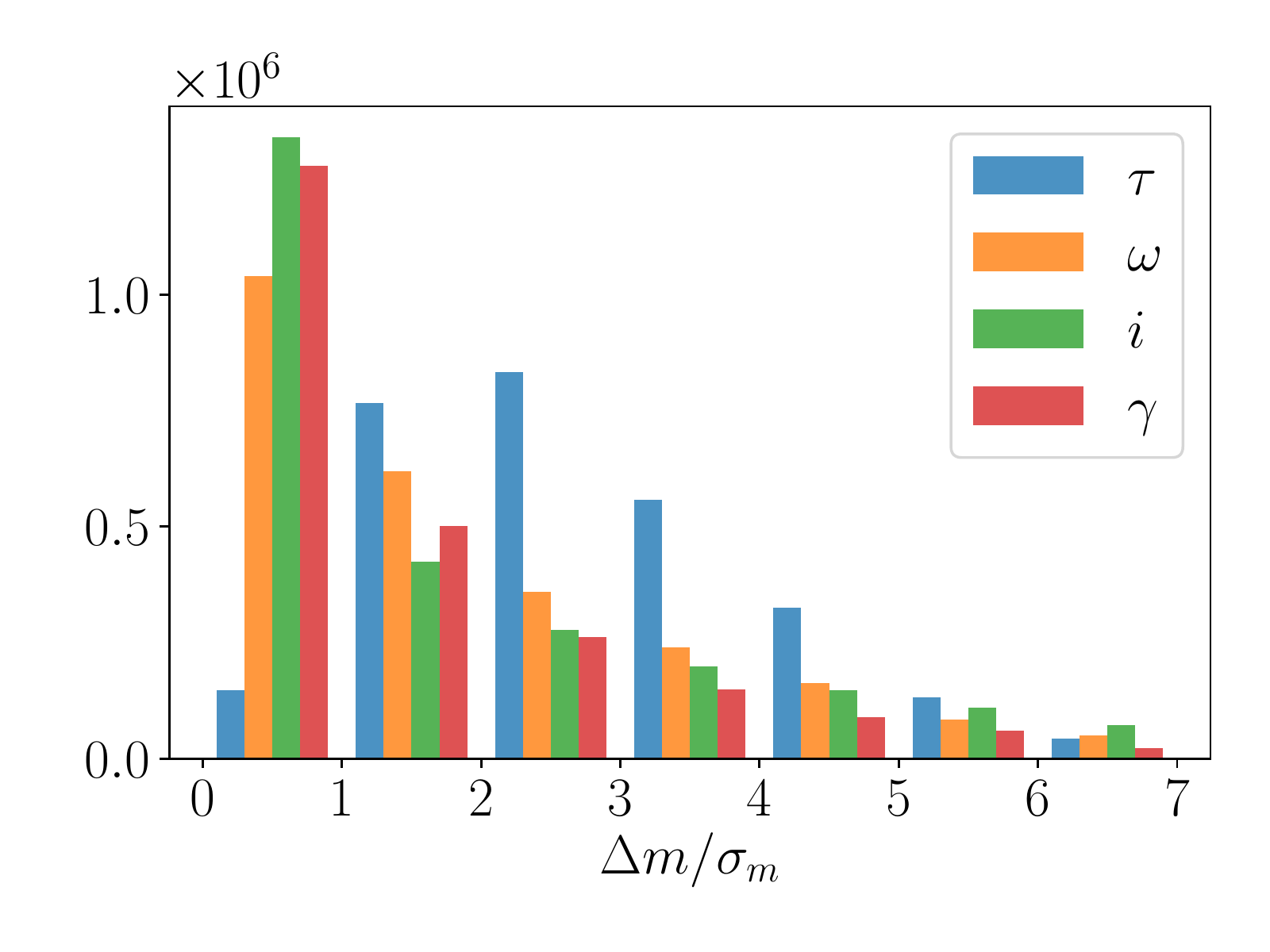}
\caption{Histograms of the magnitude differences in units of minimum error ($\sigma_m = 0.01\,{\rm mag}$) between adjacent models on the entire \texttt{PARS} grid in Section \ref{pars_grid}. Differences are taken along each of the four grid dimensions, indicated in the legend. Most differences are only a few $\sigma_m$. Assuming that the magnitude function does not deviate significantly from linearity on this scale, interpolation on the \texttt{PARS} grid should be very accurate.}
\label{fig:pars_diff}
\end{figure}

To assess the accuracy of interpolations within our model grid, we take each $\texttt{PARS}$ grid parameter and calculate the magnitude differences between any two adjacent values of that parameter, with all other parameters fixed. Most of these differences are only a few minimum magnitude errors $\sigma_m$, or a few hundredths of a magnitude, as demonstrated in Figure \ref{fig:pars_diff}. Assuming that the magnitude function does not deviate significantly from linearity on the scale of a few $\sigma_m$, interpolation on the \texttt{PARS} grid should be very accurate.

\subsection{Calculation of Observables} \label{calc_obs}

The previous section describes the calculation of magnitudes for individual stars.
We also allow for the possibility that a star in our data set is actually an unresolved, non-interacting binary system, consisting of a rotating primary and a non-rotating companion that do not eclipse each other. Allowing for the rotation of the secondary would increase the dimensionality of model space from 5 to 7.  At the same time, this change would only have an effect for stars whose companions lie above the Kraft break, around 20\% of binaries, assuming a turnoff mass of $\approx$1.6$M_\odot$ and a flat companion mass function.  The effects of rotation would further be subdominant to those of binarity even for these stars.  The binary fraction of the cluster is estimated to be $\approx$6\% from independent measurements.  With $\sim$1000 stars on the turnoff above the Kraft break, we therefore expect to be neglecting a subdominant effect for $\sim$10 stars (comparable to the effect of our neglect of the subgiant branch).  

We assume that the companion's initial mass $M_{\rm Ci}$ does not exceed the primary's initial mass $M_{\rm i}$, so that the binary mass ratio $r \equiv M_{\rm Ci} / M_{\rm i} \in [0, 1]$. We combine the magnitude of a primary $m_{\rm p}$ with that of its companion $m_{\rm c}$ as follows:
\begin{linenomath*}\begin{equation} \label{eq:binary_combine}
    m = -2.5\, \log{\left(10^{-m_{\rm p}/2.5} + 10^{-m_{\rm c}/2.5}\right)}.
\end{equation}\end{linenomath*}

\begin{figure}[ht]
\vspace{10pt}
\includegraphics[width=\linewidth]{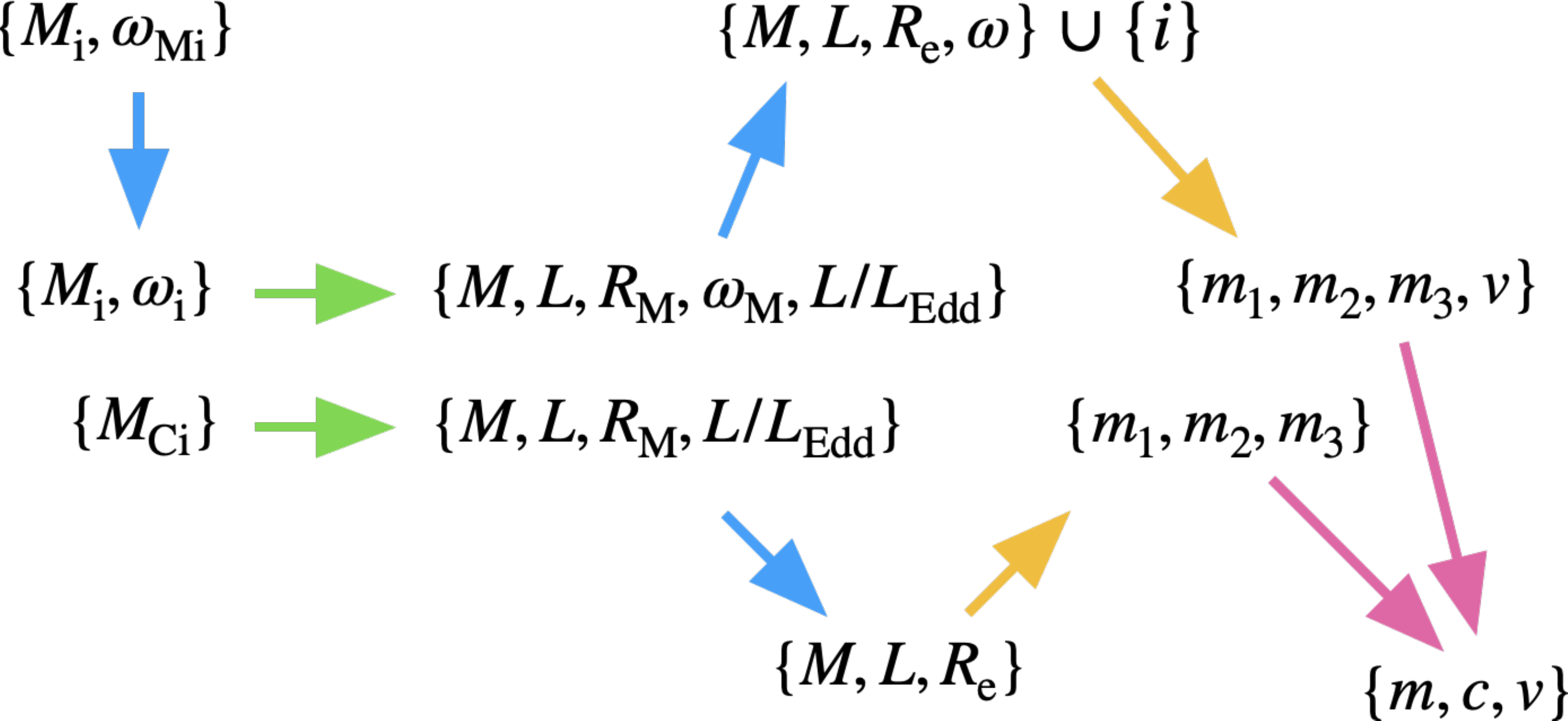}
\caption{Schematic of our procedure for the calculation of the observables $\{m, c, v\}$ at constant age $t$. The top branch of the schematic pertains to the rotating primary, the bottom -- to the non-rotating companion. The combined models are parametrized by initial mass $M_{\rm i}$, binary mass ratio $r \equiv M_{\rm Ci} / M_{\rm i}$, initial rotational speed $\omega_{\rm i}$, and rotational axis inclination $i$. Blue arrows indicate conversion of rotational speeds and radii (Section \ref{rot_conv}), green -- interpolation in the \texttt{MIST} grid (Section \ref{evolution}), yellow -- interpolation in the \texttt{PARS} grid (Section \ref{synth_mag}), and pink -- combination of the primary's and its companion's magnitudes.}
\label{fig:observables}
\end{figure}

We now define the initial stellar parameters ${\bm \theta'} \equiv (M_{\rm i}, r, \omega_{\rm i}, i)$, as well as the full set of parameters ${\bm \theta} \equiv (M_{\rm i}, r, \omega_{\rm i}, i, t)$, where $\omega_{\rm i}$ is the initial dimensionless rotation rate of the primary and $i$ is the primary's inclination.  
We wish to obtain the observables on grids of ${\bm \theta}$. Towards this end, we first interpolate dependent model parameters between original \texttt{MIST} ages at constant initial rotation rate $\omega_{\rm Mi}$ and constant equivalent evolutionary phase, ${\rm EEP}$ (see the latter portions of Section \ref{evolution}). The rest of the procedure, outlined in Figure \ref{fig:observables}, happens at constant age. It starts with the conversion of $\omega_{\rm Mi}$ to $\omega_{\rm i}$ (see Section \ref{rot_conv}), proceeds to the interpolation of model parameters in $M_{\rm i}$ and $\omega_{\rm i}$, includes the interpolation of magnitudes in the \texttt{PARS} grid, and concludes with the combination of the primary's and the companion's magnitudes. Figure \ref{fig:mist} presents the observables that result from the procedure outlined in this subsection for a subset of unary (single, non-binary) stars on the original \texttt{MIST} grid. Here, $v_{\rm e}\sin{i}$ is calculated from a model's current parameters $\omega$, $M$, $R_{\rm e}$, and $i$ as
\begin{equation}
    v_{\rm e}\sin{i} = \Omega R_{\rm e} \sin{i} = \omega \sqrt{\frac{G M}{R_{\rm e}}} \sin{i},
\end{equation}
where we have made use of the expression for Keplerian velocity $\Omega_{\rm K}$ in Equation \eqref{eq:oK} and the definition $\omega\equiv\Omega/\Omega_{\rm K}$.

\section{Probabilities of Observables} \label{probabilities}

Section \ref{stellar} describes the procedures that map stellar model parameters to observable space. The \texttt{MIST} model grid is discrete, with substantial separations in mass and rotation rate between neighboring models.  Figure \ref{fig:mist} shows the discrete colors and magnitudes corresponding to the \texttt{MIST} grid at two fixed inclinations. Na\"ively, such discrete distributions suggest zero probability of stars existing between the discrete locations. To use these observables for statistical inference, we must instead construct {\it continuous} distributions in color-magnitude space.
Colors and magnitudes can change steeply with the initial mass of a stellar model, especially as a star approaches the end of its main sequence life.  For combined accuracy and computational efficiency, we seek a grid of model parameters ${\bm \theta}$ that maps onto a nearly uniform grid in observable space ${\bm x}$.  This grid will be coarse in ${\bm \theta}$ near parameters for which observables change slowly, and fine where observables change sharply.  In this  section, we state our priors on model parameters ${\bm \theta}$, describe the calculation of a suitable ${\bm \theta}$ grid, our subsequent calculation of continuous distributions in color-magnitude-$v\sin i$ space, and finally the integrations over these distributions that allow us to interpret them as probability densities.

\subsection{Cluster Model} \label{cluster}

In this section, we state our prior distributions on stellar parameters ${\bm \theta}$. The star-by-star posterior distributions that we obtain are the product of these priors and the likelihood, integrated to unit probability.  Some of the priors are themselves parametrized by what are more properly called hyperparameters, i.e., parameters associated with the cluster as a whole. We adopt parametrized descriptions of the rotation rate distribution and the age distribution and fit for those hyperparameters in later sections.  Here, we begin by describing our model for the distribution of initial rotation rates before discussing our models and priors on binarity, mass, and age.

We wish to construct a model of the rotational distribution that has reasonable and sufficient degrees of freedom. \citetalias{Kamann_2020MNRAS} find evidence for a bimodal $v\equiv v_{\rm e}\sin{i}$ distribution in NGC 1846, with about 55\% of the observed stars clustered near $v = 140\,{\rm km\,s^{-1}}$ and the rest -- near $v = 60\,{\rm km\,s^{-1}}$. There is additional evidence of bimodal rotational distributions in clusters \citep{DAntona_2017NatAs,Gossage_2019ApJ}. We add an extra degree of freedom and assume three rotational populations: one with a maximum probability density at zero rotation, one with maximum density at critical rotation, and one with an intermediate maximum-probability rotation. We assume that each population has a Gaussian distribution of initial rotation rates, truncated at $\omega=0$ and $\omega=1$. 

We choose parameters for the three Gaussians so that their best-fit amplitudes result in all three distributions contributing a nonzero fraction of the cluster's stars.  Many sets of parameters result in all, or nearly all, stars being attributed to only two of these rotational populations.  Future work will explore the robustness of our results to different parametrizations of the rotation rates and to changes in the stellar models.  For the present work, we use standard deviations of 0.6 and 0.15 in $\omega$ for the slow (mean $\omega=0$) and fast (mean $\omega=1$) rotating populations.  We then find the intermediate rotation rate where the slow and fast rotating populations contribute equally.  We adopt this rotation rate, $\omega=0.696$, for our intermediate rotators, with a narrow standard deviation of 0.05.  The {\tt MIST} model library only extends to $\omega = 0.86$; we use these models for all stars with $0.86 < \omega < 1$.

Our choice of rotational distribution allows for distinct populations of rotators that concentrate at critical, zero, and intermediate rotation, in accordance with the qualitative evidence of such concentrations \citep{Kamann_2020MNRAS,DAntona_2017NatAs,Gossage_2019ApJ}. 

Multiplicity of stellar systems significantly affects the CMD of a cluster. Similarly to rotation, it can alter both the evolutionary trajectory of a star system (through binary evolution) and its present-day spectrum (by combining the light of the two stars). In the present analysis we include unresolved binarity (a single point source comprising the light of two stars) but neglect the effects of binary evolution. A radial velocity variation technique in Section 4.4 of \citetalias{Kamann_2020MNRAS} \citep[see also][]{Giesers_2019A&A} estimates that the unresolved binary fraction of NGC 1846 is $\sim6\%$. This is similar to estimates of unresolved binary fractions in Galactic globular clusters \citep{Milone_2012A&A}. Although \citetalias{Kamann_2020MNRAS}'s binary fraction for NGC 1846 is lower than the estimate of this parameter for the LMC by \citet{Moe_2013ApJ}, at least part of this difference may be due to the fact that the latter authors work with field stars as opposed to globular cluster stars.  \citetalias{Kamann_2020MNRAS} argue that the small binary fraction that they find cannot lead to the much larger fraction of slowly rotating stars in NGC 1846, supporting the idea that binary interactions are generally unlikely to play a significant role in the evolution of stellar rotation rates in this cluster.  

We therefore treat each star as either single or as an unresolved binary, with $b$ denoting the hyperparameter for the binary fraction. 
 
For the present analysis, we adopt a uniform prior on the binary fraction $b$ and the simple uniform prior $r \sim U(0, 1)$ for the binary mass ratio, although there is some evidence of relative dearth in the middle of $r$'s range. Specifically, \citet{Raghavan_2010ApJS} say that the mass ratio distribution is approximately uniform for Solar type stars, with a bit of an excess towards equal masses.  Other recent papers suggest that the binary mass ratio prefers lower-mass companions, with a bit of an excess towards equal mass companions \citep{Moe_2013ApJ,Chulkov_2021MNRAS}.

We assume that the cluster stars have a lognormal distribution in age, with logarithmic mean age $\mu_t$ and logarithmic standard deviation $\sigma_t$.  A coeval cluster would have $\sigma_t = 0$, while a cluster with an age dispersion, as has been suggested for LMC clusters \citep[e.g.,][]{Goudfrooij_2011ApJ_4}, would have a significantly nonzero $\sigma_t$.  We adopt uniform priors on the hyperparameters $\mu_t$ and $\sigma_t$.  This favors younger ages, but given the few percent precision of the age that we derive for NGC 1846, it has a negligible effect on our results.

We adopt the Salpeter IMF, $\pi(M_{\rm i}) = M_{\rm i}^{-2.35}$, as the prior on the initial mass of the primaries, as well as a prior on inclination that corresponds to an isotropic distribution, $\pi(i) = \sin{i}$.

Finally, we  introduce $q$, the fraction of stars in the CMD that are described by our cluster model.  We assume that the rest of the stars, a fraction $1-q$, come   from a population of stars that we haven't modeled. This population could contain stars that are not in the cluster or stars that are in the cluster but aren't described by our stellar model---they exist in regions of the CMD that should be empty. For this background population,  we utilize a probability distribution that is uniform over observable space. Our overall cluster model, then, is parametrized by the hyperparameters ${\bm \phi} \equiv \{w_0, w_2, \mu_t, \sigma_t, b, q\}$.

\subsection{Probability Density for a Given Population} \label{prob_density}

We next aim to calculate the probability density of a star at each point ${\bm x}$ in observable space.  This is the convolution of the probability density of stars given by the stellar model with that particular star's error kernel.  The probability density without observational error would be the same for all stars, but non-uniform uncertainties in color, magnitude, and $v$ break this symmetry. 

We define the error kernel with width $\bm{\sigma_{{\bm x}{\bm p}}}$ for a set of observable deviations ${\bm \Delta x}$ as  
\begin{linenomath*}\begin{equation}
    G({\bm \Delta x}; \bm{\sigma_{{\bm x}{\bm p}}}) \equiv G(\Delta m; \sigma_{mp})\,G(\Delta c; \sigma_{cp})\,G(\Delta v; \sigma_{vp}),
\end{equation}\end{linenomath*}
where $G(\Delta y; \sigma)$ is the Gaussian distribution in $\Delta y$ with mean 0 and standard deviation $\sigma$, $p$ is the data point index, and the other subscript on the components of $\bm{\sigma_{{\bm x}{\bm p}}} \equiv (\sigma_{mp}, \sigma_{cp}, \sigma_{vp})$ specifies observable type. This subscript is either $m$ for magnitude, $c$ for color, or $v$ for $v_{\rm e}\sin{i}$. Thus, the  probability of a data point with observables ${\bm x_{\bm p}} \equiv (m_p, c_p, v_p)$, given stellar parameters ${\bm \theta}$, can be written as $G({\bm x_{\bm p}} - {\bm x}({\bm \theta}); {\bm \sigma_{{\bm x}{\bm p}}})$. Here, ${\bm x}({\bm \theta}) \equiv \left(m({\bm \theta}), c({\bm \theta}), v({\bm \theta})\right)$, and $m({\bm \theta})$, for example, is the magnitude of a star with parameters ${\bm \theta}$ according to the stellar model.

For each combination of rotational population $j$, multiplicity $k$, data point $p$, and age distribution parameters $\mu_t$ and $\sigma_t$, we wish to compute $\rho_{jkp}({\bm x_{\bm p}}; \mu_t, \sigma_t)$, the theoretical probability density evaluated at ${\bm x_{\bm p}}$, where
\begin{linenomath*}\begin{equation}
    \rho_{jkp}({\bm x}; \mu_t, \sigma_t) = \frac{1}{Z} \int {\rm d {\bm \theta}}\,\pi_{jk}({\bm \theta}; \mu_t, \sigma_t) \,\,G({\bm x} - {\bm x}({\bm \theta}); {\bm \sigma_{{\bm x}{\bm p}}}),
\label{eq:rho_jkp}
\end{equation}\end{linenomath*}
\begin{linenomath*}\begin{equation} \label{eq:dtheta}
    {\rm d}{\bm \theta'} = {\rm d}M_{\rm i}\,{\rm d}r\,{\rm d}\omega_{\rm i}\,{\rm d}i, \quad {\rm d}{\bm \theta} = {\rm d}{\bm \theta'}\,{\rm d}t,
\end{equation}\end{linenomath*}
\begin{linenomath*}\begin{equation} \label{eq:pi_thetap}
    \pi_{jk}\left({\bm \theta'}\right) = \pi_j\left(\omega_{\rm i}\right)\, \pi_k\left(r\right)\, \pi(M_{\rm i})\, \pi(i),
\end{equation}\end{linenomath*}
\begin{linenomath*}\begin{equation} \label{eq:pi_theta}
    \pi_{jk}({\bm \theta}; \mu_t, \sigma_t) = \bar{\pi}\left(t; \mu_t, \sigma_t\right)\, \pi_{jk}\left({\bm \theta'}\right),
\end{equation}\end{linenomath*}
the Gaussian $G({\bm \cdot})$ and the priors $\pi({\bm \cdot})$ on the different components of ${\bm \theta}$ are given in Section \ref{cluster}, and $k$ = 0 and 1 correspond to unary and binary populations, respectively. The integral is over all ${\bm \theta}$, though it is finite for a given set of observables ${\bm x}$, since $\pi({\bm \theta})$ is finite and the error kernel at ${\bm x}$ is non-zero on a finite volume of ${\bm \theta}$-space. Furthermore, the normalization constant $Z$ is chosen so that probability density $\rho_{jkp}({\bm x}; \mu_t, \sigma_t)$ integrates to one on our region of interest in ${\bm x}$. Equation \eqref{eq:rho_jkp} represents a five-dimensional integral for each star.  In the following sections, we describe our approach for evaluating this integral to an acceptable accuracy using as computationally efficient a method as possible.

\subsection{Stellar Model Grid Refinement} \label{model_refinement}

\begin{figure*}[ht]
\includegraphics[width=0.5\linewidth]{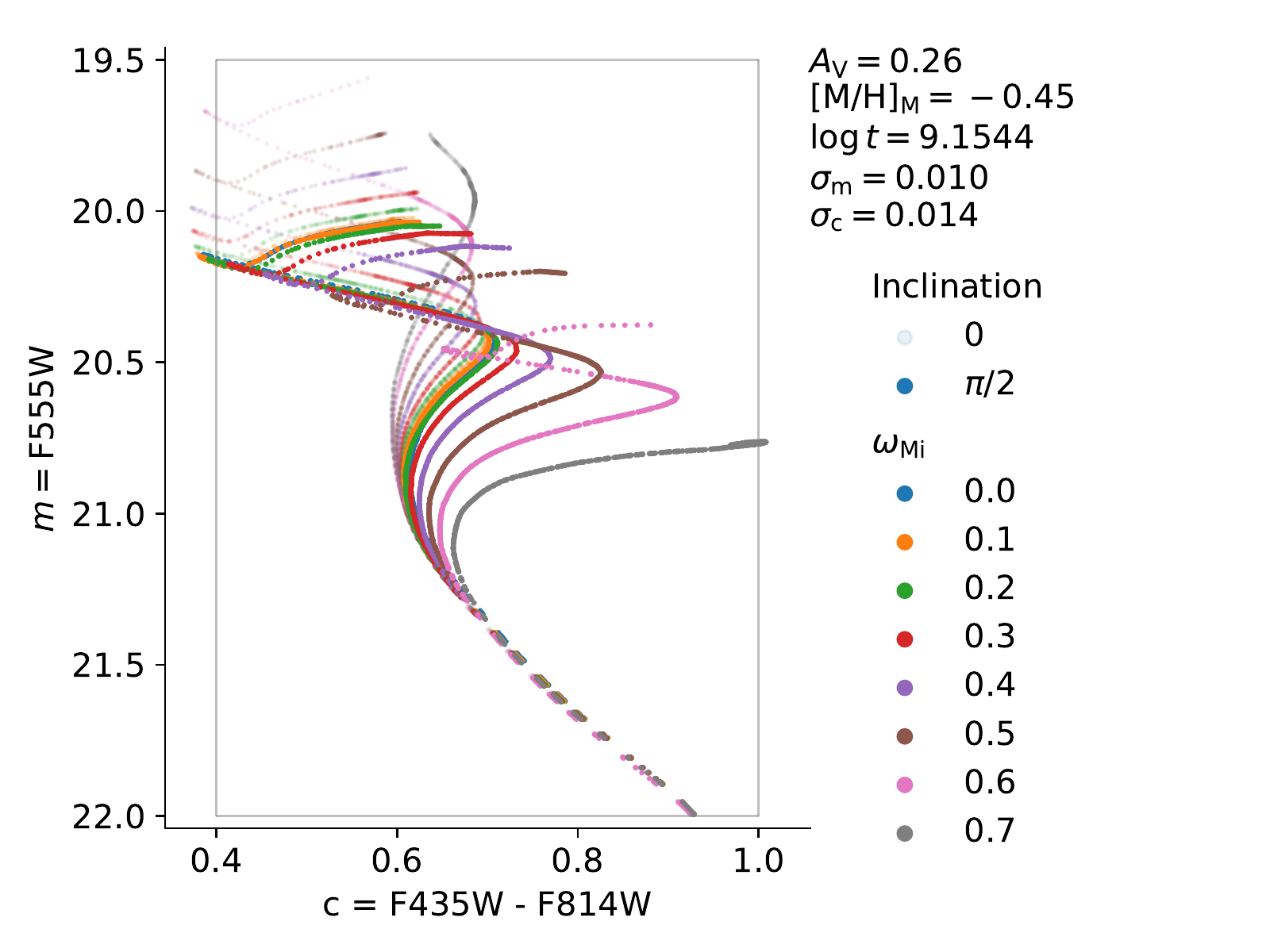}
\includegraphics[width=0.5\linewidth]{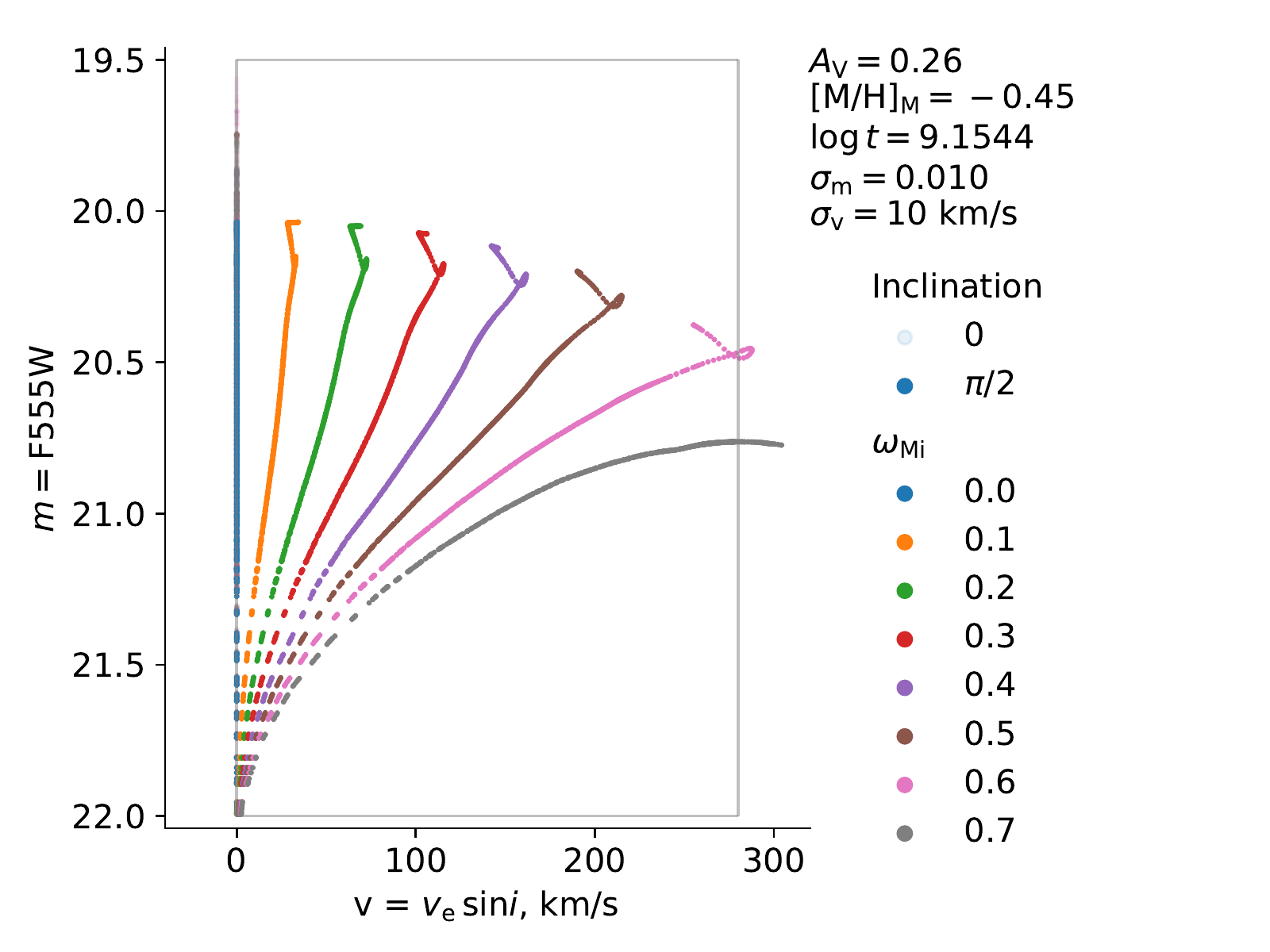}
\caption{\texttt{PARS} observables of the original \texttt{MIST} models at $\log{t} = 9.1544$ and two inclinations. At a given $\omega_{\rm Mi}$, decreasing $m$ generally corresponds to increasing $M_{\rm i}$. Marker size is approximately equal to $\sigma_m$. Spaces between models in the $m$ dimension are frequently larger than $\sigma_m$. The same can be said for the $c$ and $v$ dimensions, though the effect is most pronounced for $m$. Thus, these discrete models predict no stars in much of the empty space between the markers. This is in contrast with the underlying theory, which is continuous in $M_{\rm i}$ and $\omega_{\rm Mi}$, and thus does predict stars everywhere between the markers.}
\vspace{20pt}
\label{fig:mist}
\end{figure*}

The original \texttt{MIST} model grid in Section \ref{evolution} is too coarse in mass, age, and rotation rate to accurately integrate in Equation \eqref{eq:rho_jkp}. Figure \ref{fig:mist} shows the {\tt MIST} models at a particular age and composition.  These models should produce a continuous probability density in mass/color/$v_e \sin i$ space, but the discrete nature of the grid remains obvious.  
In Appendix \ref{appendix_refinement}, we motivate and describe our interpolation within the {\tt MIST} models, which generates a grid that is sufficiently fine to produce continuous probability densities.  Our approach balances the need to remove discretization artifacts with the need to keep the entire procedure computationally feasible.

The above-mentioned grid refinement procedure requires interpolating within the {\tt MIST} model grid.  We perform these interpolations---in mass, rotation, and age---by first converting mass to EEP as described in Section \ref{evolution}, then treating EEP, age, and rotation as the independent stellar parameters. This allows us to infer mass, equatorial radius, luminosity, and rotation from the {\tt MIST} grid, and to use these to interpolate within the {\tt PARS} grid via the equations of Section \ref{param_mag}.  
We numerically integrate according to Equation \eqref{eq:rho_jkp} on the resulting model grid. Figures \ref{fig:cmd} and \ref{fig:vmd} show that the resulting probability densities are free from artifacts of model discretization. In the following section, we describe our integration approach in detail.

\subsection{Integration Procedure} \label{integration}

In Sections \ref{prob_density} and \ref{model_refinement}, we state the integral that we wish to compute in model space, in order to obtain probability densities in observable space. We also outline the production of model grids that allow for accurate integration with minimum computational cost. In this section, we detail our integration procedure, which utilizes a number of additional techniques that ensure accuracy and speed. 

\subsubsection{Minimum-Error Densities} \label{minimum_error}

Equation \eqref{eq:rho_jkp} integrates over the 5-dimensional stellar parameter vector ${\bm \theta}$ to produce theoretical probability densities in observable space. Performing this integral successively on a grid of 3 rotational populations, 2 multiplicities, 2353 data points and a number of age prior parameter combinations is computationally prohibitive. To render the calculation tractable, we assume Gaussian errors and take advantage of the commutativity and associativity of the convolution. We first compute synthetic probability densities of observables (color, magnitude, projected rotational velocity) by integrating Equation \eqref{eq:rho_jkp} over five stellar parameters assuming a single set of uncertainties that we term the {\it minimum errors}. We may then obtain the integrals for a star with larger uncertainties from these integrals by a convolution in three observable dimensions.  By decreasing the dimensionality from five to three, and because a Gaussian falls so quickly to zero, this approach speeds the computation by orders of magnitude without sacrificing accuracy.

 Our formal approach begins by writing the convolution of two Gaussians with parameters $\{\mu_1, \sigma_1\}$ and $\{\mu_2, \sigma_2\}$ as another Gaussian with parameters $\{\mu_1 + \mu_2, \sqrt{\sigma_1^2 + \sigma_2^2}\}$. 
We compute Equation \eqref{eq:rho_jkp} for a fixed set of minimum observational uncertainties, which we take to be 0.01\,mag in magnitude, 0.014\,mag in color, and 10\,km\,s$^{-1}$ in projected rotational velocity.  We represent these minimum uncertainties by ${\bm \sigma_{\bm x}}$.  We then introduce

\begin{figure*}[ht]
\includegraphics[width=0.5\linewidth]{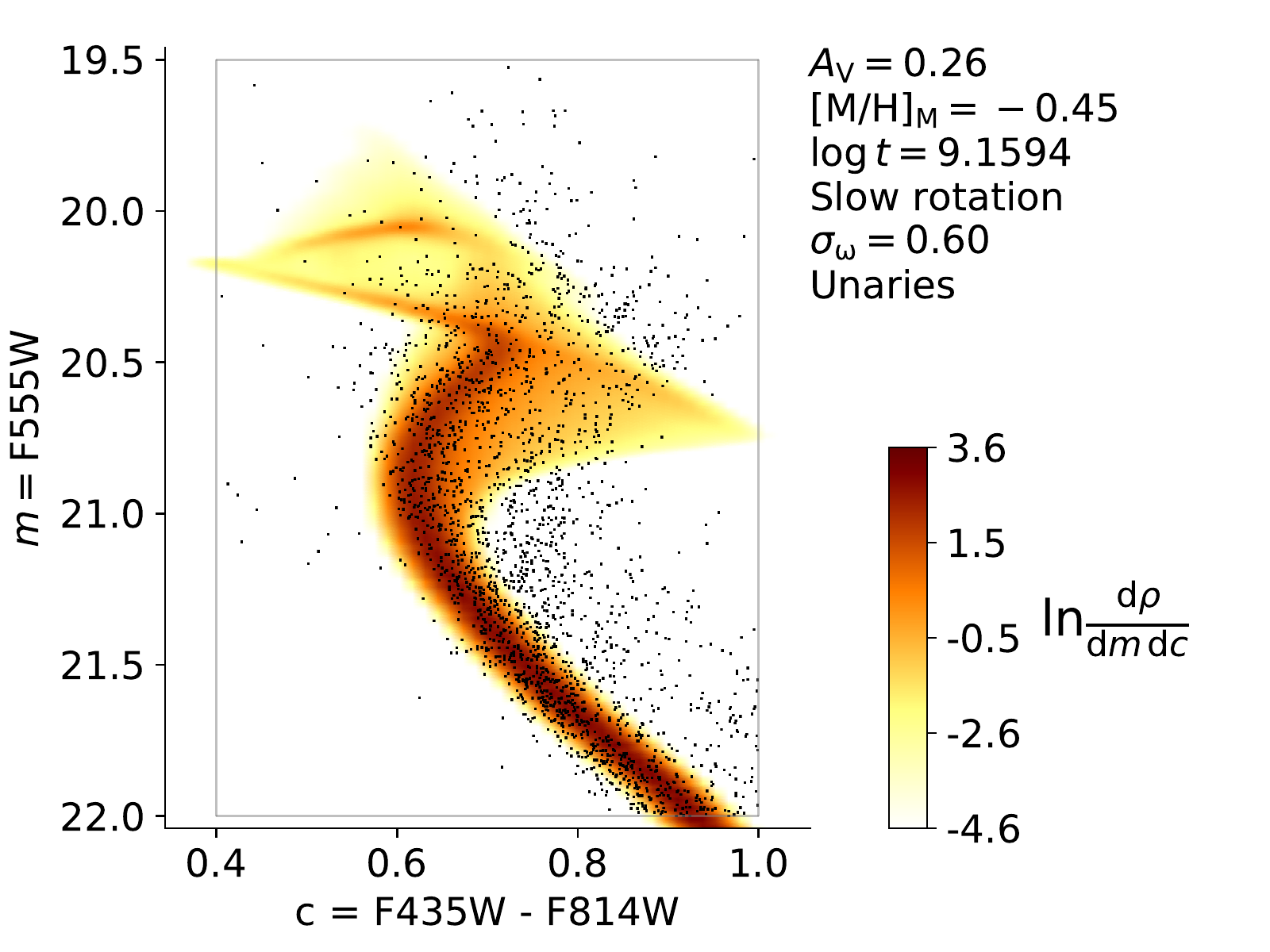}
\includegraphics[width=0.5\linewidth]{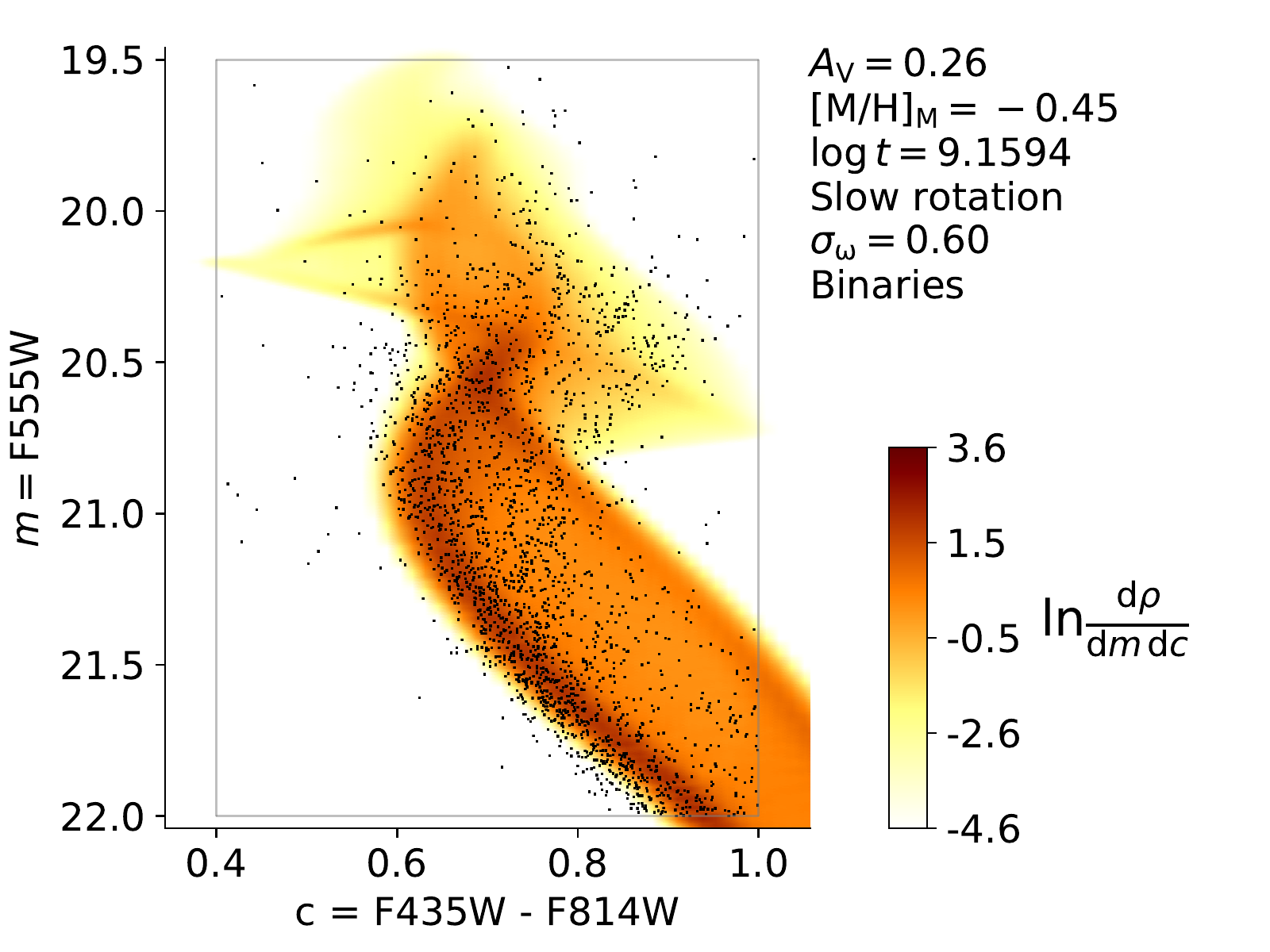}
\newline
\includegraphics[width=0.5\linewidth]{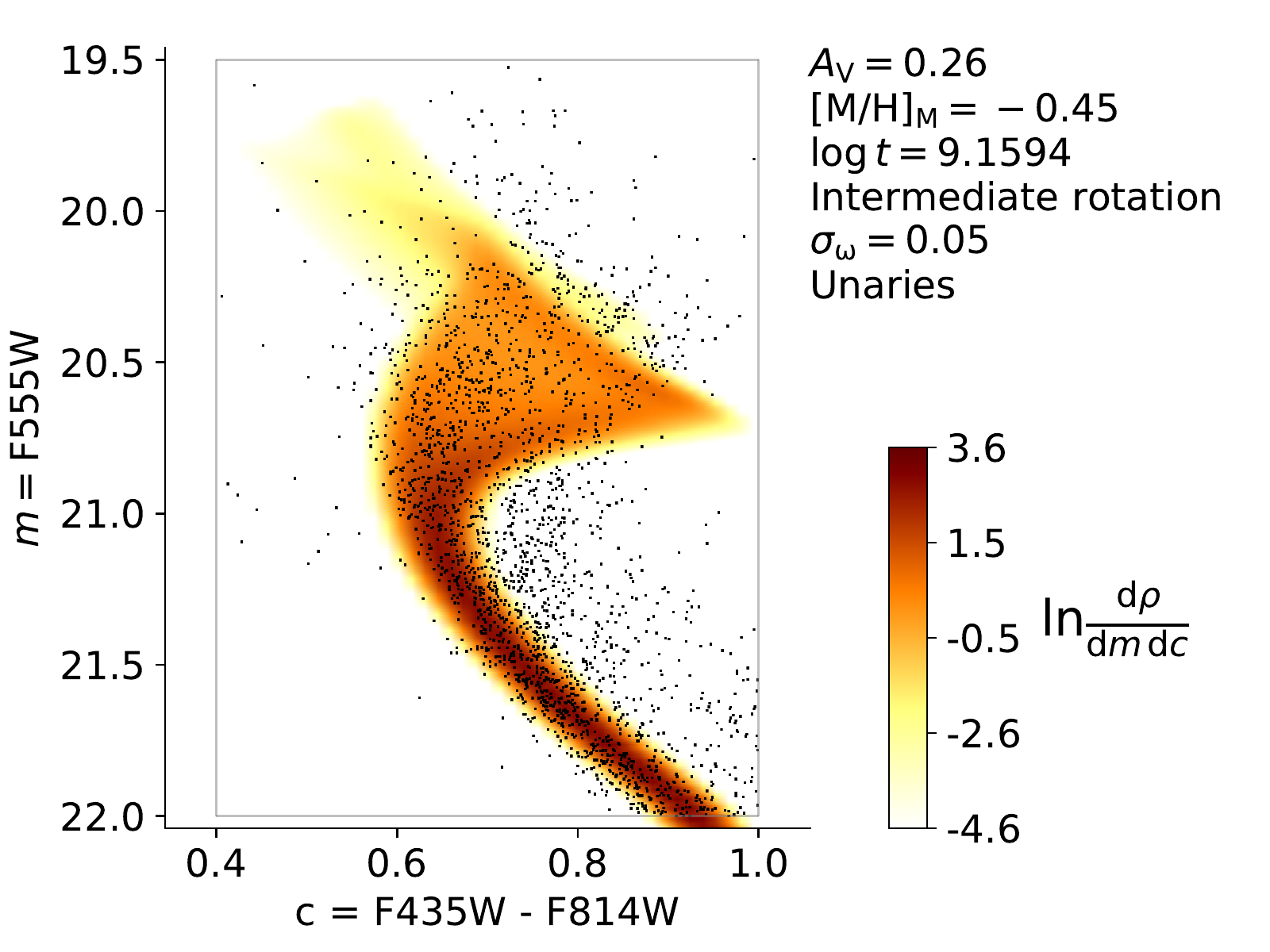}
\includegraphics[width=0.5\linewidth]{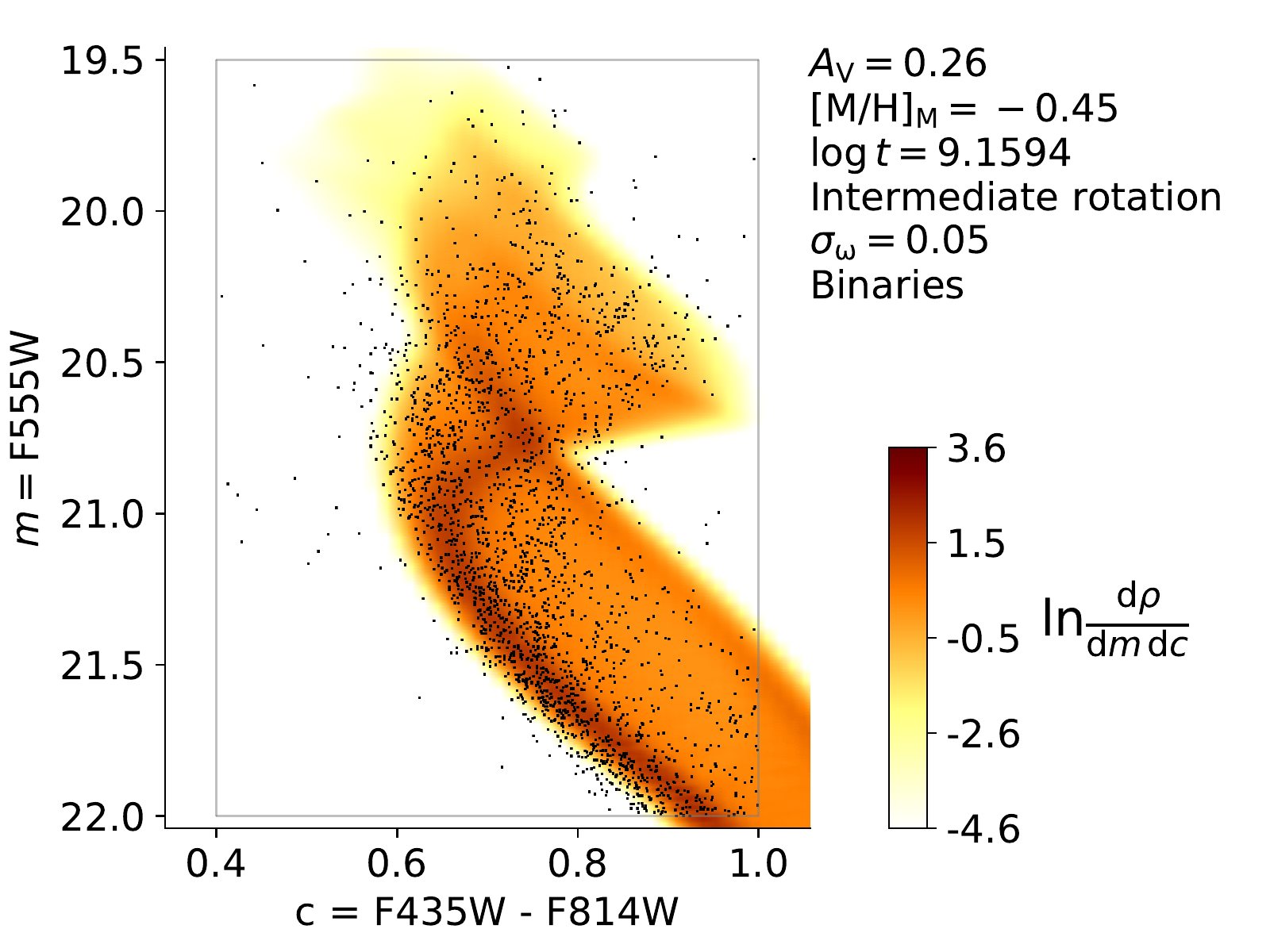}
\newline
\includegraphics[width=0.5\linewidth]{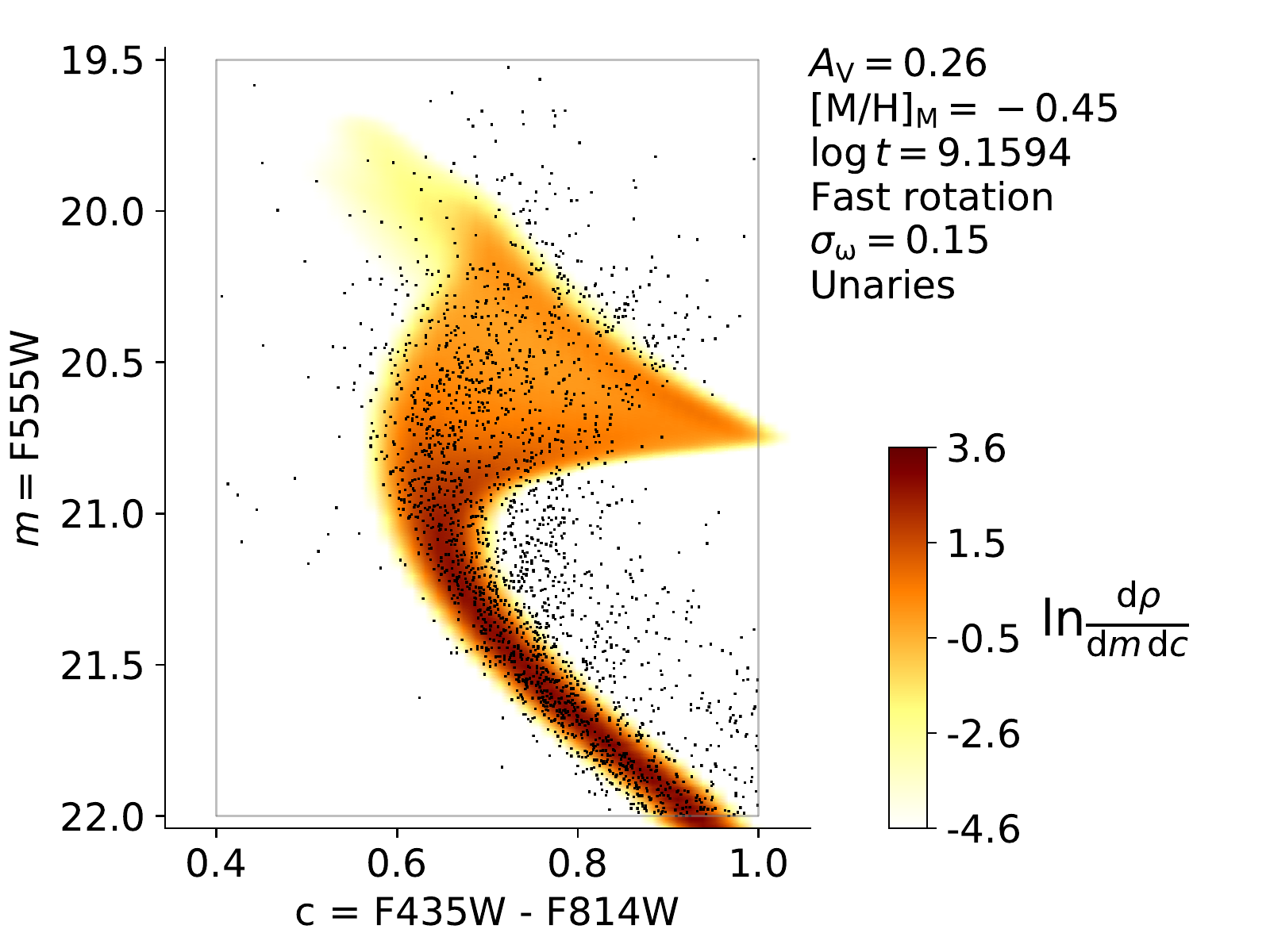}
\includegraphics[width=0.5\linewidth]{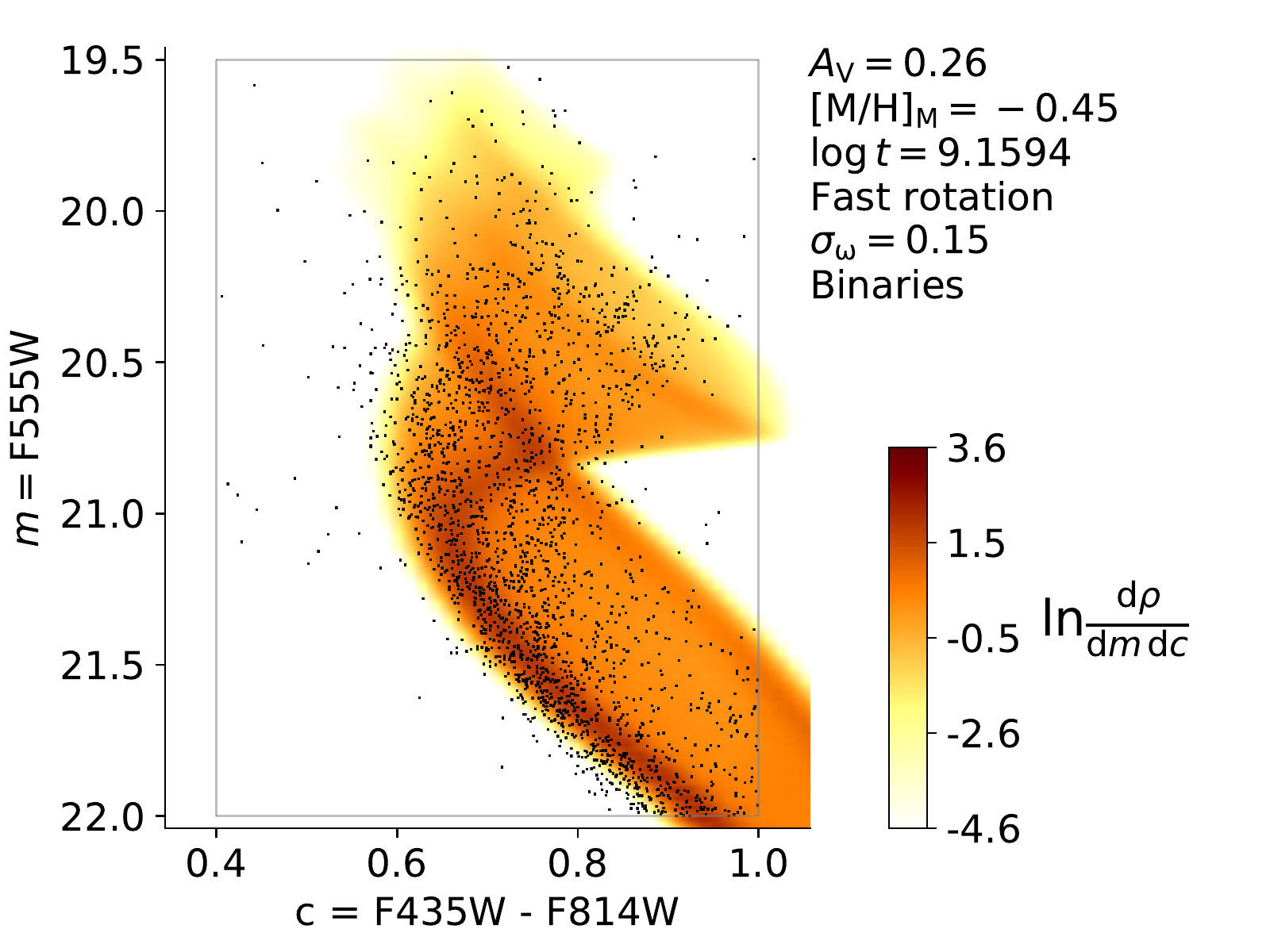}
\caption{ Theoretical probability densities at minimum error ($\sigma_m=0.01$, $\sigma_c=0.014$) and a specific age, marginalized over projected rotational speed $v$. These densities are introduced in Section \ref{minimum_error} as $\rho_{jk}(m, c; t)$. The shading interpolates between the points where density is evaluated,  illustrating that we can accurately calculate  continuous probability densities. Grey lines delimit the region of interest (ROI) that we use. In each panel, black dots denote our entire subset of the NGC 1846 data from \citetalias{Kamann_2020MNRAS}. Probability densities for single stars (unaries) are shown on the left and densities for unresolved, non-interacting binaries on the right, for three rotational populations: slow rotators (top), intermediate rotators (middle) and nearly critical rotators (bottom).}
\label{fig:cmd}
\end{figure*}

\begin{linenomath*}\begin{equation} \label{eq:rho_jk}
    \rho({\bm x};t) \equiv \rho_{jk}({\bm x}; t) = \int {\rm d}{\bm \theta'}\,\pi_{jk}({\bm \theta'})\,G\left({\bm x} - {\bm x}[{\bm \theta'}; t]; {\bm \sigma_{\bm x}}\right),
\end{equation}\end{linenomath*}
which is the minimum-error version of the probability density in Equation \eqref{eq:rho_jkp_t}. Figure \ref{fig:cmd} shows these densities for a single age and composition, for the three different rotational populations and for both single and binary stars.  In the following, we describe our approach to compute these probability densities.  We again suppress some of the arguments and subscripts in order to describe the computation of the integral in Equation \eqref{eq:rho_jk}.

We begin by constructing a fine grid in observable space $\bm{x}$ to store the probability density given by Equation \eqref{eq:rho_jkp}.  We ensure that this grid  extends well outside the ROI on the CMD and well outside the rectangular volume circumscribed by the data points. This allows us to convolve the probability density with error kernels for each star without introducing artifacts from the finite extent of the ROI. 
Our grid of $\bm{x}$ is regular and relatively fine, with spacing between neighboring ${\bm x}$ values $\approx{\bm \sigma_{\bm x}} / 3$, where ${\bm \sigma_{\bm x}}$ is the vector of minimum error standard deviations.

We then weight  $\pi({\bm \theta'})$, the prior on initial stellar parameters ${\bm \theta'}$, according to the composite multi-dimensional trapezoidal rule with variable steps. For example, let us say that we have obtained discrete values of the prior $\pi_h$ at inclinations $i_h$ with $h \in [1, H]$, $i_0 = 0$, and $i_H = \pi / 2$, where $\pi$ without subscripts indicates the mathematical constant. We designate the differences between neighboring values of inclination by $\Delta i_h = i_{h+1} - i_h$, with $h \in [1, H-1]$. Then, the weighted priors are $\frac{1}{2} \left(\Delta i_1\right) \pi_1$, $\frac{1}{2} \left(\Delta i_1 + \Delta i_2\right) \pi_2$, $\ldots$ $\frac{1}{2} \left(\Delta i_{H-1} + \Delta i_{H-2}\right) \pi_{H-1}$, $\frac{1}{2} \left(\Delta i_{H-1}\right) \pi_H$. We extend this weighting to all parameters in ${\bm \theta'}$ and place the resulting weighted prior on the ${\bm x}$-grid according to ${\bm x}\left({\bm \theta'} \right)$. Calculation of ${\bm x}\left({\bm \theta'} \right)$ is detailed in Section \ref{stellar}.

\begin{figure*}[ht]
\includegraphics[width=0.5\linewidth]{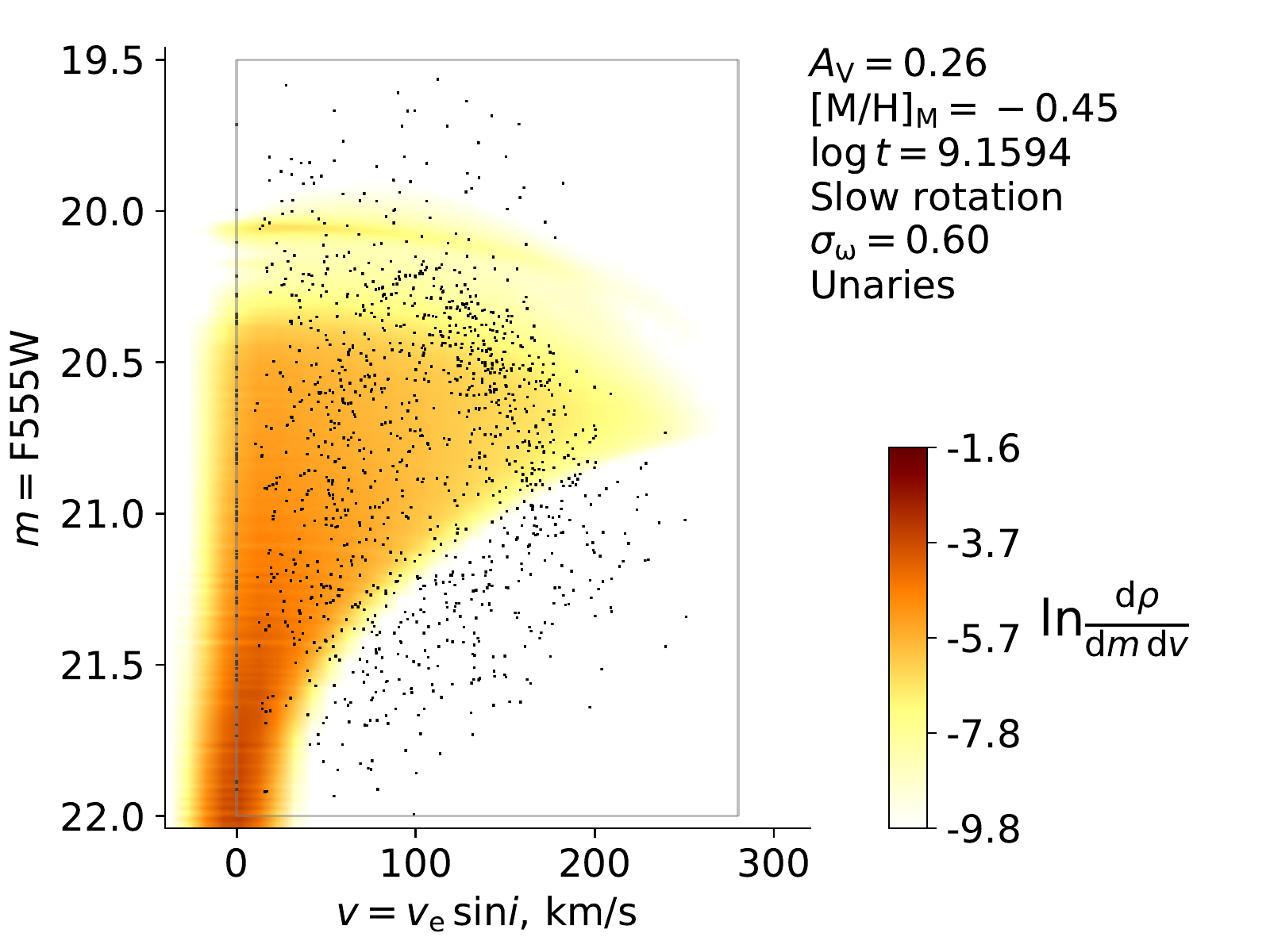}
\includegraphics[width=0.5\linewidth]{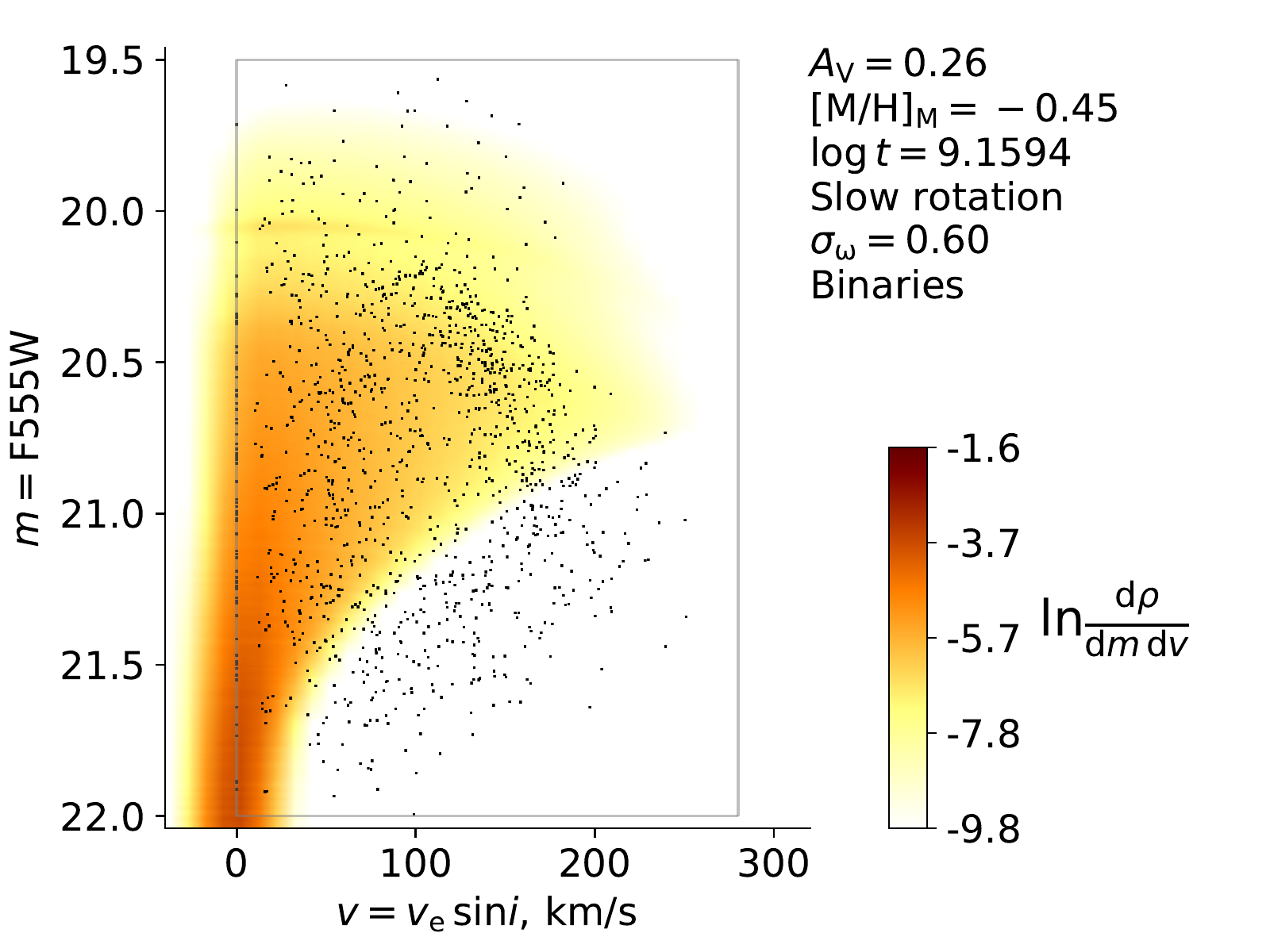}
\newline
\includegraphics[width=0.5\linewidth]{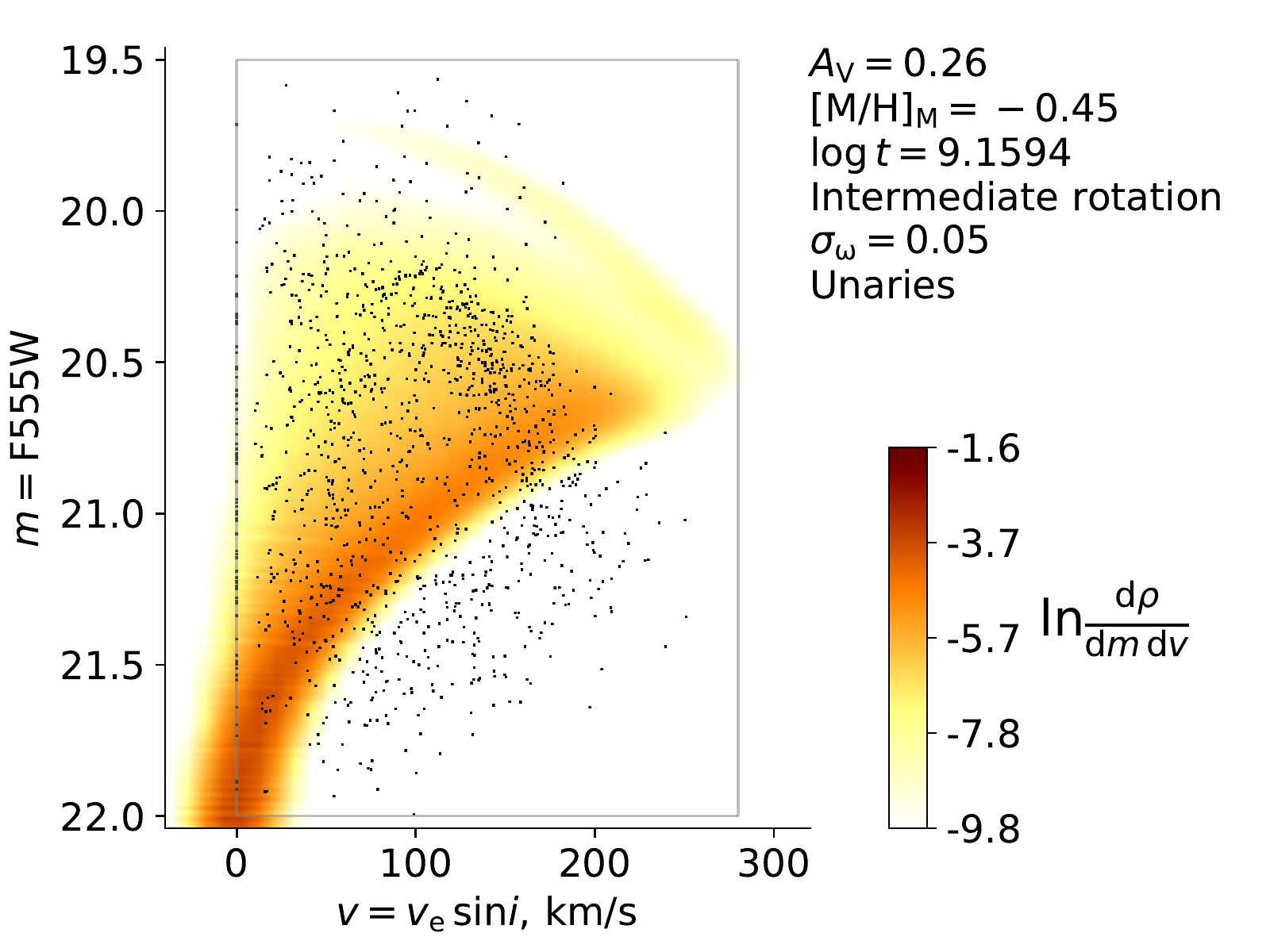}
\includegraphics[width=0.5\linewidth]{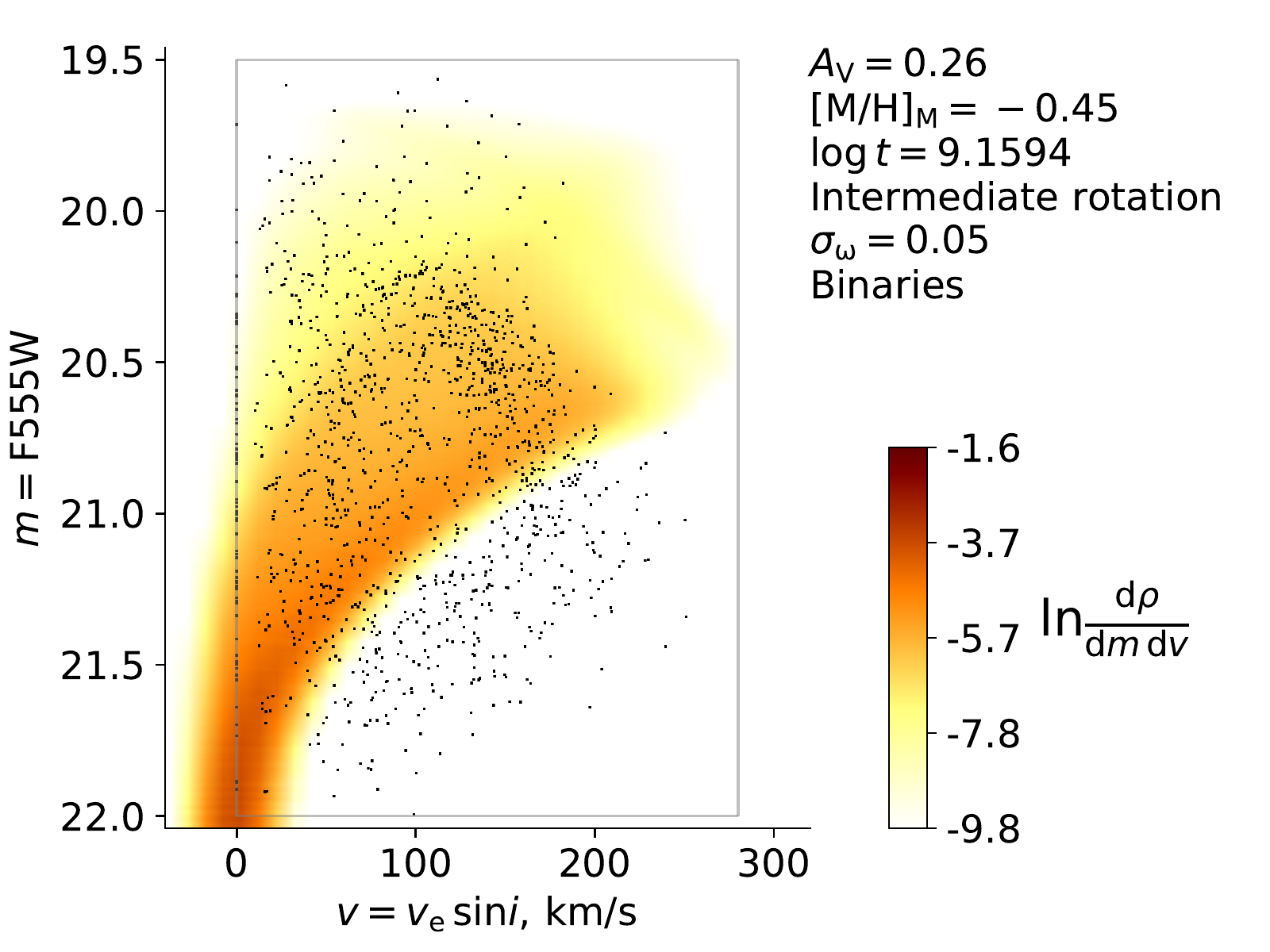}
\newline
\includegraphics[width=0.5\linewidth]{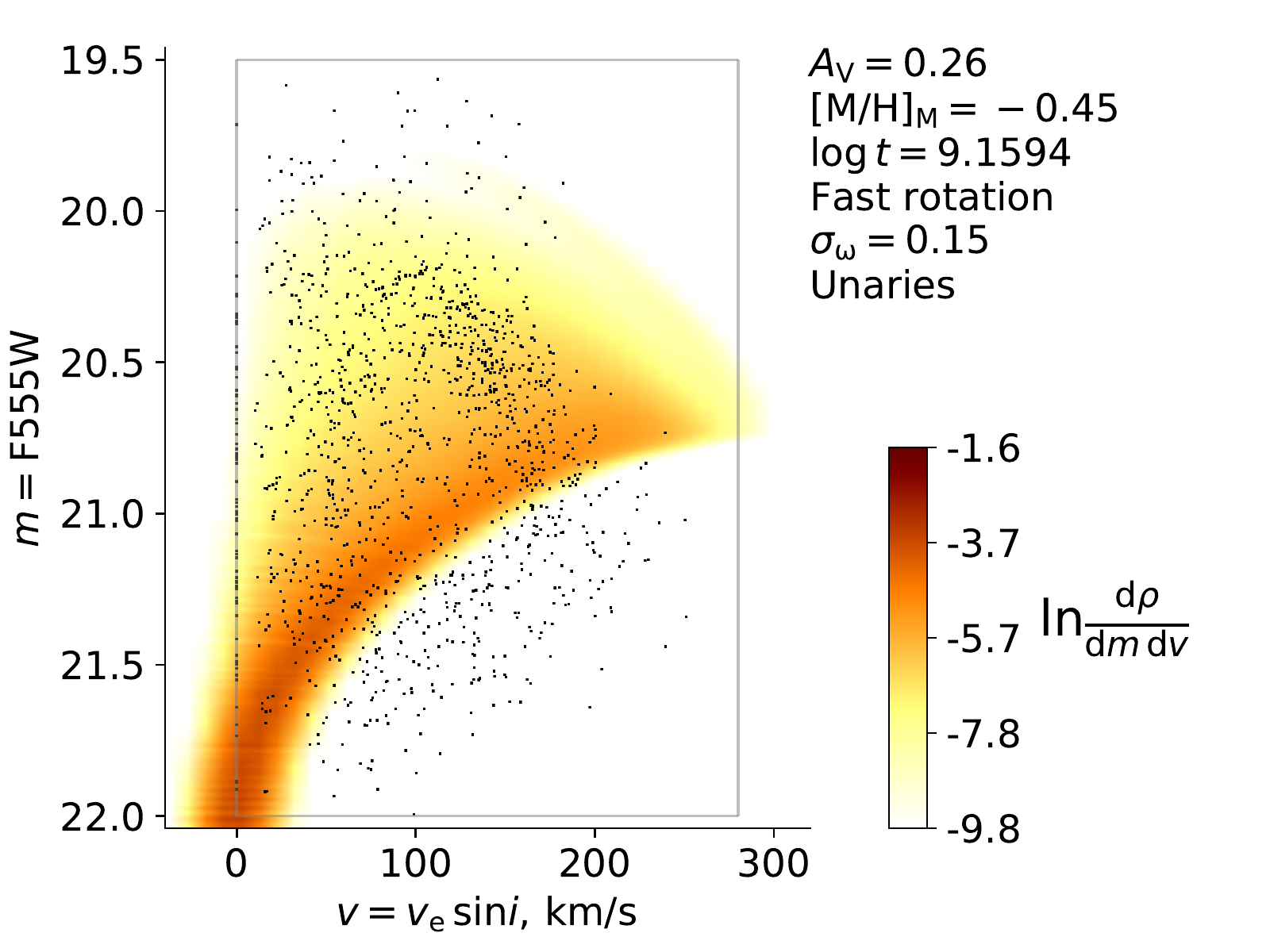}
\includegraphics[width=0.5\linewidth]{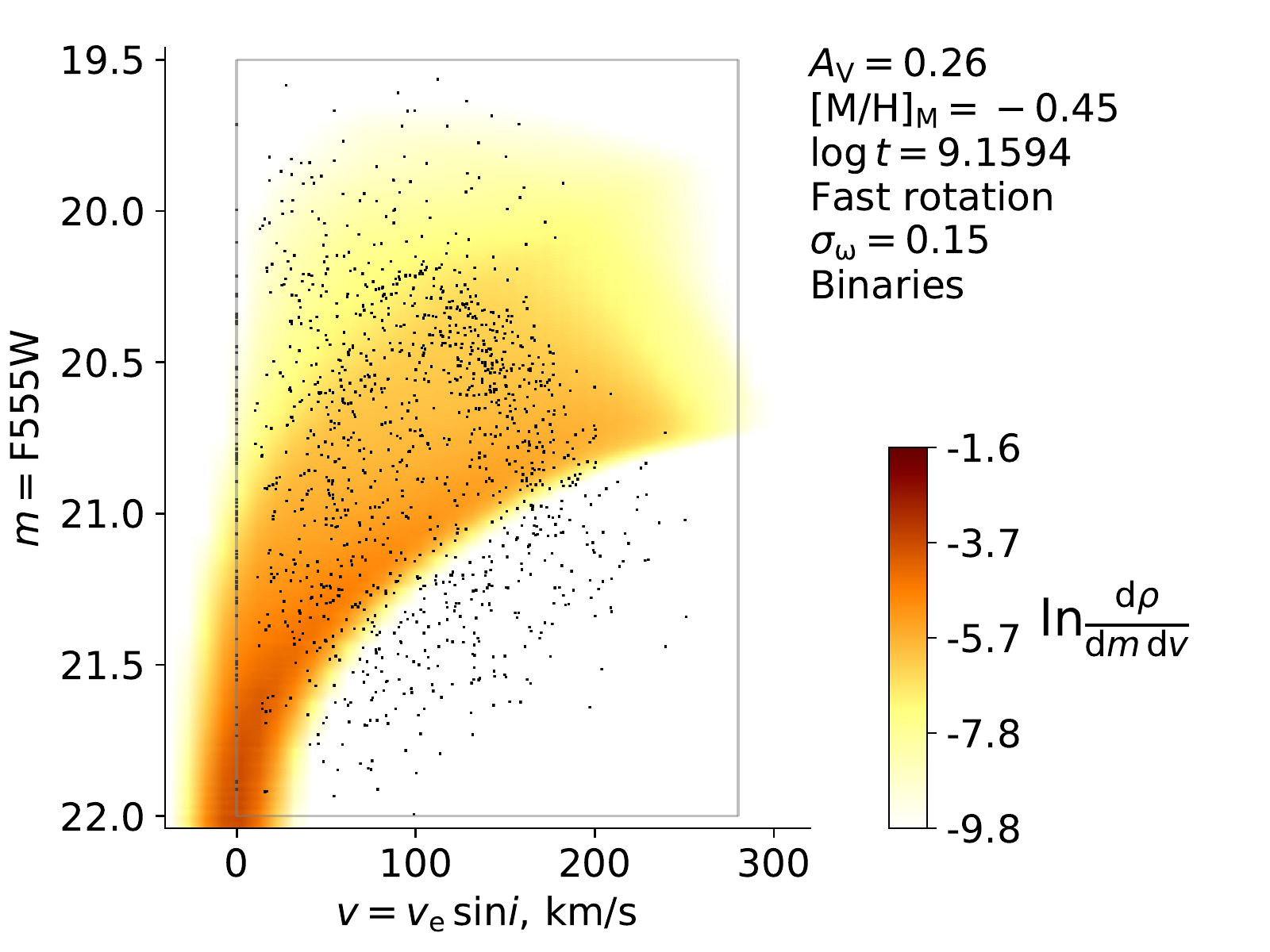}
\caption{Same as Figure \ref{fig:cmd}, except the probability densities are marginalized in color $c$. These densities are introduced in Section \ref{minimum_error} as $\rho_{jk}(m, v; t)$.}
\label{fig:vmd}
\end{figure*}

The density computation described above is not convolved with the error kernel, and shows artifacts of discretization.  Convolving with the minimum error kernel completes the calculation of Equation \eqref{eq:rho_jk} and removes these discretization artifacts.
 We first perform this task only in the $m$ dimension, i.e., magnitude, simultaneously down-sampling to a coarser grid, with spacing between neighboring $m$ values equal to $\sigma_m$.  Repeating the procedure in the $c$ and $v$ dimensions, i.e., color and $v_{\rm e}\sin{i}$, takes successively less time, since in each case all previous dimensions have been down-sampled. Afterwards, we normalize each resulting probability density $\rho_{jk}({\bm x}; t)$ as a function of ${\bm x}$ on the ROI. We marginalize the density in $v$ to obtain the two-dimensional version $\rho_{jk}(m, c; t)$ and in $c$ to obtain $\rho_{jk}(m, v; t)$, then re-normalize both. Figures \ref{fig:cmd} and \ref{fig:vmd} show the respective marginalized probability densities at $\log{t} = 9.1594$. Figure \ref{fig:vmd} is an example of a $v_{\rm e}\sin{i}$-magnitude diagram (VMD), by analogy with the CMD.

The above procedure, which starts by placing the prior onto ${\bm x}$-space, is faster than the direct integration in initial model parameter ${\bm \theta'}$-space implied by Equation \eqref{eq:rho_jk}.  The computational cost would be similar if we evaluated the likelihood for only one star, with a single location in ${\bm x}$-space.  However, Equation \eqref{eq:rho_jk} represents a five-dimensional integral for every point in the three-dimensional ${\bm x}$-space.  Having performed this integral once, we need only integrate the product of the result and a three-dimensional error kernel for each star.  Furthermore, a Gaussian has appreciable support over a limited range of $\bm{x}$, which also reduces the evaluation cost.

\begin{figure*}[ht]
\includegraphics[width=0.5\linewidth]{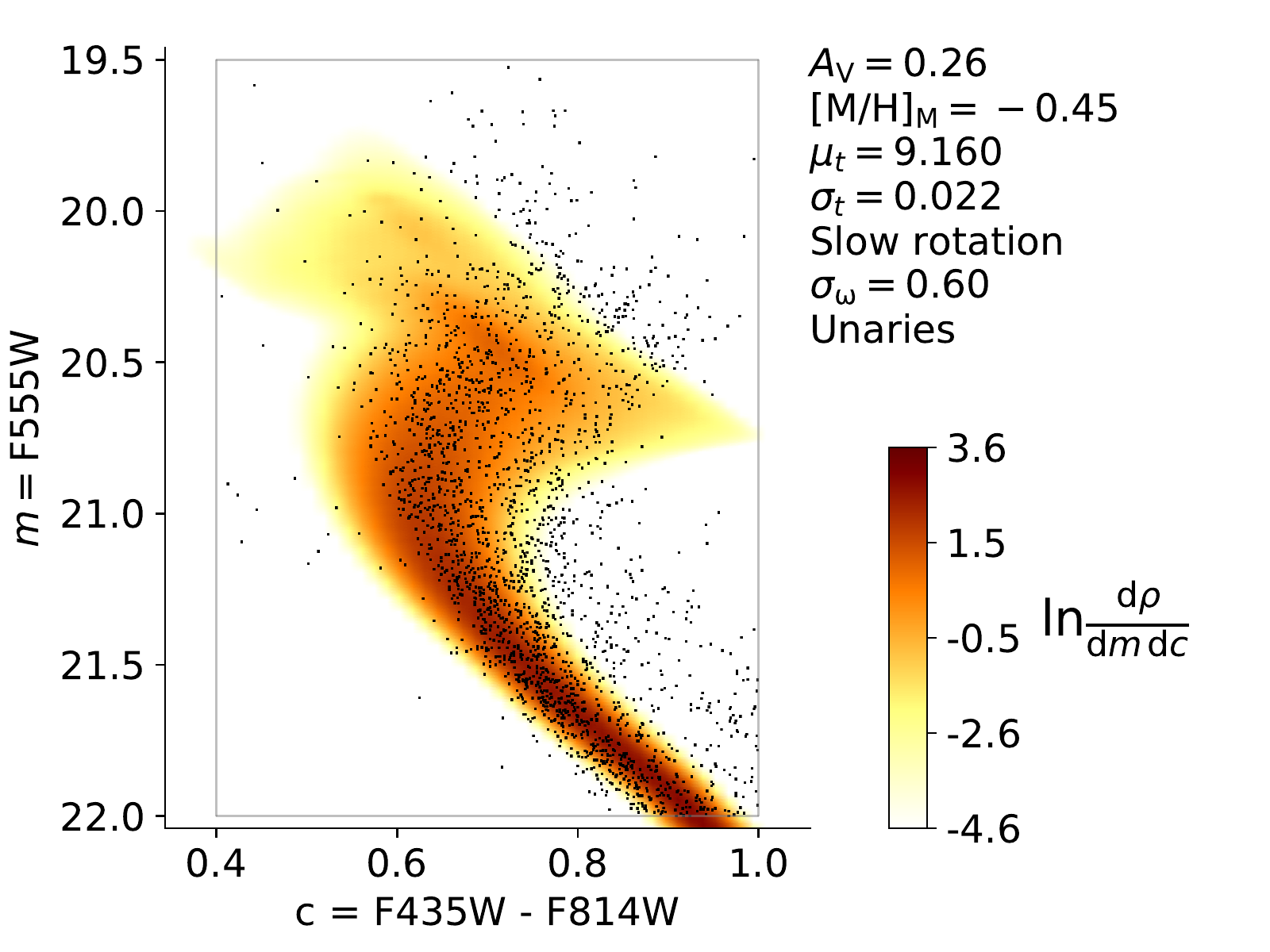}
\includegraphics[width=0.5\linewidth]{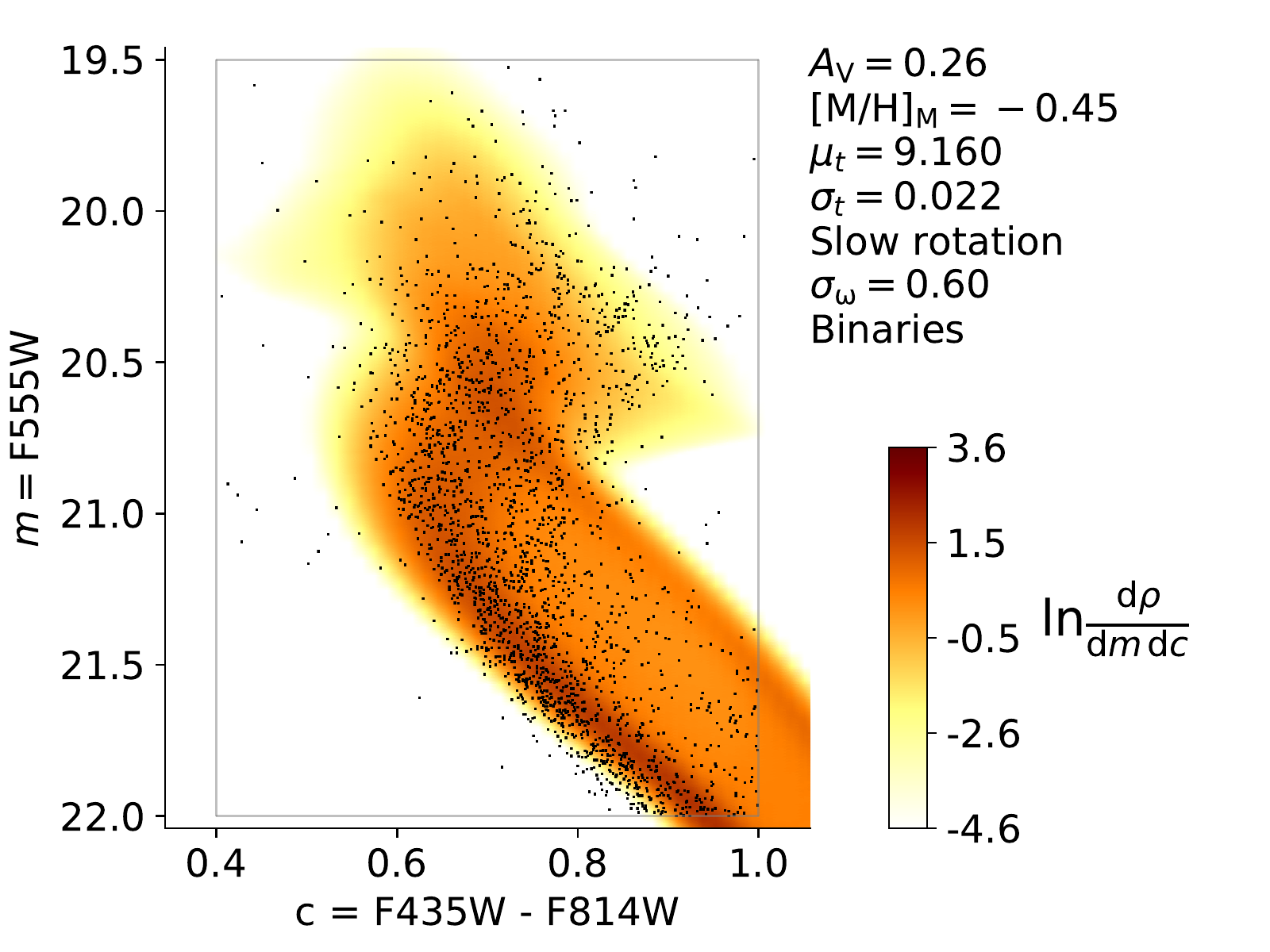}
\newline
\includegraphics[width=0.5\linewidth]{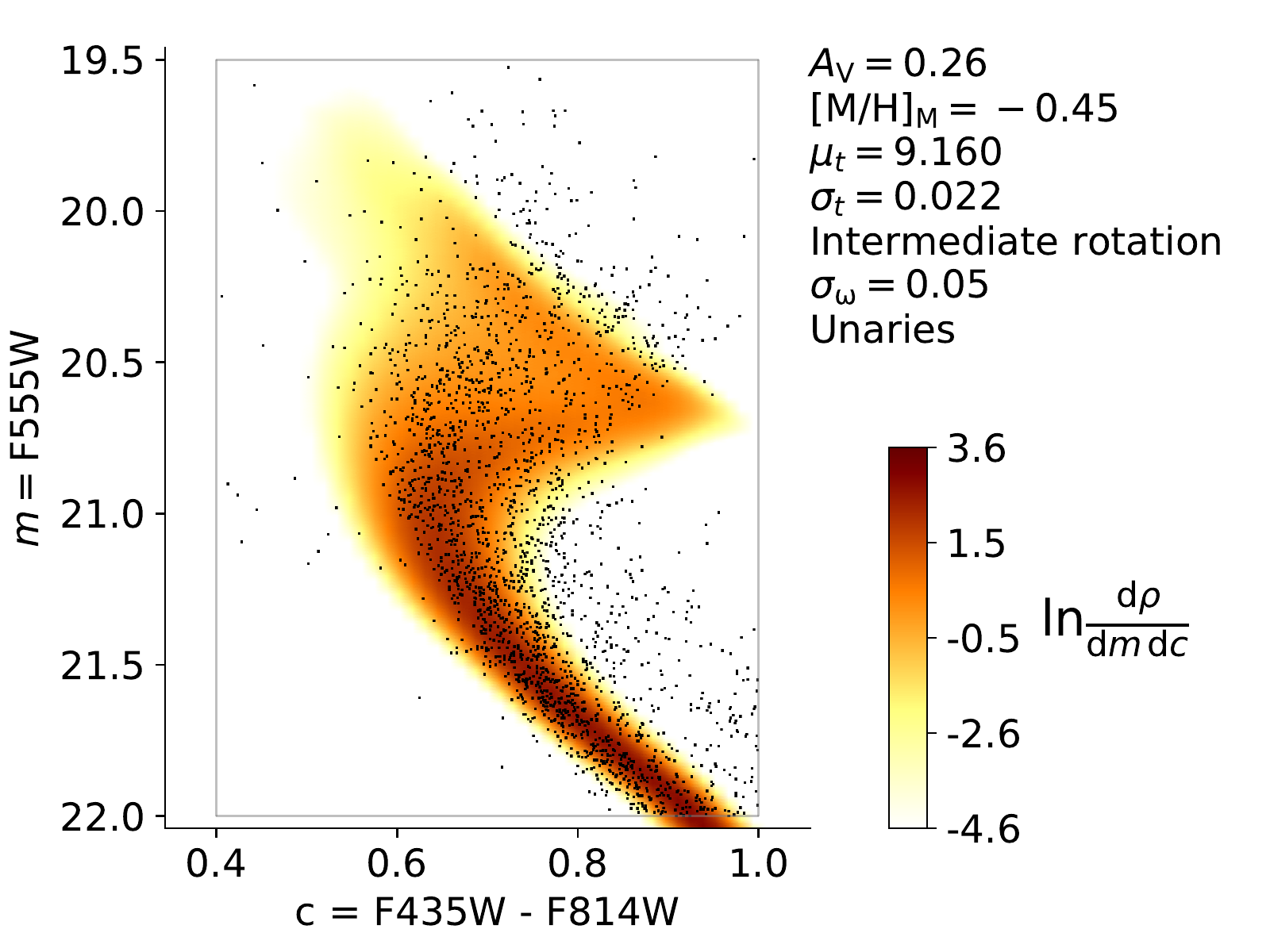}
\includegraphics[width=0.5\linewidth]{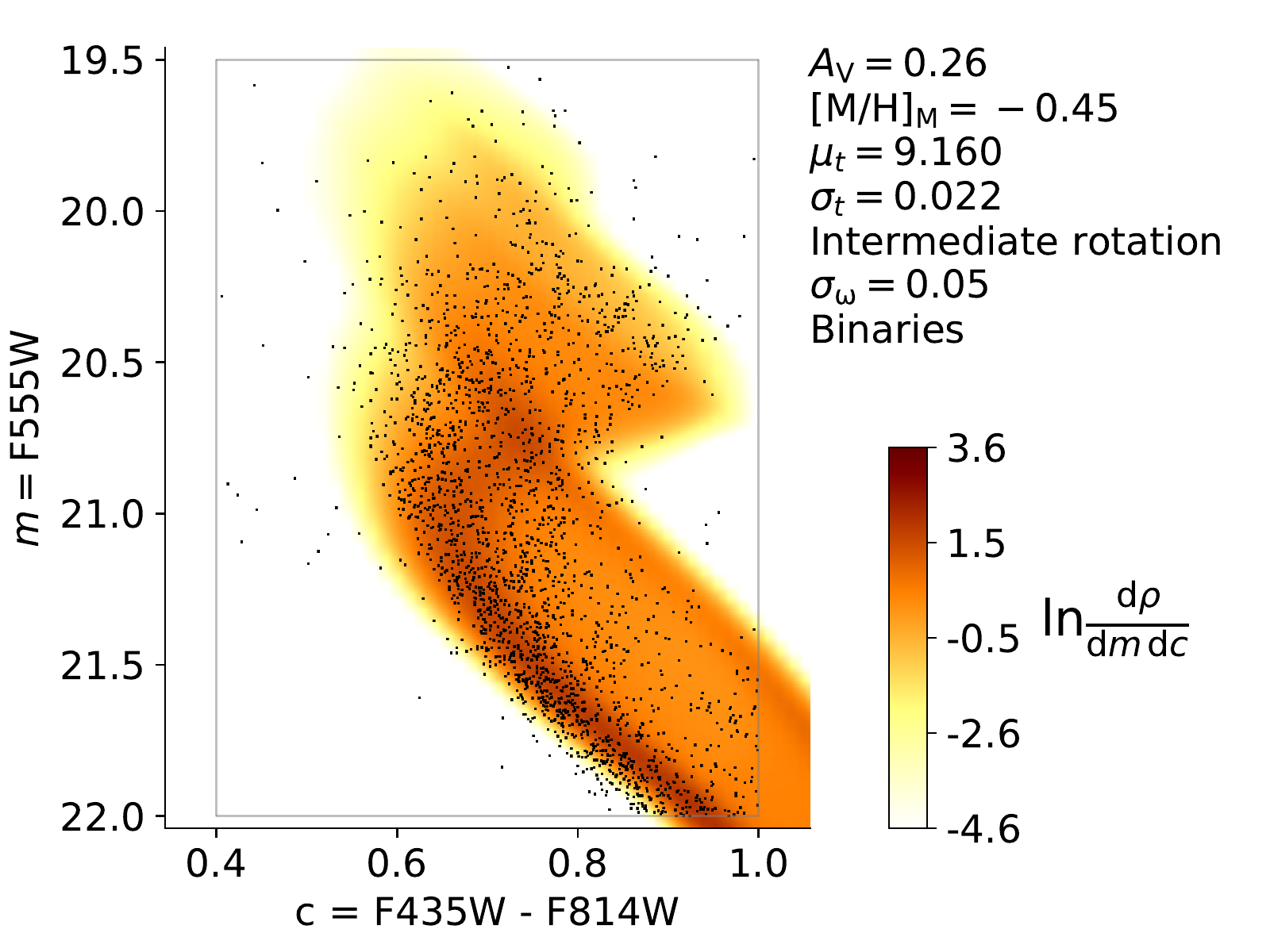}
\newline
\includegraphics[width=0.5\linewidth]{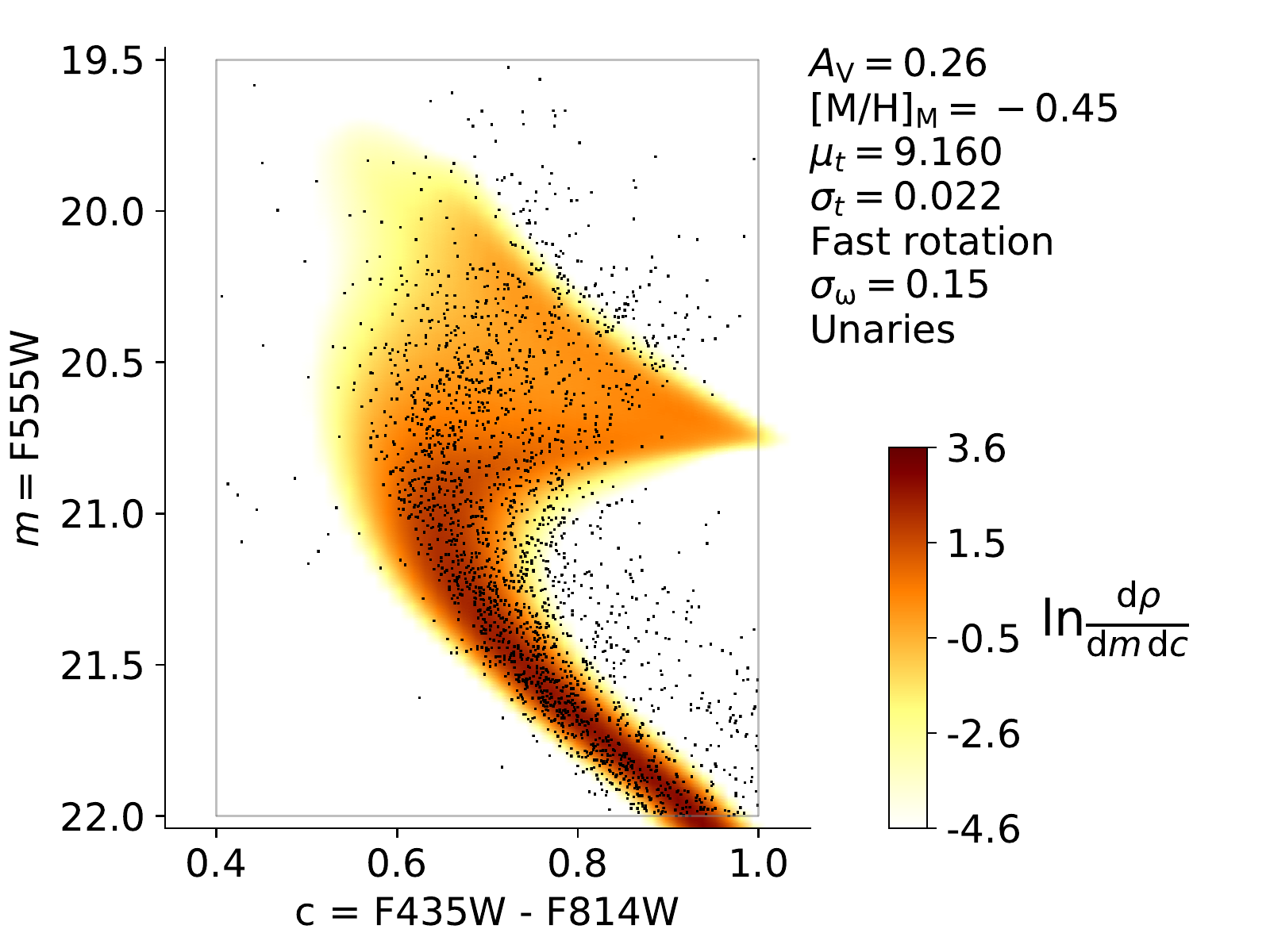}
\includegraphics[width=0.5\linewidth]{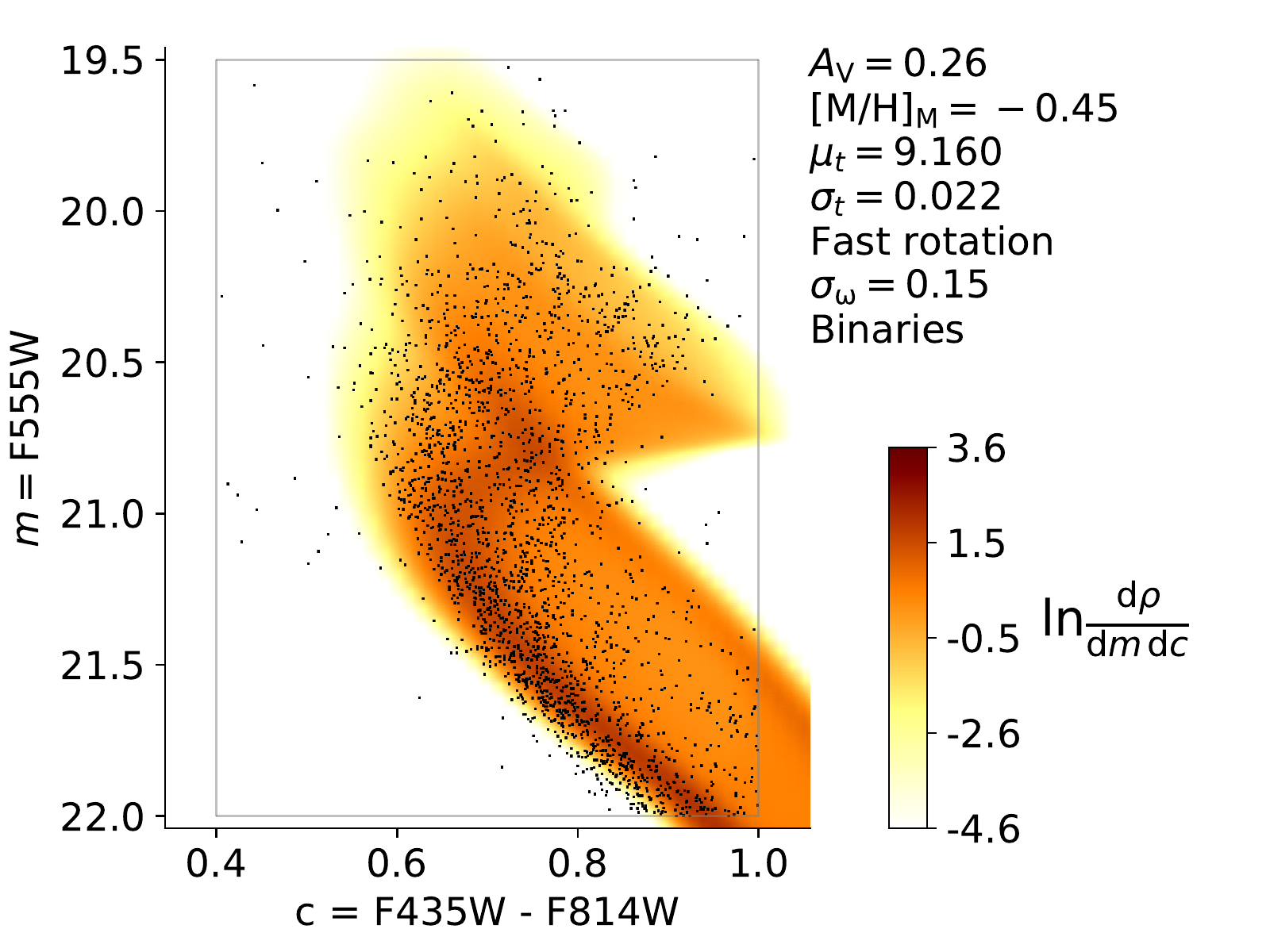}
\caption{Same as Figure \ref{fig:cmd}, except the probability densities incorporate a relatively narrow lognormal age prior. These densities are $\rho_{jk}({\bm x}; \mu_t, \sigma_t)$ in Equation \eqref{eq:rho_agepr_minerr}, marginalized in $v$.}
\label{fig:cmd_all_ages}
\end{figure*}

Our ensuing main integration procedure multiplies minimum-error densities $\rho_{jk}({\bm x}; t)$ by the residual error kernel, integrates, then multiplies by the log-normal age prior.
On the other hand, for diagnostic purposes, we can  immediately multiply the minimum-error densities by the age prior $\bar{\pi}(t; \mu_t, \sigma_t)$, then integrate the result. This procedure yields $\rho_{jk}({\bm x}; \mu_t, \sigma_t)$, the  minimum-error probability density that incorporates the age prior:
\begin{linenomath*}\begin{equation} \label{eq:rho_agepr_minerr}
    \rho_{jk}({\bm x}; \mu_t, \sigma_t) = \int {\rm d}t\,\bar{\pi}(t; \mu_t, \sigma_t) \,\rho_{jk}({\bm x};t).
\end{equation}\end{linenomath*}
Figure \ref{fig:cmd_all_ages} shows densities $\rho_{jk}({\bm x}; \mu_t, \sigma_t)$ for one combination of age mean $\mu_t$ and age dispersion $\sigma_t$, marginalized over $v$. This figure shows no artifacts of discretization in age, which supports the idea that the spacing requirement on the age grid in Appendix \ref{spacing_age} has been met. The densities also do not show any age discretization artifacts when marginalized in color $c$.

\subsubsection{De-Normalization Correction} \label{denormalization}

The previous section described the computation of probability densities $\rho_{jk}({\bm x}; t)$ for the minimum observational uncertainties.  We multiply these by the residual error kernel  
with standard deviation $\sigma'_{{\bm x}{\bm p}}$, associated with each data point $p$, and integrate to obtain a probability density for each star's observed properties.  This probability density is evaluated at the star's observables ${\bm x_{\bm p}}$. The integral that calculates $\rho_{jkp}({\bm x}; t)$ for any observable vector ${\bm x}$ and finite error standard deviation on $v_{\rm e}\sin{i}$, $\sigma_{vp}$, is 
\begin{linenomath*}\begin{equation} \label{eq:rho_jkpt}
    \rho_{jkp}({\bm x}; t) = \int\,{\rm d}{\bm x'}\,\bar{G}\left({\bm x} - {\bm x'}; {\bm \sigma'_{{\bm x}{\bm p}}}\right)\,\rho_{jk}({\bm x'};t).
\end{equation}\end{linenomath*}
Here, ${\bm \sigma'_{{\bm x}{\bm p}}}$ is the vector of residual standard deviations for data point $p$ and $\bar{G}({\bm \cdot})$ is the normalized Gaussian error kernel. Unlike $\rho_{jk}({\bm x'};t)$, we only need $\rho_{jkp}({\bm x}; t)$ at a single value of ${\bm x}$, namely ${\bm x_{\bm p}}$. 

Convolving the minimum error probability density with a normalized error kernel preserves its normalization.  However, it does not necessarily preserve its normalization over a restricted subset of the domain, e.g., our ROI.  In order to treat the result of Equation \eqref{eq:rho_jkpt} as a probability density, we must therefore ensure it remains normalized over the ROI.  This section describes the procedure we follow in order to make sure that this requirement is met. We term this procedure denormalization correction.  

Applying an additional error kernel can denormalize the probability density
in magnitude $m$ and color $c$, the dimensions where the ROI is finite. If the probability density has a nonzero gradient across the boundary of the ROI, a convolution will move different amounts of probability from inside to outside the ROI as vice versa.  

In particular, at $m$ just inside the ROI, there is a contribution to $\rho_{jkp}(m; t)$ due to $\rho_{jk}(m'; t)$ at $m'$ outside the ROI. In other words, some amount of probability leaks into the ROI. Similarly, at $m$ just outside the ROI, some amount of probability leaks out. In general, the amount that leaks in is not equal to the amount that leaks out, so that some net leakage occurs.  This is a form of selection bias, where the selection is applied to the observed, rather than the intrinsic, properties of each star.

\begin{figure*}[ht]
\includegraphics[width=0.5\linewidth]{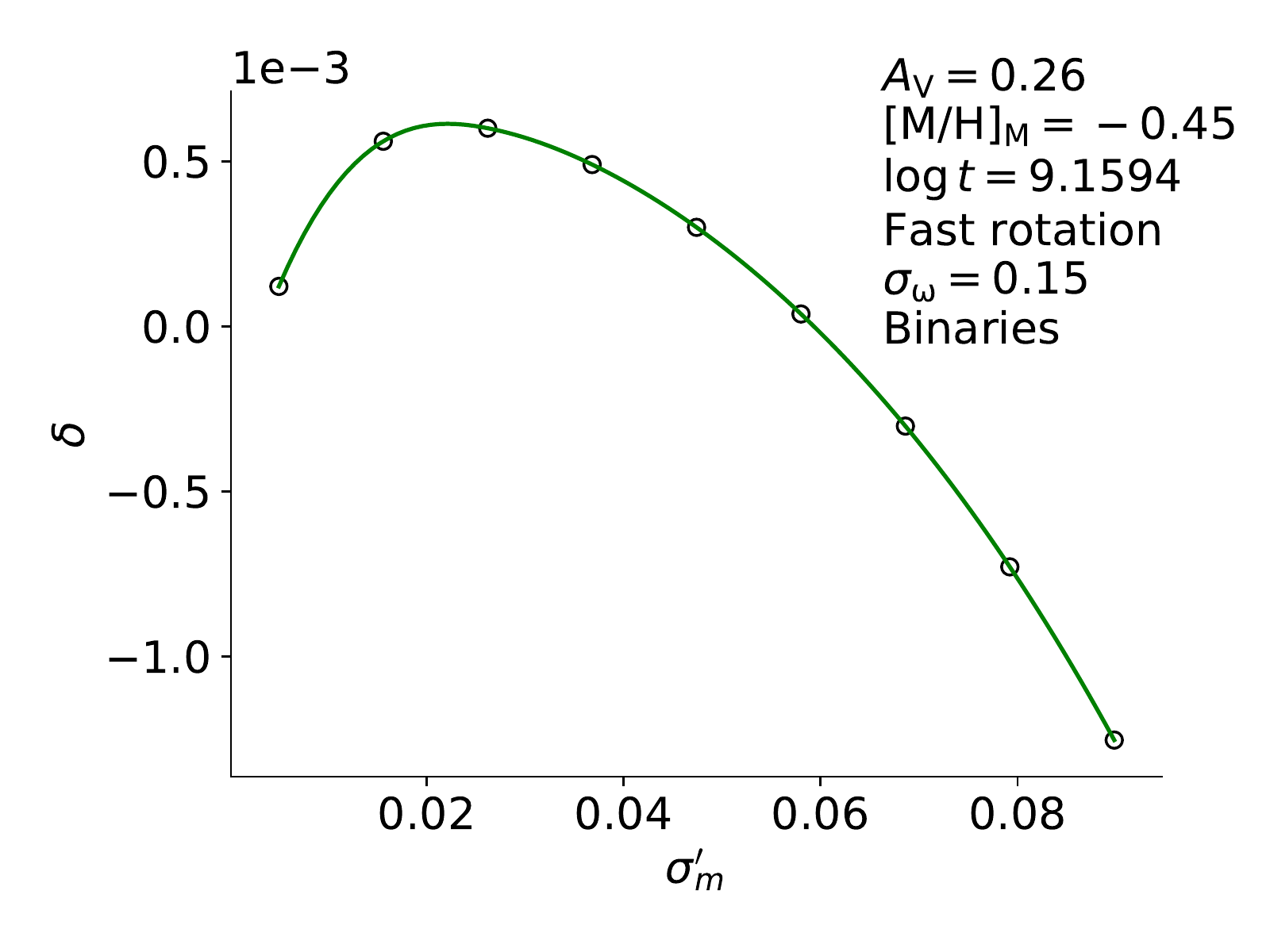}
\includegraphics[width=0.5\linewidth]{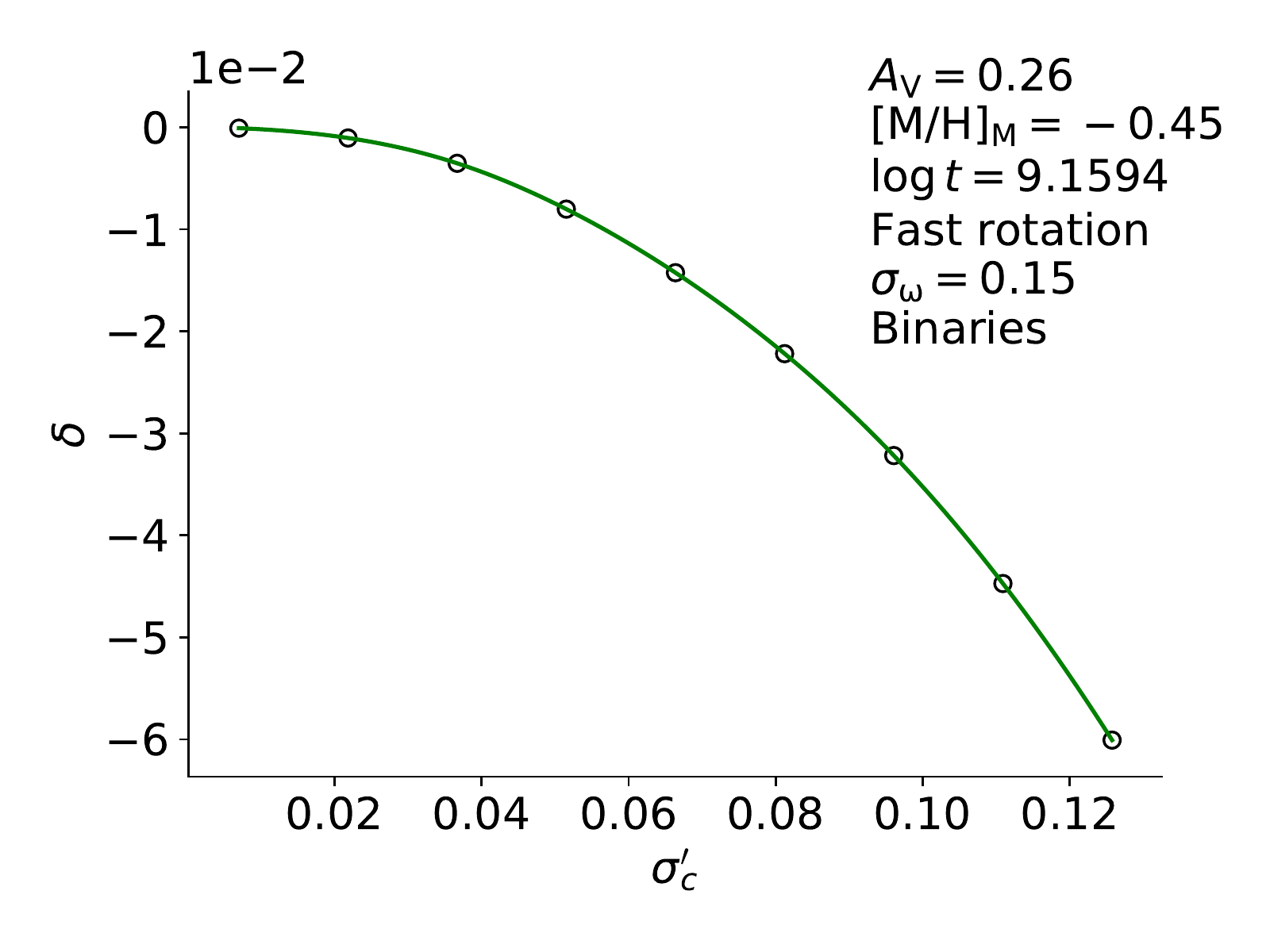}
\caption{De-normalization, i.e., the net probability that leaks into the region of interest upon the convolution of a minimum-error probability distribution with a residual error kernel, versus the kernel's standard deviation $\sigma'$. Negative values correspond to net leakage out of the region. The left panel is due to convolution in the magnitude dimension, right panel -- in the color dimension. Black open circles indicate the $\sigma'$ grid where we calculate the de-normalization precisely, with grid range similar to that of the $\sigma'$ values among the data points. Green lines indicate the cubic interpolant of the function, which we use to approximate the de-normalization between the grid points. The specific combination of age, rotational population, and multiplicity population are the same as in the lower right panel of Figure \ref{fig:cmd}. De-normalization due to integration in magnitude remains on the order of 0.1\%. On the other hand, that due to color can be on the order of several percent, indicating that the corresponding correction is necessary. These orders of magnitude for the de-normalization correction are typical for the two observables among the probability distributions.}
\vspace{20pt}
\label{fig:delta}
\end{figure*}

Accordingly, before performing the calculation of probability densities in Equation \eqref{eq:rho_jkpt} at ${\bm x} = {\bm x_{\bm p}}$ for each data point, we compute the net leakage of probability into or out of the ROI in the course of such calculation as a function of ${\bm \sigma'_{{\bm x}{\bm p}}}$. We perform this calculation separately for the integration in $m$ and in $c$. For example, for integration in $m$, rotational population $j$, and multiplicity population $k$, we compute the de-normalization correction $\delta^{m}_{jk}$ for a given age $t$, as a function of residual standard error $\sigma'_m$:
\begin{linenomath*}\begin{multline} \label{eq:delta}
    \delta^{m}_{jk}(\sigma'_m; t) = \int_{\rm ROI}\,{\rm d}{\bm x}\,\int {\rm d}m'\,\bar{G}\left(m - m'; \sigma'_m\right)\times\\
    \rho_{jk}(m', c, v; t) - 1,
\end{multline}\end{linenomath*}
Once we obtain $\delta^{m}_{jk}(\sigma'_m; t)$ on a discrete grid of $\sigma'_m$, we approximate the corresponding continuous function via cubic interpolation, extrapolating when $\sigma'_{mp}$ is outside the grid. We obtain this function for $c$ in addition to $m$ in a similar fashion. The result, for both observables and a particular combination of $t$, $j$ and $k$, is shown in Figure \ref{fig:delta}. The discrete grids of $\sigma'_m$ and $\sigma'_c$ are identical for all combinations of these parameters; for the combinations where the maximum absolute value of de-normalization drops below $10^{-5}$ on the grids, we set the function equal to zero.

\subsubsection{Rotational Measurement Status} \label{status}

All of our stars have measured colors and magnitudes.  Some have positive measured values for $v_e \sin i$, while others either have no measurement due to inadequate signal-to-noise, or have upper limits on their $v_e \sin i$, with a reported $v_e \sin i = 0$.  Each of these cases must be treated differently.  

When a data point $p$ includes a positive $v_{\rm e}\sin{i}$ measurement, i.e., when $v_p > 0$, the probability density associated with the point is 3-dimensional, given by Equation \eqref{eq:rho_jkpt}. Recall that we evaluate the integral in this equation, and thus the resulting density, only at the data point's observable vector, ${\bm x} = {\bm x_{\bm p}}$. 

Even though a star cannot have $v_{\rm e}\sin{i} \equiv v < 0$, density $\rho_{jkp}({\bm x}; t)$ can be non-zero at negative $v$ as a result of measurement error. This does not affect the densities for data points with $v_p > 0$, since these points never result from $v < 0$ instrument readings. However, a $v_p = 0$ measurement corresponds to $v \le 0$ instrument readings. Thus, the 2-dimensional probability density at age $t$ for such a measurement, $\rho^{v0}_{jkp}(m, c; t)$, results from integration over $v \le 0$:
\begin{linenomath*}\begin{equation} \label{eq:rho_v0_jkpt}
    \rho^{v0}_{jkp}(m, c; t) = \int_{-\infty}^{0}{\rm d}v \, \rho_{jkp}({\bm x}; t),
\end{equation}\end{linenomath*}
which we evaluate before applying the de-normalization correction, and only at the data point's observables $m = m_p$ and $c = c_p$. 

About half the stars in the ROI have no measurements of $v$.  For these stars, the appropriate probability density is integrated over $v$ and becomes a two-dimensional density $\rho_{jkp}(m, c; t)$ in $m$ and $c$, evaluated
at $m = m_p$ and $c = c_p$.

Consequently, theoretical probability takes the form of the 3-dimensional density function $\rho_{jkp}({\bm x}; t)$ only for $0 < v < v_1$. Here, $v_1 = 280\,{\rm km\,s^{-1}}$ is an upper limit on $v_{\rm e}\,\sin{i}$, which is larger than any of the $v_{\rm e}\,\sin{i}$ measurements. The theoretical probability density is 2-dimensional at $v = 0$ and at $v = v_1$. In particular, the density at $v = 0$ is $\rho^{v0}_{jkp}(m, c; t)$ and we do not need to calculate the density at $v = v_1$. The sum of the integral of the 3-dimensional density and the integrals of the 2-dimensional densities over the functions' respective domains equals 1.

The discussion of de-normalization in Section \ref{denormalization} applies to the 2-dimensional probability densities $\rho^{v0}_{jkp}(m, c; t)$ and $\rho_{jkp}(m, c; t)$ the same way it applies to the 3-dimensional density $\rho_{jkp}({\bm x}; t)$. Accordingly, for example, $\rho_{jkp}(m, c; t)$ is multiplied by the de-normalization correction factor,
\begin{linenomath*}\begin{equation}
    C_{jk}(\sigma'_{mp}, \sigma'_{cp}; t) = \frac{1}{1+\delta^{m}_{jk}(\sigma'_{mp};t)} \times \frac{1}{1+\delta^{c}_{jk}(\sigma'_{cp};t)},
\end{equation}\end{linenomath*}
where functions of the form $\delta^{m}_{jk}(\sigma'_{mp};t)$ are defined in Equation \eqref{eq:delta}.

In the limit of a narrow residual error kernel, the kernel acts as a Dirac delta function.  Multiplying with this error kernel and integrating simply picks out a value within the minimum error probability density.  We therefore linearly interpolate within the minimum error density for the dimension(s) in which the residual error $\sigma'$ is smaller than $\frac{1}{2}$ the minimum error $\sigma$. Otherwise, we integrate the product of the minimum uncertainty density and the error kernel using a Riemann sum.

Once we have evaluated the probability densities $\rho_{jkp}({\bm x}; t)$, $\rho^{v0}_{jkp}(m, c; t)$, and $\rho_{jkp}(m, c; t)$ at ${\bm x} = {\bm x_{\bm p}}$, we can compute the counterparts of these densities that take the age prior into account: $\rho_{jkp}({\bm x}; \mu_t, \sigma_t)$, $\rho^{v0}_{jkp}(m, c; \mu_t, \sigma_t)$, and $\rho_{jkp}(m, c; \mu_t, \sigma_t)$, which are similar to the minimum-error density $\rho_{jk}({\bm x}; \mu_t, \sigma_t)$ in Equation \eqref{eq:rho_agepr_minerr}. For example, we can evaluate the following at ${\bm x} = {\bm x_{\bm p}}$:
\begin{linenomath*}\begin{equation} \label{eq:rho_agepr}
    \rho_{jkp}({\bm x}; \mu_t, \sigma_t) = \int {\rm d}t\,\bar{\pi}(t; \mu_t, \sigma_t) \,\rho_{jkp}({\bm x};t),
\end{equation}\end{linenomath*}
where $\bar{\pi}(t; \mu_t, \sigma_t)$ is the normalized age prior.

\subsection{Background Densities} \label{background}

We do not expect our cluster model to describe all the stars in the ROI.  Some stars will be interlopers physically unassociated with the cluster.  Others will be poorly fit by the cluster model, whether because of neglected binary interactions, imperfect treatment of relevant physics in the stellar model, or something else.  We include a background population to account for all of these stars. 

We model the distribution of these data points in the space of observable vectors ${\bm x}$, instead of model space:
\begin{linenomath*}\begin{equation}
    \pi({\bm x}) = H(v),
\end{equation}\end{linenomath*}
where $v \equiv v_{\rm e}\sin{i}$ and $H(v)$ is the Heaviside step function, with $H(0) = 1$. In other words, we take these background data points to be uniformly distributed over color, magnitude, and $v$, subject to the constraint that  $v \ge 0$. The densities, after convolving with an error kernel, remain uniform in $c$ and $m$, but are not uniform in $v$ because of the physical constraint that $v \ge 0$.  The background density becomes

\begin{linenomath*}\begin{multline} \label{eq:rho_b}
    \rho_{{\rm b}p}({\bm x})= \int {\rm d}{\bm x'} \,\pi({\bm x})\,G({\bm x} - {\bm x'}; {\bm \sigma_{{\bm x}{\bm p}}}) =\\
    \frac{1}{V_x} \times \frac{1}{2} \left(1 + {\rm erf}\left[\frac{v}{\sigma_{vp}\sqrt{2}}\right]\right),
\end{multline}\end{linenomath*}
where ${\rm erf}$ is the error function, $G({\bm \cdot})$ is the appropriate Gaussian error kernel, $\sigma_{vp}$ is the error in $v$ for point $p$, and $V_x$ is a normalization constant. The density $\rho_{{\rm b}p}({\bm x})$ in Equation \eqref{eq:rho_b} plays the same role for the background population as density $\rho_{jkp}({\bm x}; \mu_t, \sigma_t)$ in Equations \eqref{eq:rho_jkp} and \eqref{eq:rho_agepr} for the modeled population. A key difference is in the treatment of the upper boundary at $v = v_1$. In the case of the modeled population, we had allowed for the possibility of data points with $v$ measurement $v_p > v_1 = 280\,{\rm km\,s^{-1}}$, even if no such points were realized in our data set. For the background population, we assume that all data points with $v_p > v_1$ are ignored, so that no integrated probability value accumulates at this boundary, and we set $V_x$ so that $\rho_{{\rm b}p}({\bm x})$ integrates to 1 on the ROI that is restricted to $v \le v_1$. With $v_1$ taken to be much greater than $\sigma_{vp}$ for all $p$, we obtain $V_x = (m_1 - m_0)(c_1 - c_0)v_1$.

On the other hand, we treat the $v = 0$ boundary for the background population density the same way we treat it in the case of modeled population densities. Thus, similarly to the manner of Section \ref{status}, we calculate the respective uniform background probability densities relevant for the data points with $v_p = 0$ and $\sigma_{vp} = \infty$ as
\begin{linenomath*}\begin{equation}
    \rho^{v0}_{{\rm b}p}(m, c) = \int_{-\infty}^0 \rho_{{\rm b}p}({\bm x}) = \frac{1}{V_{x}} \frac{\sigma_{vp}}{\sqrt{2 \pi}}
\end{equation}\end{linenomath*}
and
\begin{linenomath*}\begin{equation}
    \rho_{{\rm b}p}(m, c) = \int_{-\infty}^{v_1} \rho_{{\rm b}p}({\bm x}) = \frac{1}{(m_1 - m_0)(c_1 - c_0)}.
\end{equation}\end{linenomath*}

\section{Statistical Model} \label{statistical}

In this section, we describe our statistical model, which combines theoretical probability densities for different rotational and multiplicity populations  to infer the population parameters of the MSTO in NGC 1846 from the measurements of the turnoff's individual stars.

\subsection{Combined Probability Densities} \label{combined}

The cluster model in Section \ref{cluster} allows for 6 combinations of rotational population and multiplicity in the case when a data point is due to the stellar model, as well as the possibility that the datum is not due to the stellar model, but rather the background population. We now combine the probability densities for these 7 populations from Section \ref{probabilities} to obtain normalized densities for a given set of cluster parameters. For example, when a data point $p$ has rotational measurement $v_p > 0$, the combined probability density $\rho_p({\bm x}; {\bm \phi})$ is
\begin{linenomath*}\begin{multline} \label{eq:rho_p_phi}
    \rho_p({\bm x}; {\bm \phi}) = q(1-b) \sum_{i} w_{i}\, \rho_{i0p}({\bm x}; \mu_t, \sigma_t) + \\
    q b \sum_{i} w_{i}\, \rho_{i1p}({\bm x}; \mu_t, \sigma_t) + (1-q) \rho_{{\rm b}p}({\bm x}),
\end{multline}\end{linenomath*}
where the cluster parameters ${\bm \phi}$ are composed of fit quality $q$ (the fraction of stars described by the model), binary fraction $b$, rotational population proportions $(w_0, w_1, w_2)$, and parameters of the age prior $(\mu_t, \sigma_t)$.  These parameters obey $q \in (0, 1)$, $b \in (0, 1)$, $w_i \in (0, 1) \,\forall\, i$, and $w_0 + w_1 + w_2 = 1$.  Additionally, the second subscript $k$ in $\rho_{ikp}({\bm \cdot})$ is zero for the unaries and one for the binaries, ${\bm x} \equiv (m, c, v)$ is the observables vector, and $\rho_{{\rm b}p}({\bm x})$ is the background density for point $p$. We similarly obtain densities $\rho^{v0}_p(m, c; {\bm \phi})$ and $\rho_p(m, c; {\bm \phi})$, relevant for the other two cases of rotational measurement status. Much like in Section \ref{probabilities}, each probability density is only evaluated at the corresponding data point's observables, ${\bm x} = {\bm x_p}$. Additionally, we define a partial vector of cluster parameters ${\bm \phi'} = \{\mu_t, \sigma_t, w_0, w_2\}$ and, for every point $p$ with $v_p > 0$, likelihood factor 
\begin{linenomath*}\begin{equation} \label{eq:varrho}
    \varrho_p \equiv \varrho_p({\bm \phi}) \equiv \frac{\rho_p({\bm x_{\bm p}}; {\bm \phi})}{\rho_{{\rm b}p}({\bm x_{\bm p}})}.
\end{equation}\end{linenomath*}
 Quantity $\varrho_p$ is similarly defined when each relevant probability density has 2 dimensions instead of 3.

Next, we describe the statistical model that allows us to combine $\rho_p({\bm x}; {\bm \phi})$ and its lower-dimensional counterparts to obtain probabilities of all data under different cluster parameter combinations.

\subsection{The Likelihood of a Cluster Model} \label{cluster_like}

Our model of data generation assumes stars to arise as from a Poisson process.  It is closely related to an existing method for fitting data to stellar model distributions in color-magnitude space \citep{Naylor_2006MNRAS}, which was recently adapted to the space with dimensions of mass and rotational period \citepalias{Breimann_2021ApJ}.

Given cluster parameters ${\bm \phi}$, we assume that the $n_{\rm p}$ data points with positive rotational measurement $v_p > 0$ result from an $n_{\rm p}$-sized subset of $N_{\rm p} \gg n_{\rm p}$ Poisson processes, each non-homogeneous in ${\bm x}$-space and limited to the ROI. In other words, we assume a very large number of draws from underlying stellar probability distributions, a small fraction of which result in stars that appear in our data set. When we consider all the Poisson processes, we index them by $h \in \{1, \ldots, N_{\rm p}\}$. When we consider only the subset that produces data points, we use the same index we use for the points, $p \in \{1, \ldots, n_{\rm p}\}$. Let us say we have partitioned ${\bm x}$-space into a large number of bins, with widths $\Delta m$, $\Delta c$ and $\Delta v$ in each of the observable dimensions. In this case, the  expected number of stars $(\ll 1)$ resulting from process $h$ at location ${\bm x}$ is 
\begin{linenomath*}\begin{equation} \label{eq:lambda}
    \lambda_h({\bm x}) = \epsilon \,\rho_h({\bm x})\, \Delta {\bm x}, 
\end{equation}\end{linenomath*}
where $\epsilon \ll 1$, so that a given process does not produce more than one data point, $\rho_h({\bm x})$ is a probability distribution normalized on the ROI and given by Equation \eqref{eq:rho_p_phi}, and we have suppressed ${\bm \phi}$ in this distribution's definition.  In this case, $\Delta {\bm x} \equiv \Delta m \Delta c \Delta v$. 

If $k_h({\bm x})$ is the number of data points  
produced by process $h$ in a bin centered on ${\bm x}$, the probability of all data is 
\begin{linenomath*}\begin{equation} \label{eq:prob_poisson}
    \prod_h \prod_x {\frac{\lambda_h({\bm x})^{k_h({\bm x})} e^{-\lambda_h({\bm x})}}{k_h({\bm x})!}},
\end{equation}\end{linenomath*}
where $!$ represents the factorial.  The different Poisson processes indexed by $h$ are distinguished by differing uncertainties on the measured color, magnitude, and $v_e \sin i$.  In this case, since each Poisson process produces at most one star, the denominator is unity.  

If all data points had the same uncertainties, then each distribution $h$ would be identical.  In this case, the total number of stars in the bin would be a Poisson random number with expectation value
\begin{equation}
\label{eq:lambda_sum}
    \lambda({\bm x}) = \sum_h \lambda_h({\bm x})
\end{equation}
and an actually detected number of stars
\begin{equation}
\label{eq:k_sum}
    k({\bm x}) = \sum_h k_h({\bm x}).
\end{equation}

The probability of observing the data would then be
\begin{equation}
     \label{eq:prob_poisson_singleH}
    \prod_x {\frac{\lambda({\bm x})^{k({\bm x})} e^{-\lambda({\bm x})}}{k({\bm x})!}}.
\end{equation}
This differs from Equation \eqref{eq:prob_poisson}, but if all Poisson processes $h$ are identical, it differs only by a constant independent of the model that gives $\lambda$.  Specifically, the denominator in the two equations depends only on the number of stars actually observed in a given bin, and their exponential term is identical given Equation \eqref{eq:lambda_sum}.  The term $\lambda({\bm x})^{k({\bm x})}$ will differ by a constant, equal to $N_p^{k({\bm x})}$, from its corresponding term in Equation \eqref{eq:prob_poisson}.  In sum, Equation \eqref{eq:prob_poisson} is more general than Equation \eqref{eq:prob_poisson_singleH} but the former equation reduces to the latter (up to a constant) if the uncertainties on all stellar measurements are identical.

Equation \eqref{eq:prob_poisson} gives the probability of detecting a given number of stars in discrete bins of color-magnitude-$v\sin i$ space.  In color and magnitude alone, these bins form a Hess diagram \citep{Bastian_SilvaVilla_2013, Rubele_2013MNRAS}, where an integer number of stars are present in each bin.  Hess diagram approaches based on Equation \eqref{eq:prob_poisson_singleH} have often been used to infer cluster properties.  However, they cannot account for differences in uncertainties between different stars and they cannot naturally account for $v\sin i$ as the third dimension.  Our approach is different: we take the limit where
 $\Delta m \to 0$, $\Delta c \to 0$, and $\Delta v \to 0$. In this limit, with $k({\bm x})$  either 0 or 1, the probability of all data in Equation \eqref{eq:prob_poisson} becomes 

\begin{linenomath*}\begin{equation}
\label{eq:prob_limit}
    \prod_h \left(\prod_x e^{-\lambda_h({\bm x})}\,\prod_{{\bm x}:\, k_h({\bm x}) = 1} \lambda_h({\bm x})\right).
\end{equation}\end{linenomath*}
In this limit, the probability distributions are continuous rather than discrete and $v \sin i$ information can be naturally incorporated.  It does, however, require us to use the continuous probability distributions that we have computed in Section \ref{integration}.

Equation \eqref{eq:prob_limit} contains two components within the parentheses.  The first term is nontrivial for all Poisson processes indexed by $h$.  The second term, however, is unity unless Poisson process $h$ actually results in a detected star, i.e., unless $k_h({\bm x}) = 1$ for some ${\bm x}$ (otherwise $\lambda_h$ is always raised to the zero power). Consequently, for the second term in Equation \eqref{eq:prob_limit}, we  switch to indexing by $p$ to indicate the processes that produce data points.   Expression \eqref{eq:prob_limit} becomes 
\begin{linenomath*}\begin{equation} \label{eq:prob_prod}
    \left(\prod_h \prod_x {e^{-\lambda_h({\bm x})}}\right)\, \left( \prod_{p} \lambda_p\left({\bm x_{\bm p}}\right)\right),
\end{equation}\end{linenomath*}
where the right product is only over the Poisson processes that produce data points, since the product factor for other processes is equal to 1. Additionally, for now, this product is restricted to the data points with $v_p > 0$. With the help of Equation \eqref{eq:lambda}, the left product in Expression \eqref{eq:prob_prod} can be written as
\begin{linenomath*}\begin{multline} \label{eq:prob_left}
    \prod_h \exp{\left(- \epsilon \sum_{{\bm x}} \rho_h({\bm x})\,\Delta {\bm x}\right)} =\\
    \prod_h \exp{\left(- \epsilon \int_{\rm ROI} \rho_h({\bm x})\, {\rm d}{\bm x}\right)} = e^{-\epsilon N_{\rm p}},
\end{multline}\end{linenomath*}
where we have applied the limit that turns $\Delta {\bm x}$ into ${\rm d}{\bm x}$ and the sum into an integral. We also used the fact that $\rho_h({\bm x})$ is normalized on the ROI.

Application of Equation \eqref{eq:lambda} allows us to write the right product in Expression \eqref{eq:prob_prod} as
\begin{linenomath*}\begin{equation} \label{eq:prob_right}
    \prod_p \epsilon \,\rho_p\left({\bm x_{\bm p}}\right)\, \Delta {\bm x} = \epsilon^{n_{\rm p}} \left(\Delta {\bm x}\right)^{n_{\rm p}} \prod_p \rho_p\left({\bm x_{\bm p}}\right).
\end{equation}\end{linenomath*}

\begin{figure*}[ht]
\includegraphics[width=0.5\linewidth]{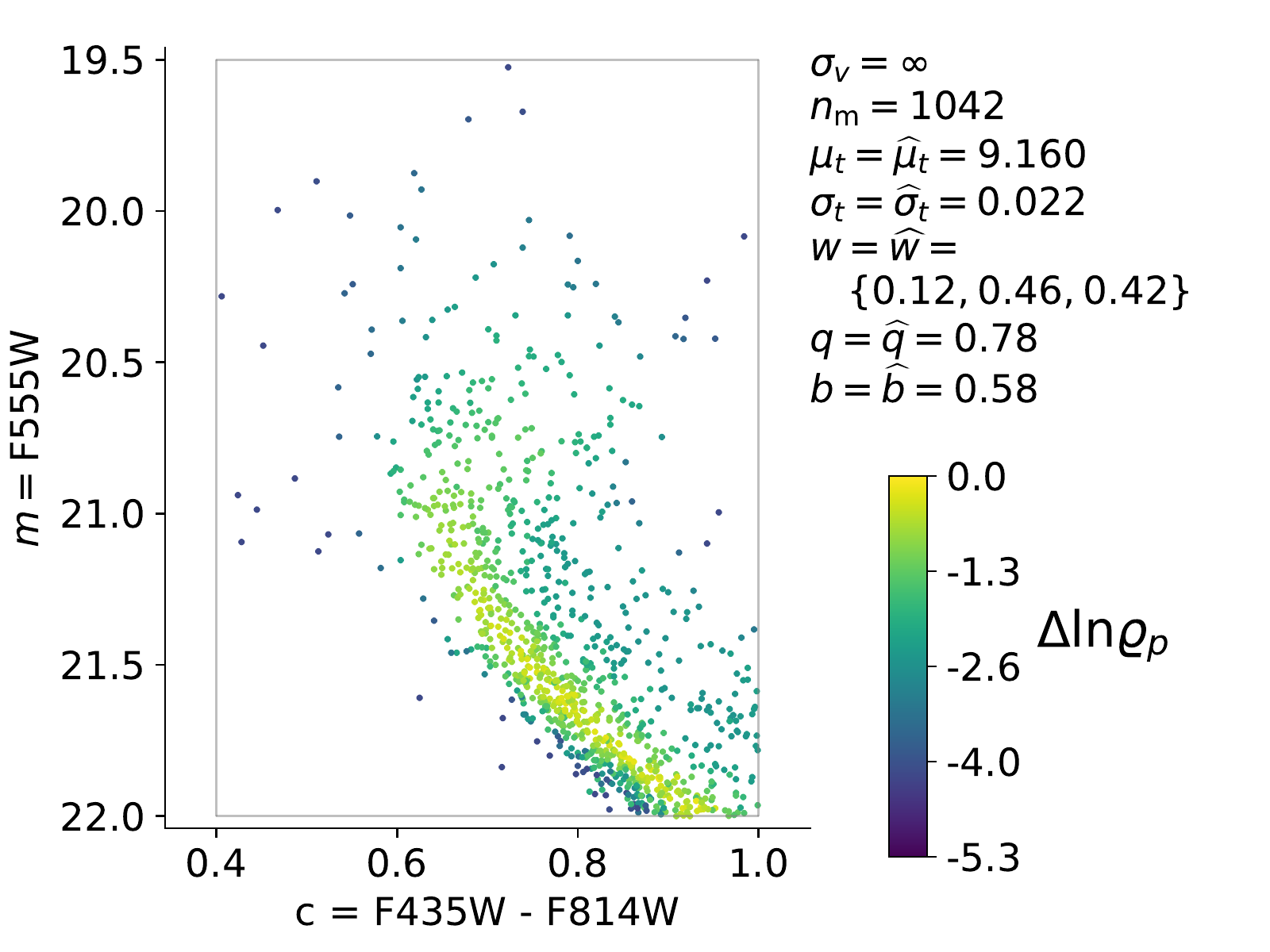}
\includegraphics[width=0.5\linewidth]{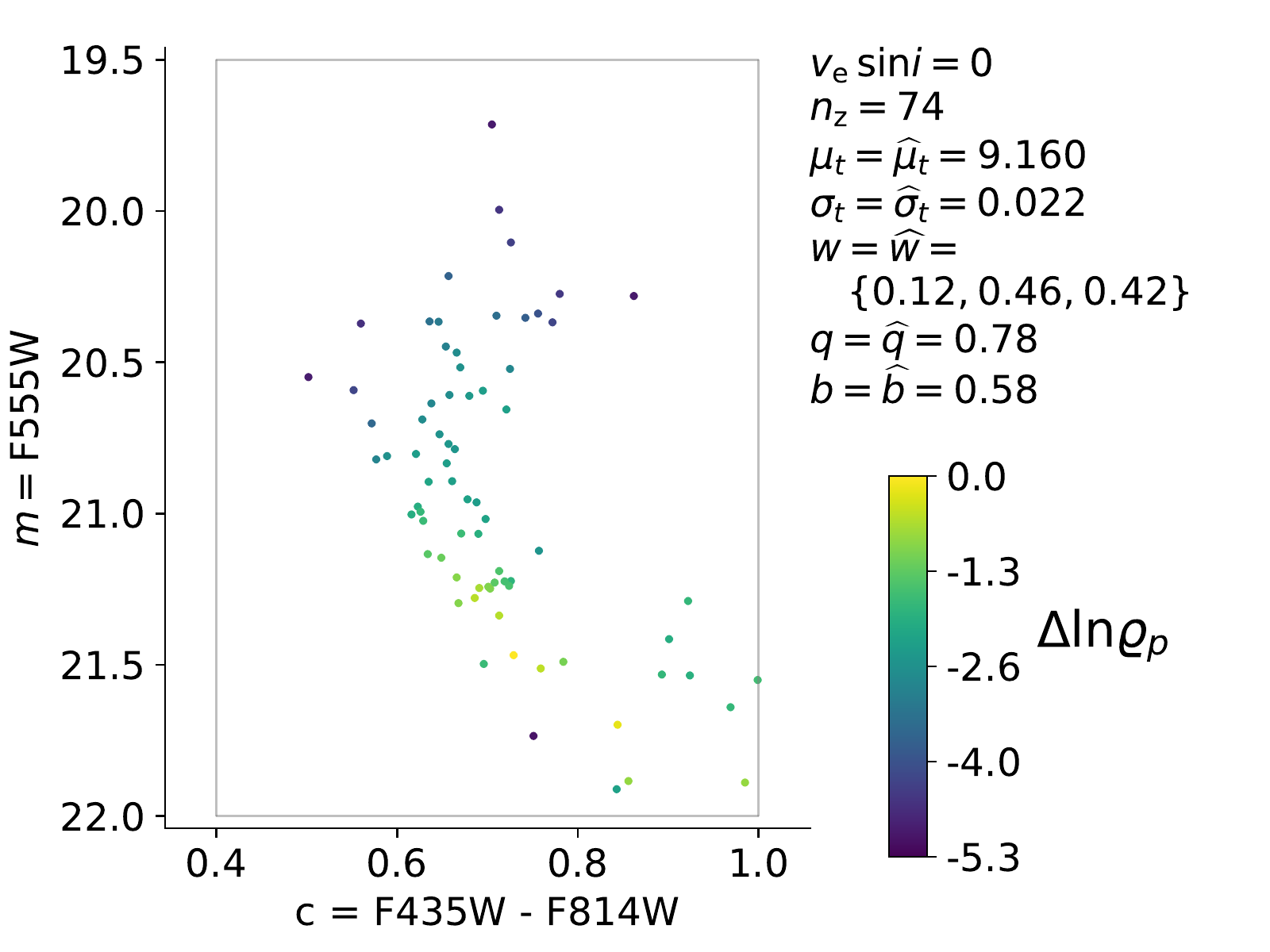}\\
\includegraphics[width=0.5\linewidth]{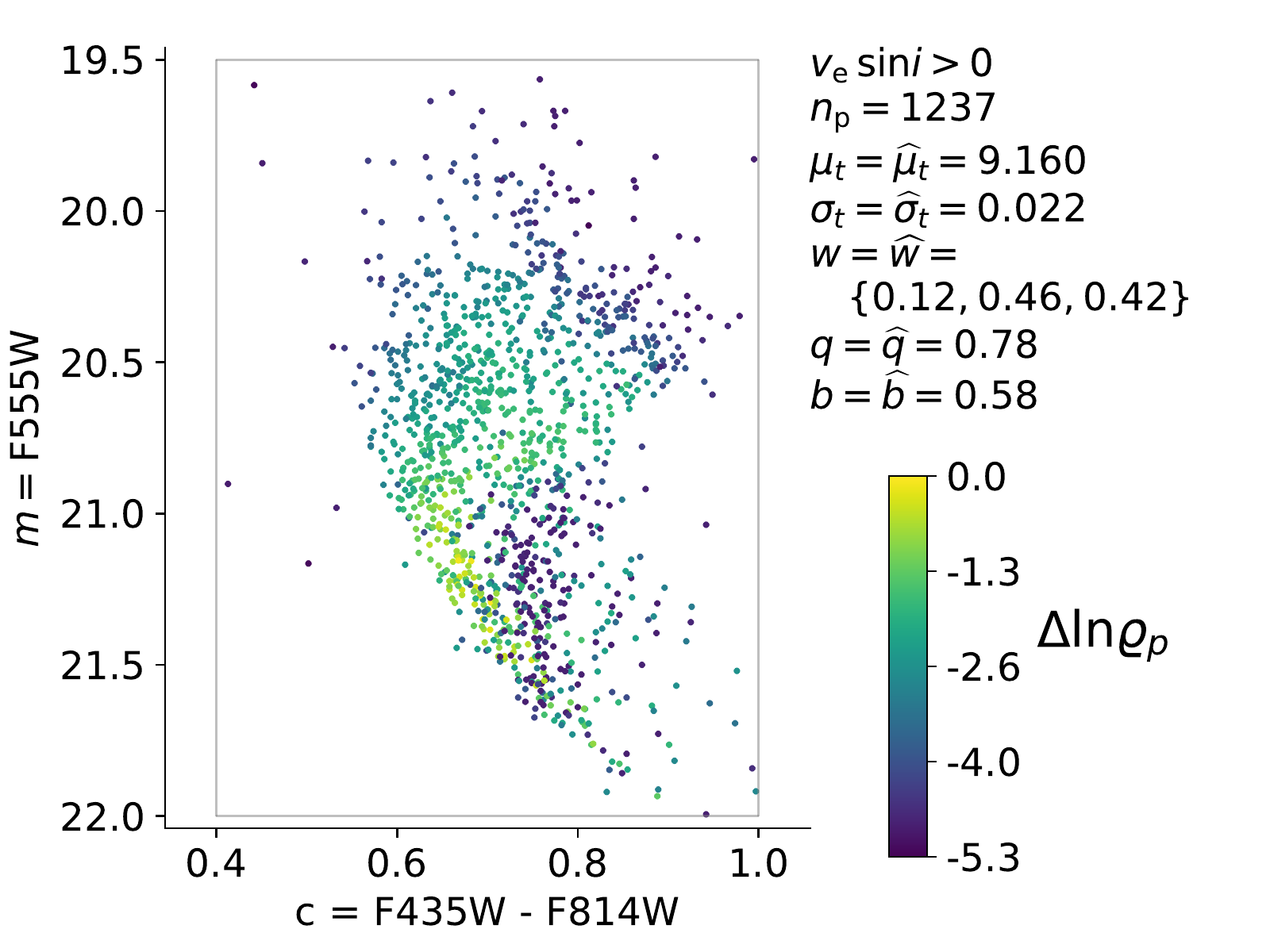}
\includegraphics[width=0.5\linewidth]{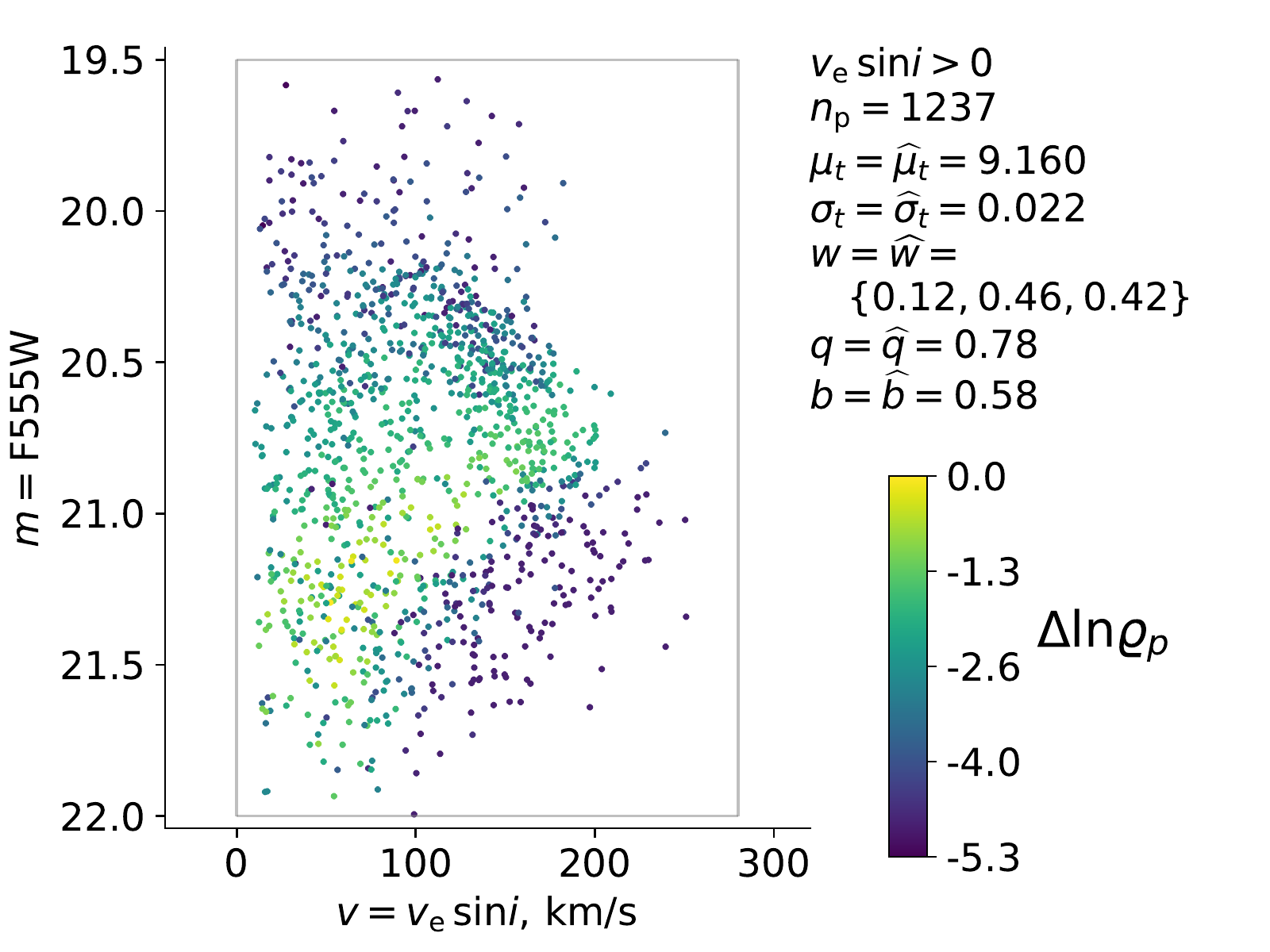}
\caption{Color scheme shows the relative likelihood factors $\Delta\ln{\varrho_p}$ for individual points (defined due to Section \ref{cluster_like}) in our portion of the data from \citetalias{Kamann_2020MNRAS}, at the maximum-likelihood cluster parameters $\hat{\bm \phi}$. Comparison of two factors is only meaningful when their rotational measurement status is the same. Thus, the top left panel compares the 1042 points without $v_{\rm e}\sin{i}$ measurements, the top right panel -- the 74 points with $v_{\rm e}\sin{i} = 0$, and the bottom two panels -- the 1237 points with $v_{\rm e}\sin{i} > 0$. Likelihood factor difference of $\Delta\ln{\varrho_p} = 2$, such as between a yellow point and a green point, indicates a 7.4-fold difference in probability density between the locations of the two points in observable space, since $e^{2}\approx7.4$. Cluster model parameters $A_V$, ${\rm [M/H]_M}$, and $\sigma_\omega$ are the same as in Figure \ref{fig:cmd_all_ages}.}
\vspace{20pt}
\label{fig:rho_p}
\end{figure*}

Multiplying Expressions \eqref{eq:prob_left} and \eqref{eq:prob_right} together, we see that the probability of the data points with $v_p > 0$, given by Expression \eqref{eq:prob_prod}, is 
\begin{linenomath*}\begin{equation} \label{eq:prob_positive}
    e^{-\epsilon N_{\rm p}} \epsilon^{n_{\rm p}} \left(\Delta {\bm x}\right)^{n_{\rm p}} \prod^{n_{\rm p}}_p \rho_p\left({\bm x_{\bm p}}\right).
\end{equation}\end{linenomath*}
We repeat the above procedure in this section for the data points with each of the remaining possibilities of the rotational measurement status, in each case substituting $\rho_p\left({\bm x_{\bm p}}\right)$ with the appropriate 2-dimensional distribution and $\Delta {\bm x}$ with $\Delta m\,\Delta c$. The probability of all data points with $v_p = 0$ turns out to be
\begin{linenomath*}\begin{equation} \label{eq:prob_zero}
    e^{-\epsilon N_{0}} \epsilon^{n_{0}} \left(\Delta m\,\Delta c\right)^{n_{0}} \prod^{n_{0}}_p \rho_p\left(m_p, c_p\right),
\end{equation}\end{linenomath*}
and similarly for the data points with $\sigma_v = \infty$ (i.e., those without measured $v \sin i$). Now, we denote $\rho_p\left({\bm x_{\bm p}}\right)$, $\rho_p\left(m_p, c_p\right)$ and $\rho_p^{v0}\left(m_p, c_p\right)$ collectively as $\rho_p$. Multiplying together Expressions \eqref{eq:prob_positive}, \eqref{eq:prob_zero} and the remaining, similar expression, we obtain the probability of all the data:
\begin{linenomath*}\begin{equation} \label{eq:prob_all}
    e^{-\epsilon N} \epsilon^n \left(\Delta v\right)^{n_{\rm p}} \left(\Delta m \, \Delta c\right)^{n} \prod^{n}_p \rho_p,
\end{equation}\end{linenomath*}
where $N$ is the total number of Poisson processes. We are free to define the likelihood of cluster parameters ${\bm \phi}$ as Equation \eqref{eq:prob_all} times any factor that doesn't depend on ${\bm \phi}$. We first retain only the right-most product over the data points indexed by $p$ in this expression, since all other factors are independent of ${\bm \phi}$. We then divide this product by the product of the appropriate 2- and 3-dimensional background densities at data point observables, which is also independent of ${\bm \phi}$. This yields the following likelihood function:
\begin{linenomath*}\begin{equation} \label{eq:like}
    {\cal L}({\bm \phi}) = \prod^n_p \varrho_p,
\end{equation}\end{linenomath*}
where $\varrho_p$ are the data point likelihood factors, defined in Equation \eqref{eq:varrho}. Appendix \ref{like_computation} describes the procedure that leads to $\hat{\bm \phi}$, the cluster parameters that maximize ${\cal L}({\bm \phi})$ in Equation \eqref{eq:like}. We split the data set into subsets that correspond to the three statuses of rotational measurement and calculate the relative differences in $\ln{\rho_p}$ at $\hat{{\bm \phi}}$ within each subset. These differences are presented in Figure \ref{fig:rho_p}.  The exponent of the sum of these over all stars gives the likelihood of the set of cluster parameters $\hat{\bm \phi}$ that maximizes the likelihood function.

\subsection{Posterior Cluster Parameters}

Equation \eqref{eq:like} gives the likelihood of a set of cluster parameters.  Our final step is to normalize the likelihood to obtain a posterior probability distribution of these cluster parameters.  We do not use MCMC, but rather directly integrate the likelihood multiplied by our adopted priors on cluster parameters ${\bm \phi}$. 
We assume log-uniform priors on $\mu_t$ and $\sigma_t$ and uniform priors on all other components of ${\bm \phi}$. This way, likelihood as a function of ${\bm \phi}$ can already be seen as the un-normalized posterior. We then define
\begin{linenomath*}\begin{equation} \label{eq:like_phi_prime}
    {\cal L}({\bm \phi'}) = \int {\rm d}q\,{\rm d}b \,{\cal L}({\bm \phi}).
\end{equation}\end{linenomath*}
Details of the integration procedure that  we use to evaluate Equation \eqref{eq:like_phi_prime} can be found towards the end of Appendix \ref{like_computation}.

If we normalize ${\cal L}$, we can interpret it as a Bayesian probability density, $P({\bm \phi}) = {\cal L}({\bm \phi}) Z_{\cal L}^{-1}$, where
\begin{linenomath*}\begin{equation} \label{eq:z_l}
    Z_{\cal L} = \int {\rm d}{\bm \phi} \,{\cal L}({\bm \phi})
\end{equation}\end{linenomath*}
is an integral over some formal region of normalization.

We wish to obtain $P({\bm \phi'}) \propto {\cal L}({\bm \phi'})$ after evaluating the likelihood over a subset of the normalization region. We can also marginalize $P({\bm \phi'})$ in $w_0$ and $w_2$:
\begin{linenomath*}\begin{equation} \label{eq:prob_t}
    P(\mu_t, \sigma_t) = \int {\rm d}w_0\,{\rm d}w_2\,P({\bm \phi'}) \propto \int {\rm d}w_0\,{\rm d}w_2\,{\cal L}({\bm \phi'}),
\end{equation}\end{linenomath*}
which would provide us with a confidence region for the age distribution parameters. Similarly, we can get a confidence region for the rotational population proportions by calculating $P(w_0, w_2)$. Appendix \ref{bayesian} describes the integration procedures that produce $P(\mu_t, \sigma_t)$ and $P(w_0, w_2)$ in the fashion suggested by Equation \eqref{eq:prob_t}.

Our final step is to assess the goodness of fit: whether the maximum likelihood cluster parameters, together with the stellar model, provide a good description of the cluster.  We assess goodness of fit by the maximum likelihood value of parameter $q$, the fraction of stars that are described by the cluster model, given the maximum likelihood values of all the cluster parameters.  The remainder of the stars, a fraction $1-q$, must be accounted for in a background population.  Our sample of stars near the main sequence turnoff is overwhelmingly dominated by real cluster members.  A formally good model, then, should have $q$ very close to one ($\gtrsim$0.95).  Lower values of $q$ indicate that many cluster stars cannot be well-fit by the stellar model, and that the rest of the cluster parameters should be interpreted cautiously.

\section{Results} \label{results}

In this section, we present the maximum-likelihood (ML) cluster parameter estimates that result from the evaluation of likelihoods that we defined in Section \ref{cluster_like}. We also offer bounds on these estimates, which are based on the integration of the likelihoods, as described in Appendix \ref{like_computation} and the integration of Bayesian probabilities in Appendix \ref{bayesian}. We caution that, due to the intermediate quality of the fit between the evolutionary model and the data, our cluster parameter estimates are only somewhat reliable. In this section and, especially, in Sections \ref{discussion} and \ref{summary}, we discuss the degree of reliability and the ways in which one might calibrate the stellar evolutionary model to better fit cluster data and consequently produce cluster parameter estimates that are more trustworthy.

Our ML estimate of the probability that an observed star is due to the evolutionary model is $\hat{q} = 0.78$.  In other words, 22\% of the stars are better explained by a uniform background distribution.  The actual fraction of contaminants is expected to be much lower \citepalias[$\approx$ 190/3189 $=$ 6\%, based on Sections 3.4 and 3.5 of][]{Kamann_2020MNRAS}.  Even though our $\hat{q}$ indicates that the stellar model can account for the observed
 photometry and $v_{\rm e}\sin{i}$ measurements of most stars in the ROI, the remainder of the stars constitute a signficant minority.
The $\sim$80\% of stars that are accounted for by the stellar model contribute to the inference of cluster parameters $w_0$, $w_2$, $\mu_t$, $\sigma_t$, and $b$ in this work.  An evolutionary model with a higher $\hat{q}$ would fit the data better, thus producing cluster parameter estimates that would be more reliable. Since $1 - \hat{q} = 0.22$ is appreciable, such new estimates could be very different from this work's. The parameter $\hat{q}$ can serve as a measure of the goodness-of-fit of the stellar model.

Roughly speaking, the non-zero value of $1 - \hat{q}$ results from 22\% of the data points with the lowest likelihood factor offset $\Delta\ln{\varrho_p}$ within each subset of the rotational measurement status in Figure \ref{fig:rho_p}.  These are the stars that are most inconsistent with our cluster model. The bottom panels of this figure present relative likelihood factors $\Delta\ln{\varrho_p}$ for individual stars with $v_{\rm e}\sin{i} > 0$ at ML cluster parameters. Of these stars, 316/1237 = 26\% have $\Delta\ln{\varrho_p} \le -4$ . We define these as the $v_{\rm e}\sin{i} > 0$ data points that poorly match the evolutionary model.  Near the middle of the turnoff, at a magnitude $m \approx 20.7$, nearly all stars are satisfactorily accounted for by the model.  At brighter magnitudes, the model predicts a smaller proportion of stars (see Figure \ref{fig:cmd_all_ages}).  At fainter magnitudes, it predicts the stars to exist only in a narrow color spread around $c = 0.75$ and at very low rotational speeds (see Figure \ref{fig:vmd}). 
 As we discuss later, in Section \ref{discussion}, it is likely that reduction in the evolutionary model's magnetic braking may significantly improve the model's match to the dimmer points.

As furthermore discussed in Section \ref{discussion}, our ML estimate of the binary fraction, $\hat{b} = 0.58$, is almost certainly higher than the parameter's real value; a  reduction in magnetic braking is likely to reduce  our estimate significantly. Thus, we do not compute the formal confidence region for $b$, although Section \ref{like_computation} shows that, generally, $\sim99.9\%$ of the integrated likelihood lies between $b_0 = 0.40$ and $b_1 = 0.76$. We similarly treat the confidence region for $q$, with the following limits from Section \ref{like_computation}: $q_0 = 0.70$ and $q_1 = 0.84$.

\begin{figure*}[ht]
\includegraphics[width=0.5\linewidth]{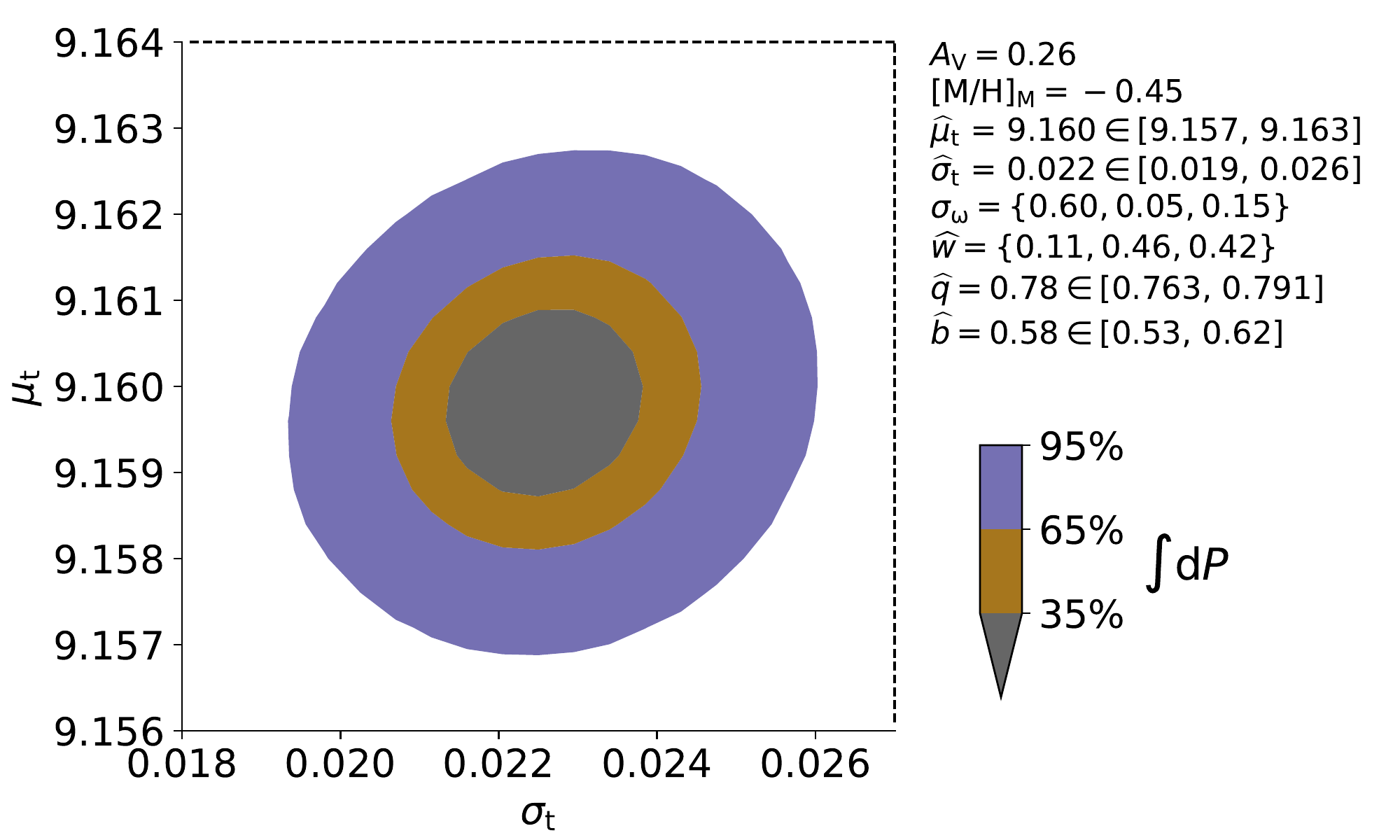}
\includegraphics[width=0.5\linewidth]{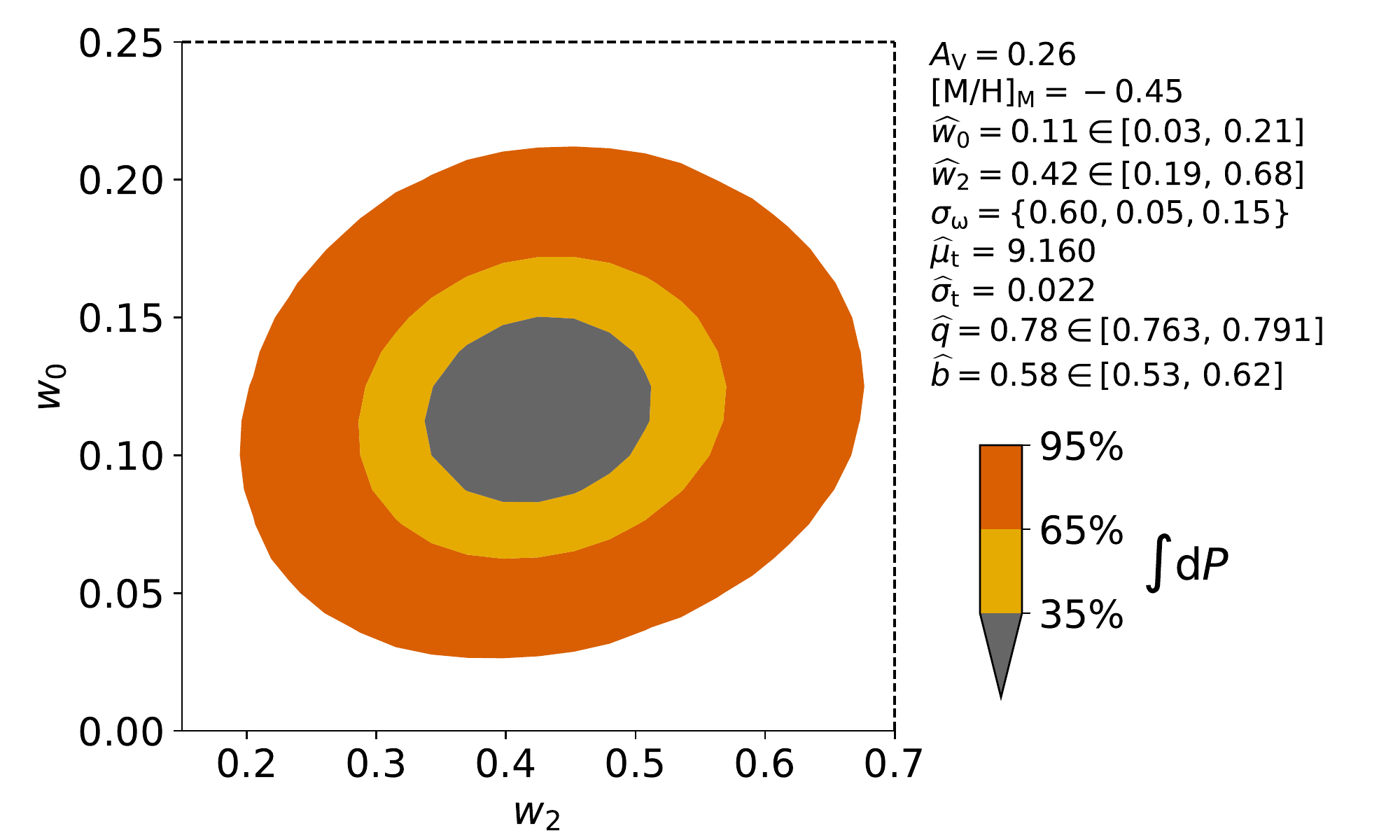}
\caption{Contours designate 35\%, 65\% and 95\% confidence regions due to marginalized Bayesian probability densities of cluster model parameters. Left panel: density over mean stellar age and age standard deviation, marginalized in all other parameters. Right panel: same, over the rotational population proportions. Intervals for $\hat{q}$ and $\hat{b}$ indicate the ranges of maximum-likelihood values of $q$ and $b$ over all combinations $(w_0, w_2, \mu_t, \sigma_t)$ in this figure.}
\vspace{20pt}
\label{fig:prob}
\end{figure*}

In Section \ref{cluster}, we state our rotational population model, with slow rotators distributed according to a wide half-Gaussian peaked at zero rotation, fast rotators -- according to a narrow half-Gaussian peaked at critical rotation, and intermediate rotators -- according to a narrow Gaussian with a peak at the location where the other two probability densities are equal. We chose the widths of the three distributions to ensure that the ML estimates of the corresponding population proportions are all appreciably greater than zero, i.e., that the data distinguish between three separate populations to a large degree. The population proportion estimates are between the corresponding 1-dimensional boundaries of the 2-dimensional 95\% confidence region in the right panel of Figure \ref{fig:prob}: $\hat{w}_0 = 0.11 \in [0.03, 0.21]$ and $\hat{w}_2 = 0.42 \in [0.19, 0.68]$ for the slow and the fast rotators, respectively. The width of the confidence region in the $w_2$ dimension is significantly larger than that in the $w_0$ dimension, indicating that the slow rotator population is more distinct from the fast and intermediate rotators than the latter are from each other. This interpretation makes sense in view of a qualitative comparison between the three populations' theoretical probability densities in Figure \ref{fig:cmd} and suggests that the true rotational distribution is bimodal instead of trimodal. 

The population proportion of the intermediate and fast rotators is $\hat{w}_1 + \hat{w}_2 = 0.89$--the vast majority of stars. This combined population is somewhat larger than the population with high $v_{\rm e}\sin{i}$ in \citetalias{Kamann_2020MNRAS} with a distribution that peaks around $v_{\rm e}\sin{i} = 140\,{\rm km\,s^{-1}}$ and a population proportion of $\sim0.55$. The correspondence is very rough, considering both the difference in the estimated proportions between the two studies and the fact that all rotational populations in this work have probability distributions that extend over most of the $v_{\rm e}\sin{i}$ range (e.g., see Figure \ref{fig:vmd}). Nonetheless, it is encouraging that our results, like those of \citetalias{Kamann_2020MNRAS}, point to a bimodal rotational distribution.

Our ML estimates of the age distribution parameters are within the 95\% confidence region in the left panel of Figure \ref{fig:prob}: $\hat{\mu}_t = 9.1600 \in [9.1569, 9.1628]$ and $\hat{\sigma}_t = 0.0225 \in [0.0193, 0.0260]$. The corresponding non-logarithmic values are $1.445 \in [1.435, 1.455]\,{\rm Gyr}$ and $75 \in [64, 87]\, {\rm Myr}$, where the non-logarithmic equivalent of a logarithmic standard deviation $\sigma_t$ is $10^{\hat{\mu}_t + \left(\sigma_t/2\right)} - 10^{\hat{\mu}_t - \left(\sigma_t/2\right)}$. Parameters $\hat{\mu}_t = 9.160$ and $\hat{\sigma}_t = 0.023$ correspond to an age distribution with high probability of $\log{t} = 9.14$, the age adopted in \citetalias{Kamann_2020MNRAS}. The left panel of Figure \ref{fig:prob} shows that the Bayesian probability distribution covariance between $\hat{\mu}_t$ and $\hat{\sigma}_t$ is small, which suggests that our age and age dispersion estimates are not greatly affected by the specific log-normal shape of the prior on stellar age. Furthermore, since the posteriors on $\mu_t$ and $\sigma_t$ are both rather narrow, we conclude that our estimates of these parameters are not greatly affected by our specific choices of their relatively uninformative priors.

\section{Discussion} \label{discussion}

Both the theory of stellar evolution and the theory of cluster formation have ingredients that are subject to considerable uncertainty. On the other hand, well-established ingredients of one of these theories could help reduce uncertainty in the other. We are specifically interested in a better understanding of the rotational and age distributions of stars within clusters, as well as the internal transport processes that are linked to the rotation and evolution of individual stars. 
Our work offers a robust statistical framework that connects the theory of rotating star evolution and the theory of cluster formation in view of spectro-photometric data from many stars in a given cluster. Much of this work builds on the studies by \citetalias{Gossage_2019ApJ} and \citetalias{Brandt_2015ApJ_58}. 

Our case study is based on the photometry and projected equatorial velocities $v_{\rm e}\sin{i}$ of main sequence turnoff (MSTO) stars in the intermediate-age globular cluster NGC 1846. We assume the \texttt{MIST} stellar evolution model and allow for rotational and age distributions in the cluster, constraining them using free parameters. We build a detailed statistical framework to obtain these constraints as posterior probability densities, but this entire framework operates under the fundamental assumption that the \texttt{MIST} models are accurate.
Our probability distributions lead to estimates of cluster parameters that are tightly linked to the particulars of the evolutionary model.

When allowing for the cluster stars to possess a range of rotation rates, we obtain an age dispersion that is about half the previous estimates due to non-rotating models. This result agrees with the conjecture that rotational variation is at least partially responsible for eMSTOs. Still, both the age dispersion and the binary fraction that we obtain are greater than those suggested by previous, independent studies. Our relatively large age variations and binary frequencies may be compensating for other sources of physical variation that are not present (or insufficiently present) in the \texttt{MIST} models.
Consideration of the fit suggests specific rotation-related processes that one may be able to tune in the model to simultaneously improve the fit to individual data, produce a lower estimate of the binary fraction, and further lower the estimated age spread.

In sum, a comparison of theoretical probability distributions to individual star data and a comparison of inferred cluster parameters to independent estimates lead to suggestions of evolutionary processes that can improve both fits. In the remainder of this section, we offer a detailed account of this reasoning process and the evolutionary model tuning that it suggests. In future work we will apply our approach in the other direction: tuning stellar evolutionary parameters to better match the properties of cluster stars. 

\subsection{Reduced Magnetic Braking Of Low-Mass Stars}

A population of single, nonrotating, coeval stars follows a line in color-magnitude space.  Rotational variation, binarity, and age distributions will all broaden the main sequence, while rotational variations and age distributions can broaden the turnoff in particular. 

As an example, consider that the single-age probability distributions in the left panels in Figure \ref{fig:cmd} are significantly narrower in color than the corresponding age-dispersed distributions in the left panels of Figure \ref{fig:cmd_all_ages}. Additionally, comparison between the left and the right panels in Figures \ref{fig:cmd} and \ref{fig:cmd_all_ages} shows that binarity could be broadening the MSTO area in its own way, as well. \citetalias{Bastian_2015MNRAS} obtain an age dispersion of 136 Myr from the eMSTO of NGC 1846 under the assumptions of zero binarity and rotation. In view of the discourse in Section \ref{introduction}, one wonders whether the inclusion of rotational and multiplicity degrees of freedom in the evolutionary model might allow for a narrower age distribution. As mentioned in Section \ref{results}, the inclusion of these degrees of freedom, in addition to the rotational information inherent in projected equatorial velocities, yields an age dispersion estimate of $75\in[64, 87]\,{\rm Myr}$. This estimate is, indeed, significantly lower than the above-mentioned estimate due to unary, non-rotating models.

A close look at Figures \ref{fig:cmd} and \ref{fig:cmd_all_ages} reveals that the unary probability distributions at $m \gtrsim21.0$ are generally narrow in color, even though they include dispersions in the initial rotation rate. This portion of the CMD contains relatively low-mass \texttt{MIST} models ($1.25\,M_\odot\lesssim M_{\rm i} \lesssim1.5\,M_\odot$) that have spun down magnetically towards zero rotation rate, so that the corresponding rotational broadening is small. Age dispersion in \texttt{MIST} also does not widen this part of the distribution to match the observed color spread, although binarity does. Hence, our high estimate of the binary fraction, $\hat{b} = 0.58$. We emphasize that this estimate is driven by the lower portion of the main sequence turnoff, with magnitudes above $21.0$ that correspond to masses below $1.5\,M_\odot$. 

According to Figure \ref{fig:vmd}, as $m$ increases from 21.0, predicted probability densities become small beyond a decreasing upper limit in $v_{\rm e}\sin{i}$, although there are many  observed stars beyond that limit, i.e., that remain rapidly rotating at the observed age of NGC~1846. 
 In other words, the \texttt{MIST} implementation leads to a Kraft break that is higher on the CMD than it ought to be. We hypothesize that there is room for reduction in the magnetic braking efficiency of the \texttt{MIST} models in the dim portions of the CMD to significantly increase the models' rotational speeds and produce better-fitting probability densities. Such calibration of magnetic braking theory to an intermediate-age massive cluster might both resemble and complement the recent calibrations of magnetic braking to open clusters by \citet{Gossage_2021ApJ} and \citetalias{Breimann_2021ApJ}. 
Since the models with reduced braking would spin faster, the probability densities would also broaden at $m \gtrsim21.0$ on the CMD, up to significant fractions of the current rotational broadening at $m \lesssim21.0$. This would, in turn, reduce the need for a high binary fraction to broaden the dimmer portion of the MSTO on the CMD.

A lower estimate of the binary fraction would better agree with independent estimates for NGC 1846 and similar clusters. For example, \citetalias{Kamann_2020MNRAS} use the radial velocity (RV) variation technique to estimate that unresolved binaries constitute $\sim$6\% of the stars in NGC 1846. Based on the CMD of this cluster's main sequence, \citet{Goudfrooij_2009AJ} estimate that its binary fraction is $\sim15\%$, which is somewhat different than \citetalias{Kamann_2020MNRAS}'s estimate but  still significantly lower than this work's value. Additionally, Galactic clusters that are similar to NGC 1846 appear to have comparatively low binary fractions, on the order of a few percent \citep{Milone_2012A&A}.

The reduction in magnetic braking that we suggest would not rotationally broaden the part of the CMD at $m \lesssim21.0$ in Figures \ref{fig:cmd} and \ref{fig:cmd_all_ages}, since the stars with corresponding masses, $M_{\rm i} \gtrsim 1.5\,M_\odot$, do not brake very much in the first place.  The poor agreement at the bright end of the MSTO must have another explanation; one possibility is

an age dispersion that is similar to this work's 75 Myr. On the other hand, several lines of evidence in Section \ref{intro_distributions} suggest that the age dispersion in NGC 1846 and similar clusters is lower than 10 Myr. In particular, \citetalias{Bastian_2015MNRAS} find that the CMD spreads of the cluster's sub-giant branch (SGB) and red clump (RC) regions are consistent with zero age spread. Furthermore, \citet{Bastian_CabreraZiri_2013MNRAS} find that clusters as young as 10 Myr and otherwise similar to NGC 1846 exhibit no evidence of on-going stellar formation. We now turn to possible physical explanations of the disagreement at the top of the turnoff other than a real age dispersion.

\subsection{Enhanced Effect of Rotation on Internal Mixing}

In looking for rotation-related processes other than magnetic braking that one could tune to further reduce the inferred age spread, we turn to the work of \citet{Brandt_2015ApJ_25}. Instead of \texttt{MIST}, these authors compare the \texttt{SYCLIST} library \citep{Georgy_2013A&A} with NGC 1846 MSTO photometry. They find that the eMSTO of NGC 1846 is qualitatively consistent with instantaneous star formation, if rotational variation is present. A comparison of \texttt{MIST} and \texttt{SYCLIST} models at the MSTO suggests that rotation increases internal mixing in \texttt{SYCLIST} more than it does in \texttt{MIST}. As a result, rotating models age more slowly in \texttt{SYCLIST}, so that a distribution of rotation rates in this model library has a greater propensity to mimic an age distribution \citep{Gossage_2018ApJ,Gossage_2019ApJ,Brandt_2015ApJ_58}. 
Future work might modify \texttt{MIST} v1.0 with enhanced internal mixing to test whether these changes to the mixing physics   can explain the extent of this MSTO without any need for an age dispersion. 
Enhanced rotational mixing, combined with decreased magnetic braking for stars $\approx$1.3\,$M_\odot$, could also modify the lower-end of the MSTO.

Future work could also
 re-construct \texttt{MIST} with a more recent version of \texttt{MESA}, a version that better incorporates the extreme effects of near-critical rotation \citep[see the introduction to Section 4 in][]{Paxton_2019ApJS}.  In addition, our statistical approach offers a path to tune the coefficients that regulate the onset and the degree of mixing due to rotation in \texttt{MESA} \citep[$f_c$ and $f_\mu$ in][]{Gossage_2018ApJ}.

\subsection{Additional Remarks}

On the whole, we have suggested several modifications to stellar evolution models, including reduction in magnetic braking for stars with $M_{\rm i} \lesssim 1.5\,M_\odot$ and enhancement of rotation's effect on internal chemical mixing. Future work will show how effectively such modifications can bring evolutionary models to an ideal fit with the data. It is possible, for example, that an increase in the models' internal chemical mixing will fail to account for the poorly matching points with magnitudes brighter than 21. In this case, an analysis such as ours will continue to largely model these points as part of the background distribution. As in all other such cases, the interpretation would be that the points are stars that are not included as a possibility in our stellar evolution model.

The current \texttt{MIST} library has an important limitation that might be affecting our conclusions: the library only allows for stellar models with initial dimensionless rotation rates $\omega_{\rm i} \le 0.8590$. If rotation rates of a sizeable fraction of stars are above this limit, inclusion of models with $0.8590 < \omega_{\rm i} < 1$ would increase rotational broadening of the MSTO and consequently may reduce the required binary fraction and age spread. We can estimate the extent of this putative effect, based on the evidence that near-critical rotation, i.e., with $\omega$ approaching 1, is quite common in clusters \citep{Townsend_2004MNRAS,Bastian_2017MNRAS}. The largest $v_{\rm e}\,\sin{i}$ measurements in our data set, $\gtrsim200\,{\rm km/s}$, likely correspond to critical or near-critical rotation \citepalias[e.g., see Section 5.5.1 and Figure A.1 in][]{Kamann_2020MNRAS}. Part of the uncertainty in this correspondence is due to the complications induced by gravity darkening \citep{Townsend_2004MNRAS}. Consequently, we can approximate the requirement that $\omega_{\rm i} > 0.8590$ with $v_{\rm e}\,\sin{i} > 200\,{\rm km/s}$. Since 32 stars in our data set meet the latter requirement, this is a rough estimate of the number of stars that would receive a more accurate treatment given models with $0.8590 < \omega_{\rm i} < 1$. This constitutes $\sim$24\% of the poorly modeled stars at $m \gtrsim 21$ and high $v_{\rm e}\,\sin{i}$ (see the bottom-right panel of Figure \ref{fig:rho_p}).

In conclusion of this section, we recall from the beginning of Section \ref{results} that better-fitting evolutionary models would produce cluster parameter estimates that would be more trustworthy and could be very different from this work's. In this section, we show how this can happen for binary fraction $b$ and age dispersion $\sigma_t$. However, the same is true for cluster age $\mu_t$ and rotational population proportions $w_0$ and $w_2$. Accordingly, we recommend that the reader treat our numerical estimates of all cluster parameters with caution, pending the creation of better-fitting evolutionary models. Furthermore, we expect likelihoods of cluster parameters to drop less steeply away from the ML values for models that fit better, resulting in parameter confidence regions that are wider than the ones in Section \ref{results}.  

\section{Summary and Future Work} \label{summary}

We jointly infer the age dispersion, the rotational distribution, and the binary fraction of the main sequence turnoff (MSTO) stars in the massive intermediate-age Large Magellanic Cloud cluster NGC 1846. This inference is based on cluster photometry and projected equatorial velocity measurements $v_{\rm e}\sin{i}$, as well as the \texttt{MIST} stellar evolution model library in combination with the \texttt{PARS} rotating star magnitude calculator. Our age dispersion estimate is $\sim70-80\,{\rm Myr}$, about half the earlier estimates due to non-rotating evolutionary models. This finding is consistent with the conjecture that rotational variation is at least partly responsible for the extended MSTO (eMSTO) in NGC 1846 and similar clusters. At the same time, independent lines of evidence indicate that the true age dispersion is probably even lower than the value we find. In addition, our binary fraction estimate is an order of magnitude higher than previous independent estimates for NGC 1846 and similar clusters.

Our methodology captures the pattern of the fit between the evolutionary model and the individual cluster stars in intricate detail. This, in combination with a poor quality of the fit, allows us to posit that a reduction in the magnetic braking of \texttt{MIST} models with initial masses between $\sim1.25\,M_\odot$ and $\sim1.5\,M_\odot$ would improve our fit to individual observed stars at magnitudes $\gtrsim21.0$, increase rotational broadening in this portion of the CMD, and subsequently remove the need for broadening by the implausibly high inferred binary fraction. 

However, due to the fact that reduction in magnetic braking would have little effect at magnitudes $\lesssim21.0$, this change would not improve the fit via increased rotational broadening in this brighter portion of the CMD and thus would probably not significantly alter the age dispersion estimate. On the other hand, our analysis, in combination with previous work, suggests that a greater enhancement of internal chemical mixing with rotation may provide the extra rotational broadening that would improve the fit throughout the CMD and would allow the inferred age dispersion to decrease. 

Consequently, a fruitful future direction would be to calibrate magnetic braking and the effect of rotation on chemical mixing to better fit the individual data points in NGC 1846. If such work were to produce an age dispersion for NGC 1846 that is significantly lower than this work's estimate, this decrease would bring the analysis of photometry and $v_{\rm e}\sin{i}$ in MSTO stars closer to concordance with the evidence of $\lesssim10\,{\rm Myr}$ age dispersion in young clusters that are similar to NGC 1846.

In this work, we have assumed a stellar model and used it to infer cluster parameters.  Future work can apply the same tools but in the other direction, or as a hierarchical model.  By tuning both the cluster parameters and the stellar evolutionary model, our approach can enable new constraints on the rotation and evolution of intermediate-mass stars.

Finally, we point out that another worthwhile future direction would be to repeat this work's analysis for additional young and intermediate-age massive clusters in the Magellanic Clouds with photometry and $v_{\rm e}\sin{i}$ measurements, in order to provide further constraints on the theory that combines stellar evolution with cluster formation.

The \texttt{Python} code that produces the analysis in this article is available for download, along with the accompanying pseudo-code and usage instructions \citep{Lipatov_calc_cluster2022}.

\begin{acknowledgements}
The authors would like to thank Nathan Bastian and Sebastian Kamann for providing helpful comments on the manuscript and for sharing the NGC 1846 data. The authors are also grateful to Aaron Dotter for helping them understand and work with the \texttt{MIST} stellar evolution models.
\end{acknowledgements}

\software{PARS \citep{Lipatov_2020ApJ}, calc\_cluster \citep{Lipatov_calc_cluster2022}, The NumPy Array \citep{vanDerWalt_2011}, Matplotlib \citep{Hunter_2007}, SciPy \citep{Virtanen_2020}.\\}

\appendix

\section{Refinement of the \texttt{MIST} Model Grid} \label{appendix_refinement}

\subsection{Initial Stellar Parameters} \label{spacing_initial}

Recall, from Section \ref{prob_density} and Equation \eqref{eq:rho_jkp}, that we wish to compute $\rho_{jkp}({\bm x}; \mu_t, \sigma_t)$, the theoretical probability density in observable space:
\begin{linenomath*}
\begin{equation}
    \rho_{jkp}({\bm x}; \mu_t, \sigma_t) = \frac{1}{Z} \int {\rm d {\bm \theta}}\,\pi_{jk}({\bm \theta}; \mu_t, \sigma_t) \,\,G({\bm x} - {\bm x}({\bm \theta}); {\bm \sigma_{{\bm x}{\bm p}}}).
\label{eq:rho_jkp_1}
\end{equation}
\end{linenomath*}

Evolutionary theory implies that the observables vector ${\bm x}$ is a continuous function of stellar model parameters ${\bm \theta} \equiv (M_{\rm i}, r, \omega_{\rm i}, i, t)$, where $M_{\rm i}$ is the initial mass of the primary, $r$ is the binary mass ratio, $\omega_{\rm i}$ is the initial rotation rate of the primary, $i$ is the inclination of its rotation axis, and $t$ is age. In practice, however, we only evaluate ${\bm x}$ at a finite set of discrete ${\bm \theta}$. This discrete evaluation approximates the continuous ${\bm x}({\bm \theta})$ sufficiently well when the ${\bm x}$ spacing between neighboring ${\bm \theta}$ is $\sim {\bm \sigma_{\bm x}}$, i.e., on the order of minimum-error vector ${\bm \sigma_{\bm x}}$. Such a spacing requirement guarantees that a discrete approximation of the integral that involves Gaussian error kernel $G({\bm \cdot})$ as one of the integrand factors in Equation \eqref{eq:rho_jkp_1} takes into account all ${\bm \theta}$ with ${\bm x}({\bm \theta})$ within about the data point error of the target observables, i.e., within $\sim {\bm \sigma_{{\bm x}{\bm p}}} \ge {\bm \sigma_{\bm x}}$ of ${\bm x}$. These are the values of ${\bm \theta}$ where $G({\bm x} - {\bm x}({\bm \theta}); {\bm \sigma_{{\bm x}{\bm p}}})$, and thus the entire integrand, is appreciable. We can state the spacing requirement in terms of the error kernel: $G({\bm x} - {\bm x}({\bm \theta}); {\bm \sigma_{\bm x}})$ at neighboring ${\bm \theta}$ have to overlap or at least come close to overlapping. This requirement is uniform in ${\bm x}$ space; however, it translates to potentially non-uniform separations in ${\bm \theta}$ space. In particular, when the derivatives of ${\bm x}({\bm \theta})$ with respect to ${\bm \theta}$ have high magnitudes, neighboring ${\bm \theta}$ have to be close.

The original set of discrete \texttt{MIST} models does not satisfy our spacing requirement. For example, Figure \ref{fig:mist} shows \texttt{PARS} observables for these models at a fixed age, two inclinations, and zero binary mass ratio. In this figure, consider the magnitude spacing between neighboring discrete ${\bm \theta}$ that differ only in $M_{\rm i}$. This spacing is often significantly greater than the magnitude uncertainty $\sigma_m$, especially at bright magnitudes, where a stellar model is likely to have high $M_{\rm i}$ and to be near the end of its main sequence life. Thus, we have to refine the $M_{\rm i}$ grid at high $M_{\rm i}$. At the same time, we may be able to coarsen the grid at low $M_{\rm i}$. Similar reasoning applies to all other components of ${\bm \theta}$.

The spacing requirement on the \texttt{MIST} grid in this section is numerically similar to the spacing characteristics of the \texttt{PARS} grid in Section \ref{pars_grid}: both the \texttt{PARS} grid and the final \texttt{MIST} grid should have observable distances between neighboring models that are on the order of observation error. These requirements, however, are distinct and have different reasons. In the former case, the requirement makes the interpolation on the \texttt{PARS} grid more accurate. In the latter case, it makes the subsequent integration, described in Section \ref{integration}, more accurate.

We aim to satisfy the above spacing requirement with a new grid of initial model parameters ${\bm \theta'} \equiv (M_{\rm i}, r, \omega_{\rm i}, i)$, but without making the grid so large as to be computationally prohibitive. To do so, we begin with a relatively sparse $\{M_{\rm i}, \omega_{\rm i}, i\}$ grid at constant $t$ and $r = 0$ and calculate ${\bm x}\left(M_{\rm i}, \omega_{\rm i}, i\right)$ via the interpolations on the \texttt{MIST} and \texttt{PARS} grids in Section \ref{stellar}. We then refine and coarsen the grid in $M_{\rm i}$, according to the following algorithm. First, for each pair of neighboring $M_{\rm i}$, we calculate the maximum absolute difference in any one observable over all $(\omega_{\rm i}, i)$, divided by three minimum-error standard deviations, $d \equiv \max{|\Delta {\bm x} / 3 {\bm \sigma_{\bm x}}|}$. We then divide the corresponding interval into $\left\lceil d \right\rceil$ equal segments. For all original intervals with a size greater than $10^{-4}\,M_\odot$, we re-calculate the observables on the new grid. In the remaining cases, we interpolate the observables in $M_{\rm i}$. We order the new, subdivided intervals according to decreasing $d$ and go through them until we can merge one with a neighboring interval without violating $d \le 1$. After the merge, we re-start at the beginning of the ordered interval list and repeat the procedure until no merges are possible. 

For $t_{\rm M}$, values of age $t$ in the original \texttt{MIST} grid, we repeat the refinement and coarsening procedure in $\omega_{\rm i}$, then go on cycling through the three elements of $\{M_{\rm i}, \omega_{\rm i}, i\}$, until the grid satisfies $d \le 1$ everywhere. Here, the threshold interval sizes are $10^{-4}$ and $10^{-4}\,{\rm rad}$ for $\omega_{\rm i}$ and $i$, respectively. Our iterative refinement procedure takes exponentially longer with the addition of each new model dimension. Thus, for $t$ not in the original \texttt{MIST} grid, we only refine and coarsen in $M_{\rm i}$, adopting the $\omega_{\rm i}$ and $i$ grids from the largest $t_{\rm M}$ that satisfies $t_{\rm M} < t$.

\begin{figure*}[ht]
\includegraphics[width=0.5\linewidth,trim=0 20 0 0, clip]{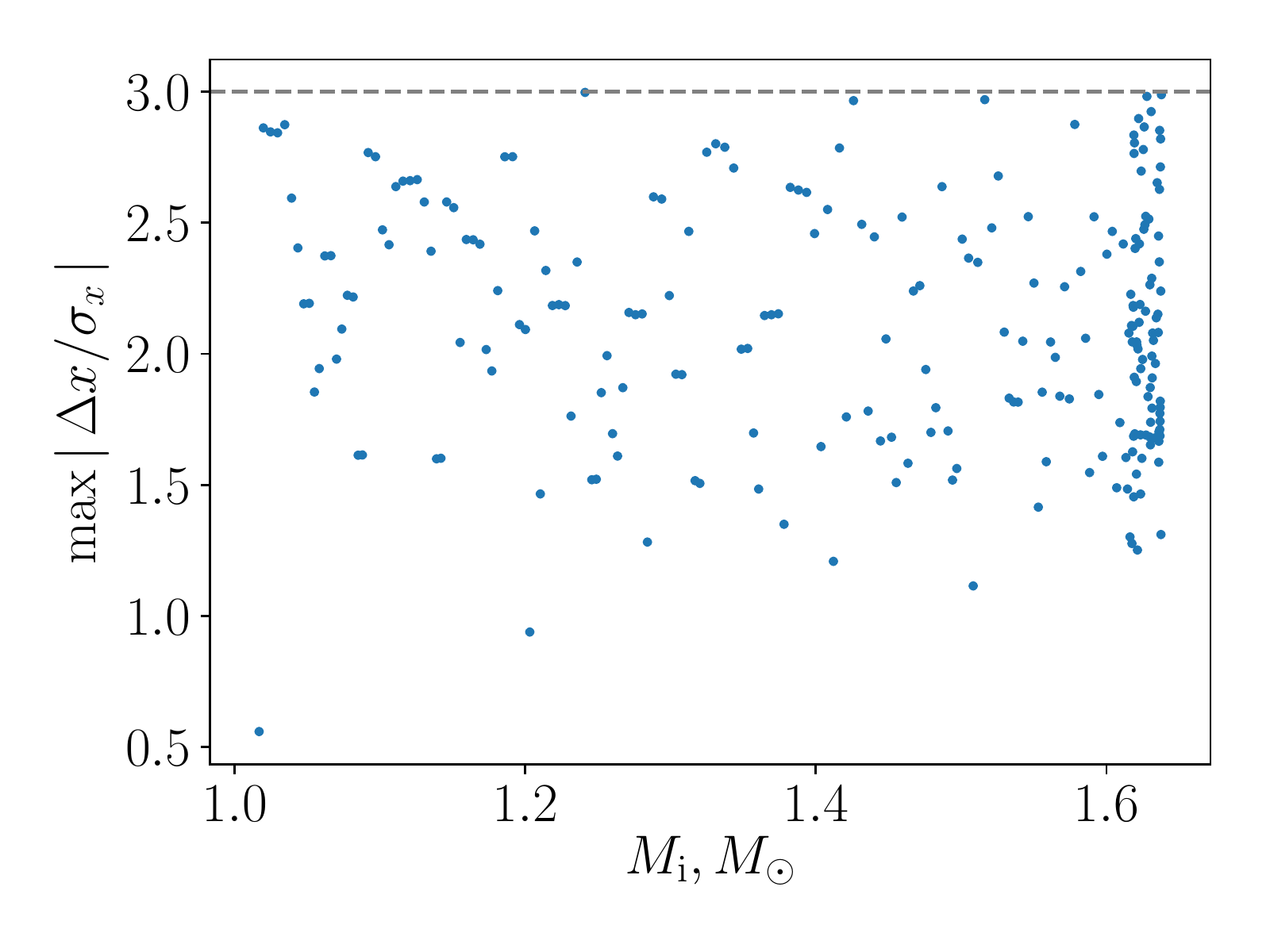}
\includegraphics[width=0.5\linewidth,trim=0 20 0 0, clip]{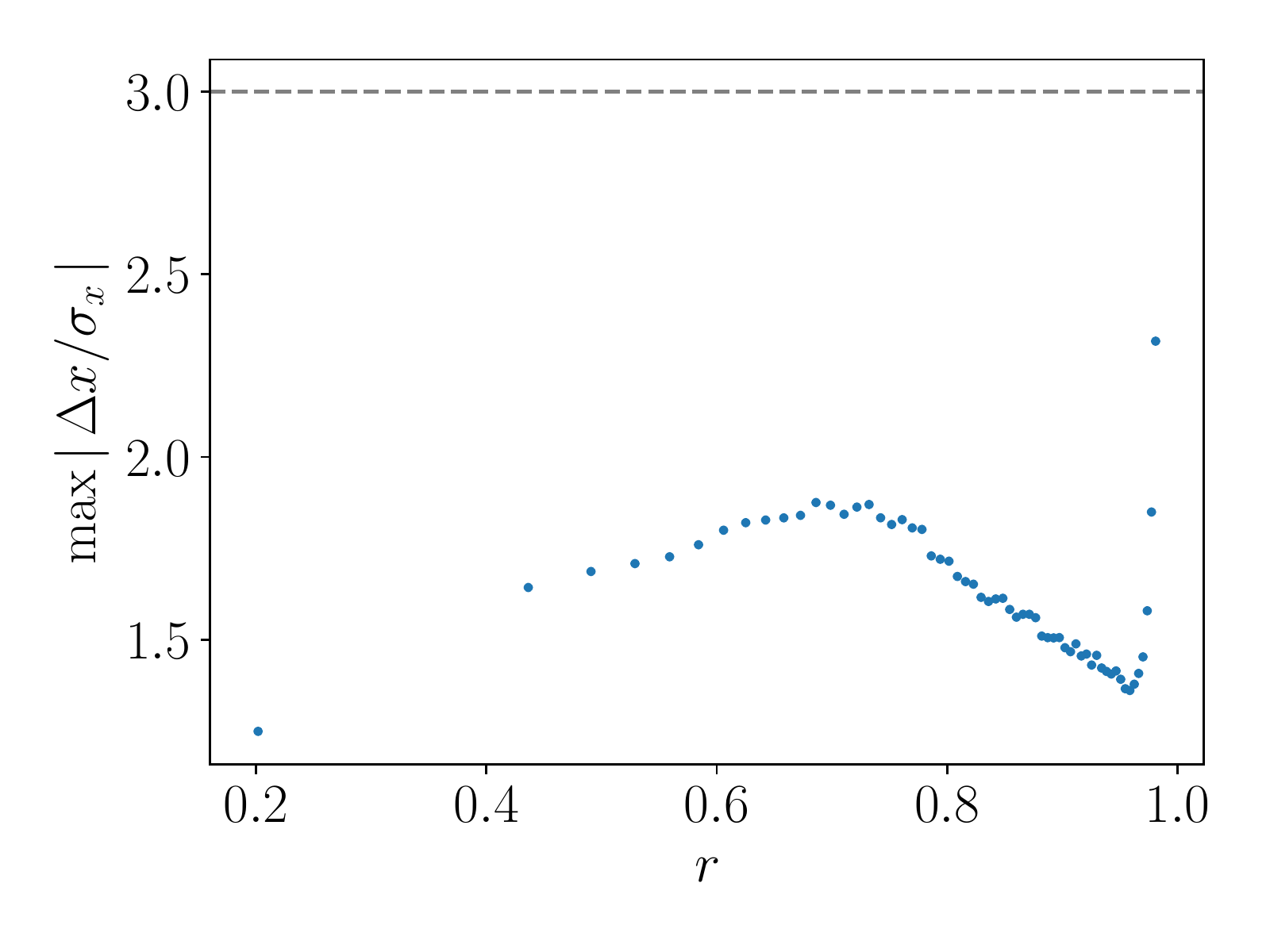}
\newline
\includegraphics[width=0.5\linewidth,trim=0 20 0 0, clip]{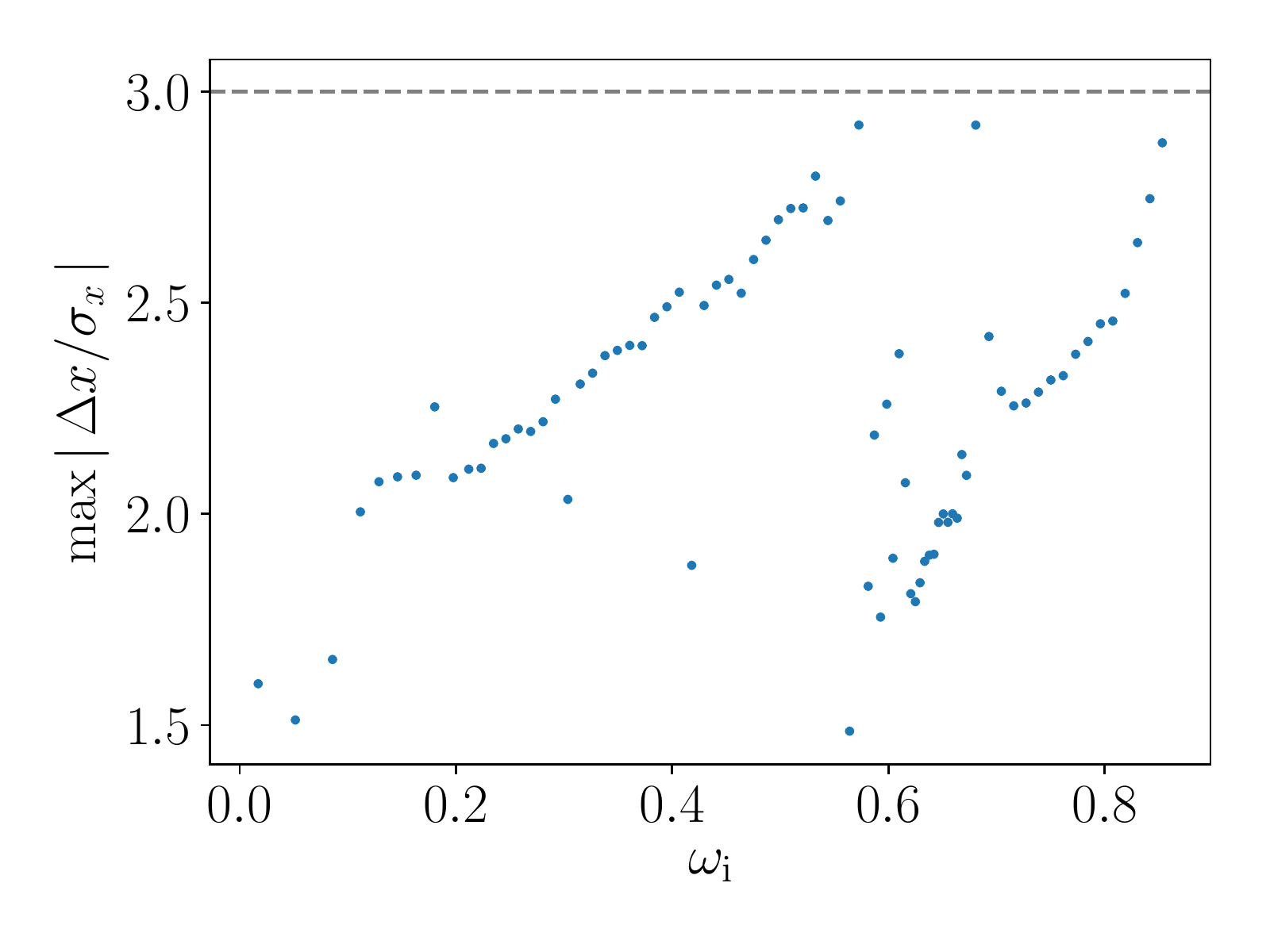}
\includegraphics[width=0.5\linewidth,trim=0 20 0 0, clip]{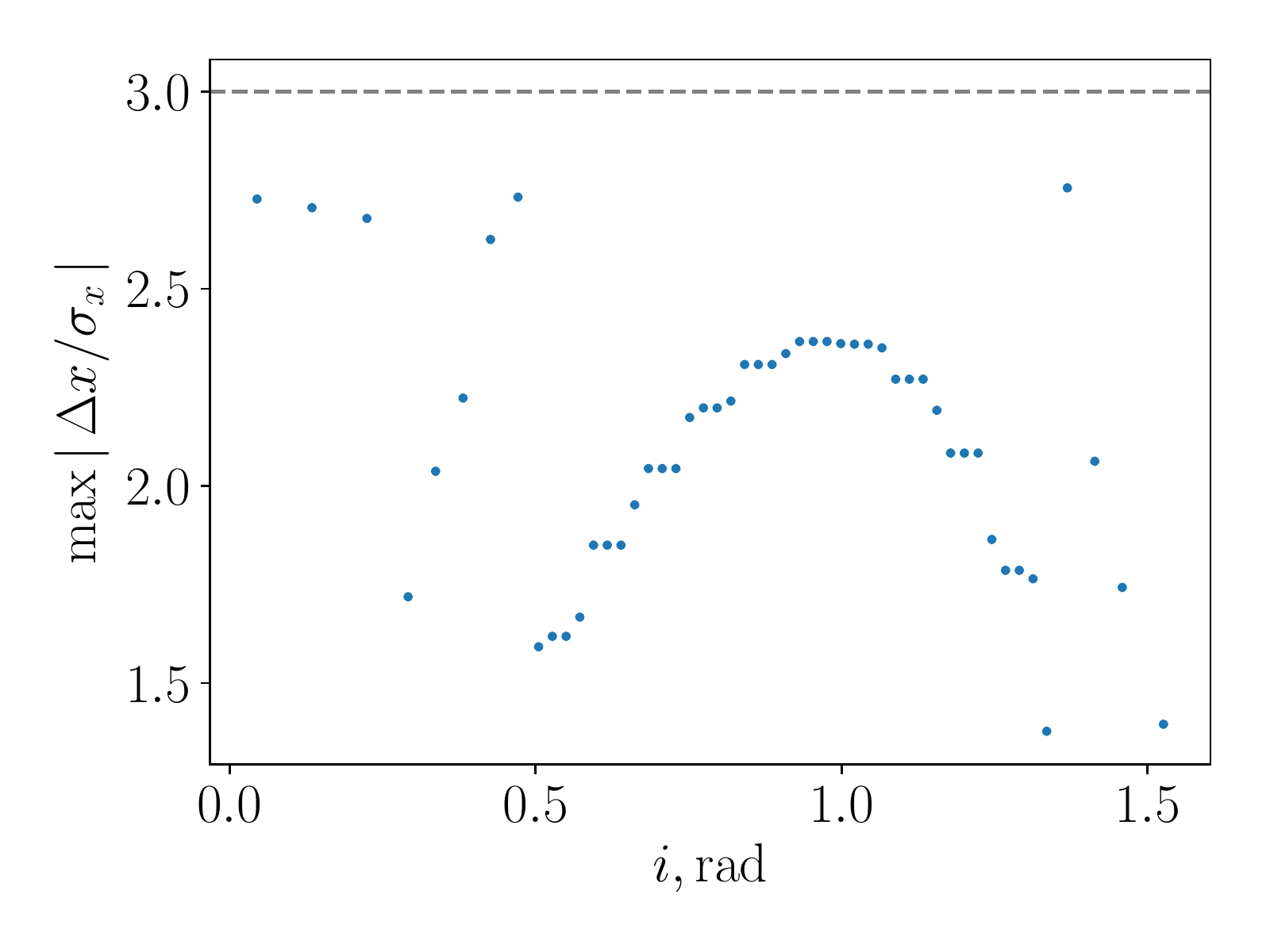}
\caption{Observable distances between neighboring models in each ${\bm \theta'}$ dimension, maximized over the remaining 3 dimensions. The corresponding 4-dimensional ${\bm \theta'}$ grid results from the model grid refinement procedure in Section \ref{model_refinement} at $\log{t} = 9.1544$, one of the ages in the original \texttt{MIST} library. Horizontal axes show interval midpoints. Note that the density of models in the $M_{\rm i}$ dimension is especially high near the maximum value of $M_{\rm i}$, where the derivatives of magnitude and color as functions of this parameter can be large. For each ${\bm \theta'}$ dimension, the distances generally do not vary by more than a factor of 2, indicating a relative uniformity in the accuracy of integration on this grid.}
\label{fig:refined}
\end{figure*}

We now propose an approximation that will allow us to refine the binary mass ratio $r$ grid independently of all other ${\bm \theta}$ grid dimensions, so that we do not have to include $r$ in the iterative refinement procedure. Specifically, solely for the purpose of $r$ grid refinement, we approximate the radiative flux from a star as proportional to its initial mass to some power $s$, e.g. ${\cal F} \propto M_{\rm i}^s$ for the primary. Under this approximation, Equation \eqref{eq:binary_combine} can be written as
\begin{linenomath*}\begin{linenomath*}\begin{equation} \label{eq:binary_comb2}
    m = -2.5\, \log{\left({\cal F}_{\rm p} + {\cal F}_{\rm c}\right)} = m_{\rm p} - \frac{2.5}{\ln{10}}\ln{\left(1 + \frac{{\cal F}_{\rm c}}{{\cal F}_{\rm p}}\right)} \approx m_{\rm p} - \frac{{\cal F}_{\rm c}}{{\cal F}_{\rm p}} = m_{\rm p} - r^s,
\end{equation}\end{linenomath*}\end{linenomath*}
where ${\cal F}_{\rm p}$, ${\cal F}_{\rm c}$, and $m_{\rm p}$ stand for the flux of the primary, the flux of the secondary, and the magnitude of the primary, respectively. Furthermore, the functionality of this procedure is not impaired when we approximate $\ln{10}$ as 2.5 and retain only the first term in the Taylor expansion of the natural logarithm around ${\cal F}_{\rm c}/{\cal F}_{\rm p} = 0$. Equation \eqref{eq:binary_comb2} suggests that if we subdivide $r^s \in [0, 1]$ into at least $1 / 3\sigma_m \approx 34$ intervals, the resulting $\{M_{\rm i}, r, \omega_{\rm i}, i\}$ grid should satisfy $d \le 1$ in the $r$ dimension. In practice, the condition is satisfied for $r \in [0, r_1]$ with 66 intervals, $s = 4.6$ and $r_1 \ge 0.98$ for all $t$. For each $t$, $r \in [0, r_1]$ becomes the range over which we integrate. Figure \ref{fig:refined} shows the maximum distances between neighboring models in each of the ${\bm \theta'}$ dimensions for the resulting grid at a specific value of $t$ that is also one of the ages in the original \texttt{MIST} library. Although parameter spacing between models is large in some parts of Figure \ref{fig:refined}, such as at low $r$, this corresponds to small enough distances in observable space that the integration remains accurate. We emphasize that we do not use approximation ${\cal F} \propto M_{\rm i}^s$ for magnitude calculations, which are outlined in Section \ref{calc_obs}.

\subsection{Refinement of Model Age} \label{spacing_age}

Our spacing requirement for the ${\bm \theta'}$ grid is rather stringent. It is necessary if one cannot assume that numerical integration in Equation \eqref{eq:rho_jkp_1} is performed sequentially in each component of ${\bm \theta'}$. We make the requirement more lenient with respect to $t$ by integrating with respect to this parameter after all the others in Equation \eqref{eq:rho_jkp_1}. To characterize the resulting requirement on spacing in $t$, we re-write Equation \eqref{eq:rho_jkp_1} as
\begin{linenomath*}\begin{equation}
    \rho_{jkp}({\bm x}; \mu_t, \sigma_t) = \frac{1}{Z} \int {\rm d}t\,\bar{\pi}(t; \mu_t, \sigma_t)\, \rho_{jkp}({\bm x}; t),
\label{eq:rho_jkp_2}
\end{equation}\end{linenomath*}
where 
\begin{linenomath*}\begin{equation}
    \rho_{jkp}({\bm x}; t) = \int {\rm d}{\bm \theta'}\,\pi_{jk}({\bm \theta'})\,G\left({\bm x} - {\bm x}({\bm \theta'}; t); {\bm \sigma_{{\bm x}{\bm p}}}\right)
\label{eq:rho_jkp_t}
\end{equation}\end{linenomath*}
is the theoretical probability density over observable space at age $t$, based on a Gaussian error kernel with width ${\bm \sigma_{{\bm x}{\bm p}}}$.

Let us define $\Theta$ as the full range of ${\bm \theta}$, $\Theta(t)$ as the subset of this range at age $t$, and $\Theta(a, t)$ as $\Theta(t)$ restricted to EEP $a$. Every $\Theta(a, t)$ extends over all $i$ and $r$, but the requirement that ${\rm EEP}$ equals $a$ selects for an age-specific range of $M_{\rm i}$ and $\omega_{\rm Mi}$. We further define $X(a, t)$ as the image of $\Theta(a, t)$ due to the function ${\bm x}({\bm \theta})$. The continous volume $X(t) = \cup_a{X(a, t)}$ in ${\bm x}$-space is an isochrone in the general sense of Section \ref{evolution}, specified for three observables, i.e., components of ${\bm x}$, and four model parameters, i.e., components of ${\bm \theta'}$. Section \ref{evolution} introduces the idea that shortest distances on the CMD between traditional isochrones, which are parametrized by $M_{\rm i}$, are at fixed EEPs. Here, we extend this concept to three observable dimensions and four isochrone parameters. Thus, we define the ${\bm x}$-space distance $\Delta {\bm x}(a, t_1, t_2)$ between the $t = t_1$ and $t = t_2$ isochrones at fixed ${\rm EEP} = a$ as the average distance between $X(a, t_1)$ and $X(a, t_2)$; the overall distance $\Delta {\bm x}(t_1, t_2)$ between these isochrones is the same average, taken across all $a$. Intuitively, accurate integration in Equation \eqref{eq:rho_jkp_2} requires the order of $\Delta {\bm x}(t_1, t_2)$ to be no larger than about the minimum error ${\bm \sigma}_{\bm x}$ for neighboring $t_1$ and $t_2$.

More formally, we require that appreciable $\rho_{jkp}({\bm x}; t)$ in Equation \eqref{eq:rho_jkp_2} at neighboring $t$ overlap in ${\bm x}$-space. This is equivalent to the error kernel formulation of the spacing requirement in Section \ref{spacing_initial}, with ${\bm \theta}$ replaced by $t$ and $G({\bm x} - {\bm x}({\bm \theta}); {\bm \sigma_{\bm x}})$ replaced by $\rho_{jkp}({\bm x}; t)$. To determine where $\rho_{jkp}({\bm x}; t)$ is appreciable, let us assume that the prior $\pi_{jk}({\bm \theta'})$ in Equation \eqref{eq:rho_jkp_t} is appreciable over the entirety of $\Theta(t)$. In this case, Equation \eqref{eq:rho_jkp_t} tells us that the locus of points in ${\bm x}$-space with appreciable $\rho_{jkp}({\bm x}; t)$ is $X(t)$ broadened by at least ${\bm \sigma_{\bm x}}$. More precisely, $\rho_{jkp}({\bm x}; t)$ is the convolution of a function that is appreciable solely over $X(t)$ and an error kernel that is at least as wide as $G\left({\bm x} - {\bm x}({\bm \theta'}; t); {\bm \sigma_{\bm x}}\right)$. Hence, we confirm our intuition that $X(t)$ at neighboring $t$ should be separated by $\sim {\bm \sigma_{\bm x}}$. This is equivalent to the formulation of the spacing requirement in Section \ref{spacing_initial}, with ${\bm x}({\bm \theta})$ replaced by $X(t)$. 

\begin{figure*}[t]
\includegraphics[width=0.5\linewidth]{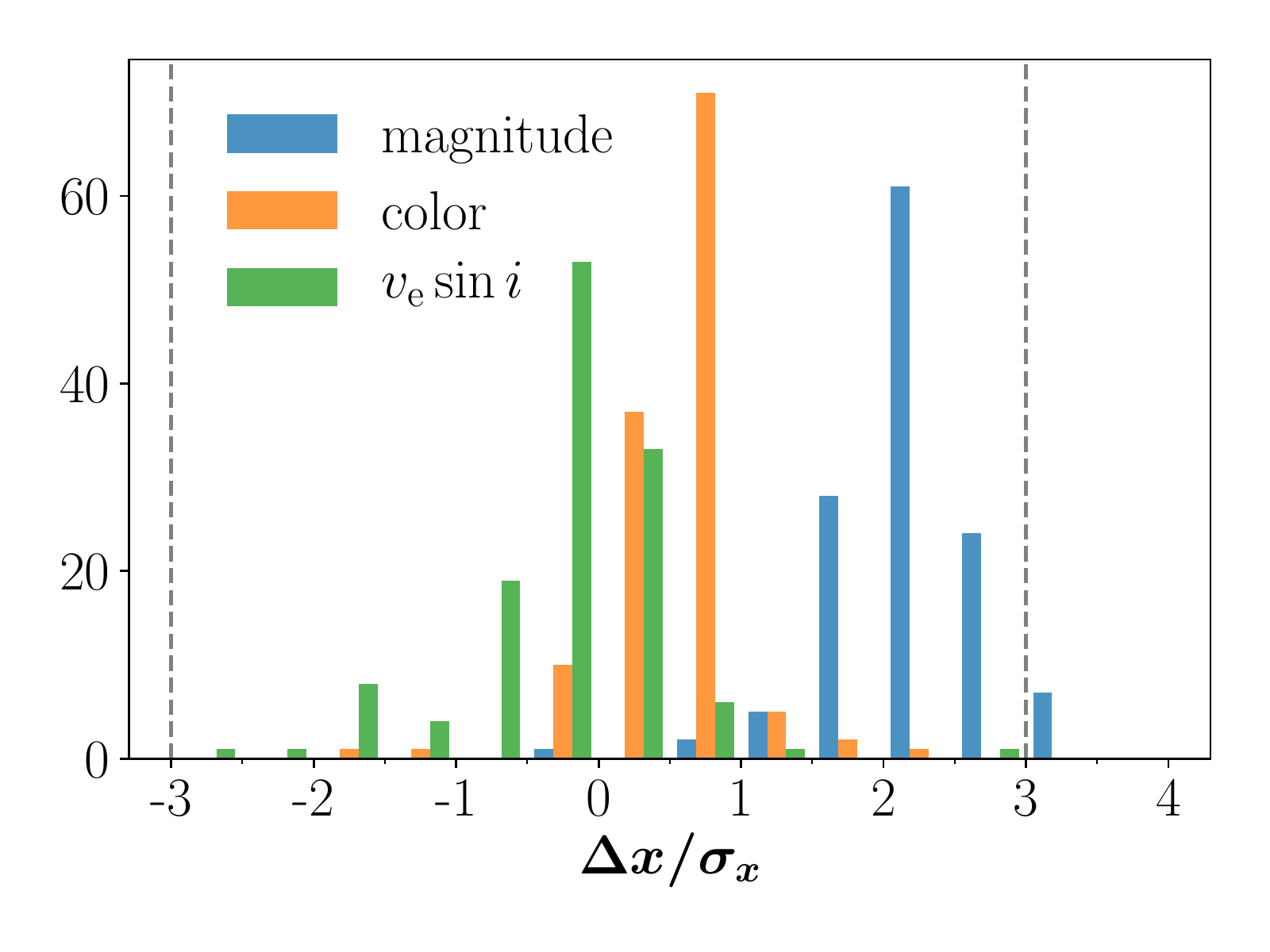}
\includegraphics[width=0.5\linewidth]{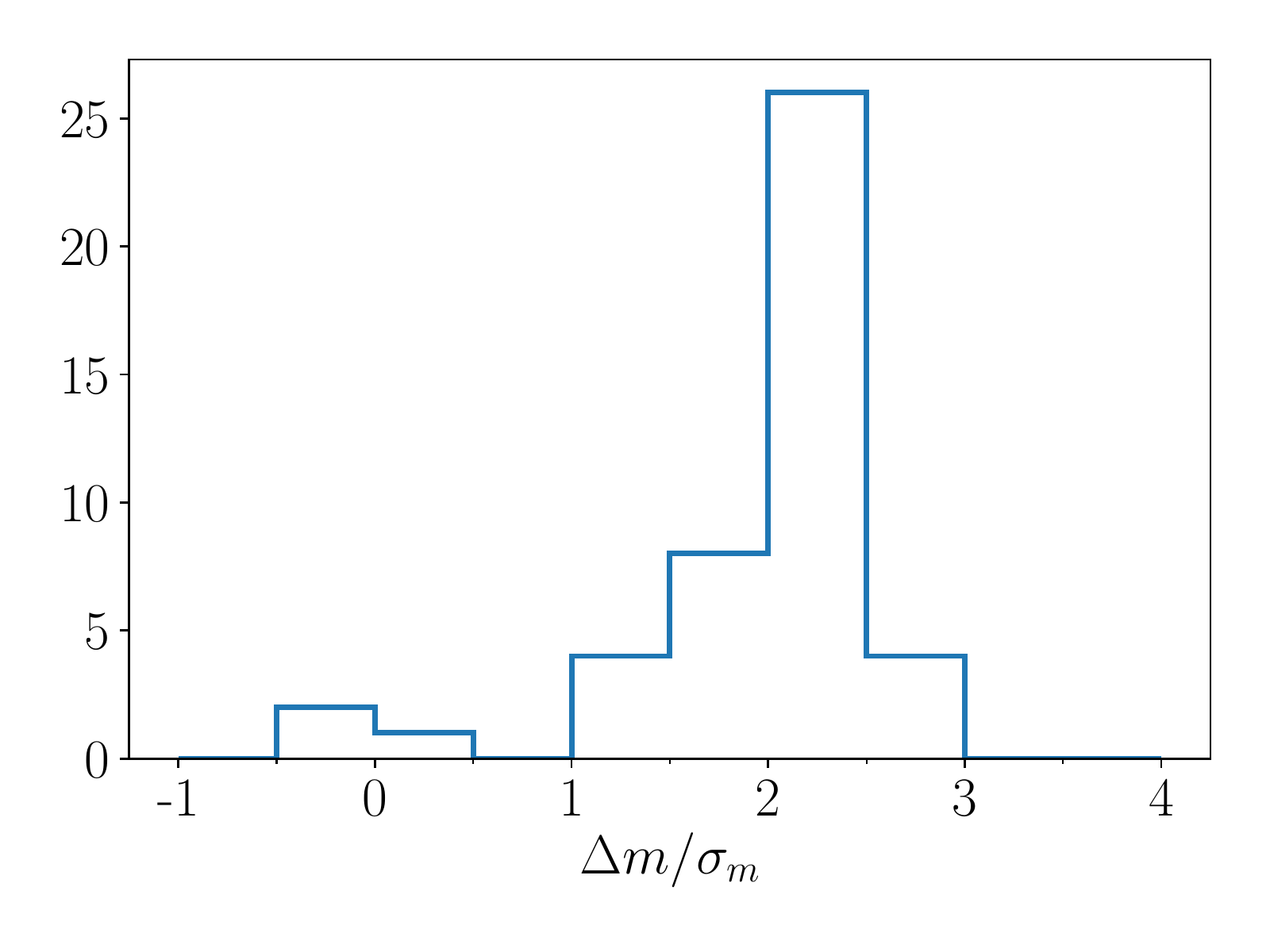}
\caption{Left panel: histogram of differences in each observable, averaged over ${\bm \theta'}$ and scaled by minimum-error standard deviation, between $\log{t} = 9.1544$ and $\log{t} = 9.1594$. Each difference is taken at one of the EEPs where each age has at least one stellar model. Most absolute differences for color and $v_{\rm e}\sin{i}$ are below 1; most differences for magnitudes are below 3. Negative values for color and $v_{\rm e}\sin{i}$ correspond to EEPs where average value of the observable decreases with increasing age. The space between vertical dashed lines is equal to the width of the Gaussian error kernels in our numerical integration procedure.  Right panel: histogram of scaled differences in magnitude (as opposed to differences in each observable) between model grids for all neighbor pairs on the age grid (as opposed to one pair, as in the left panel). Absolute differences are below 3, indicating that integration in $t$ is likely accurate.}

\label{fig:diff_EEP}
\end{figure*}

The original \texttt{MIST} age grid is spaced uniformly with $\Delta \left(\log{t_{\rm M}}\right) = 0.020$. We start with the portion of this grid between $\log{t_{\rm M}} = 9.0537$ and $\log{t_{\rm M}} = 9.2550$ and insert intermediate ages, so that the new grid is spaced uniformly with $\Delta \left(\log{t}\right) = 0.005$. Additionally, for the lowest 5 values of $\log{t_{\rm M}}$, we append the grid with $(3/4)\log{t_{\rm M}} + (1/4)\left[\log{t_{\rm M}} + \Delta \left(\log{t}\right)\right]$. The resulting grid becomes our age grid for the rest of the article. For two neighboring ages $t_1$ and $t_2$ on this grid, the left panel of Figure \ref{fig:diff_EEP} presents the distribution of $\Delta {\bm x}(a, t_1, t_2)$ across all $a$. This figure shows that most differences are no larger than a few ${\bm \sigma_{\bm x}}$, suggesting that the isochrone spacing requirement is met for $t_1$ and $t_2$. For every such pair of neighboring ages on the grid, we further obtain $\Delta {\bm x}(t_1, t_2)$ and focus on $\Delta m(t_1, t_2) / \sigma_m$, generally the largest component of $\Delta {\bm x}(t_1, t_2) / {\bm \sigma_{\bm x}}$. The right panel of Figure \ref{fig:diff_EEP} shows that the absolute value of $\Delta m(t_1, t_2) / \sigma_m$ across all neighbor pairs $(t_1, t_2)$ is less than 3, supporting the idea that the isochrone requirement is met for all age neighbor pairs.

\begin{figure*}[t]
\includegraphics[width=0.5\linewidth]{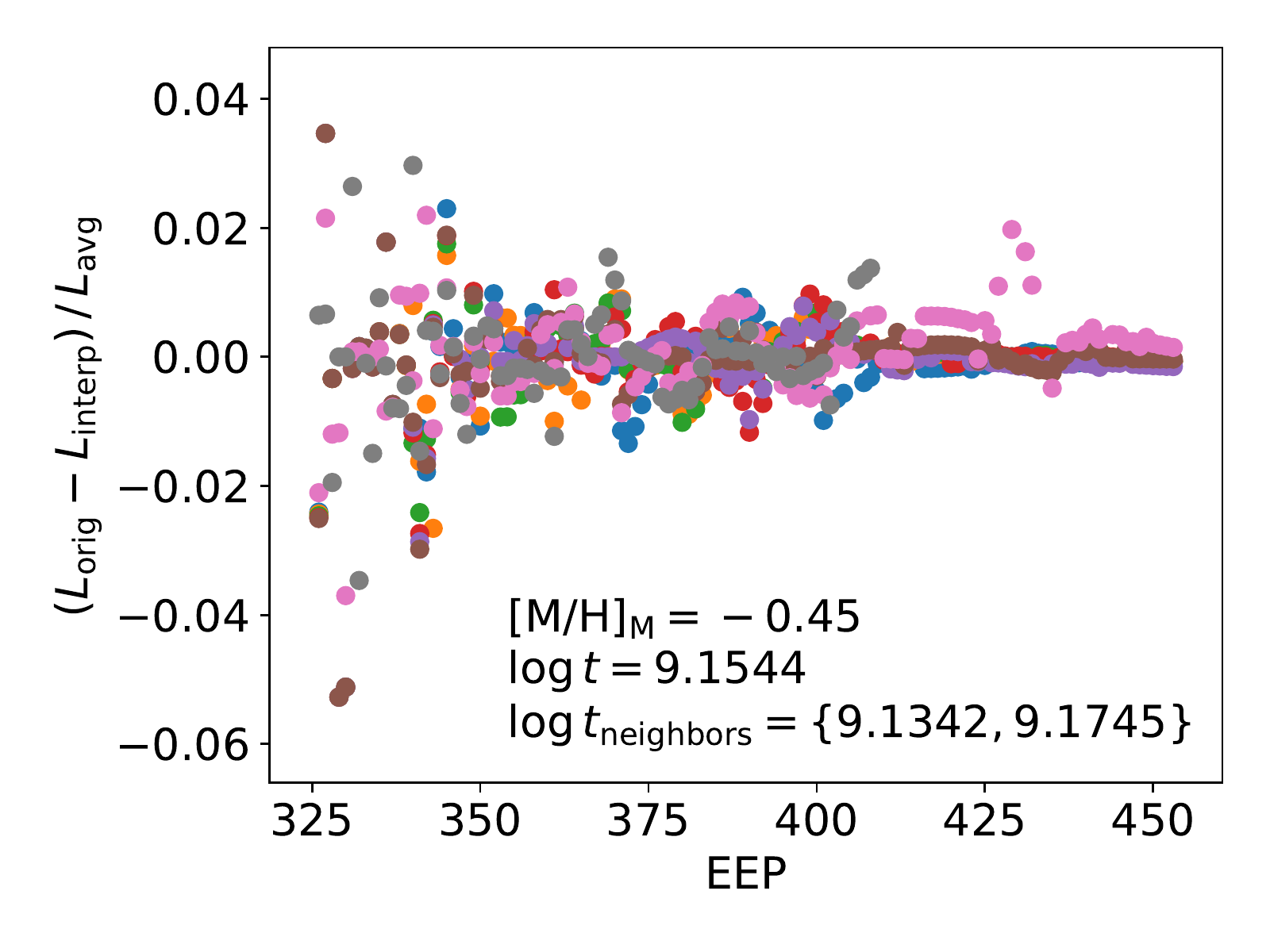}
\includegraphics[width=0.5\linewidth]{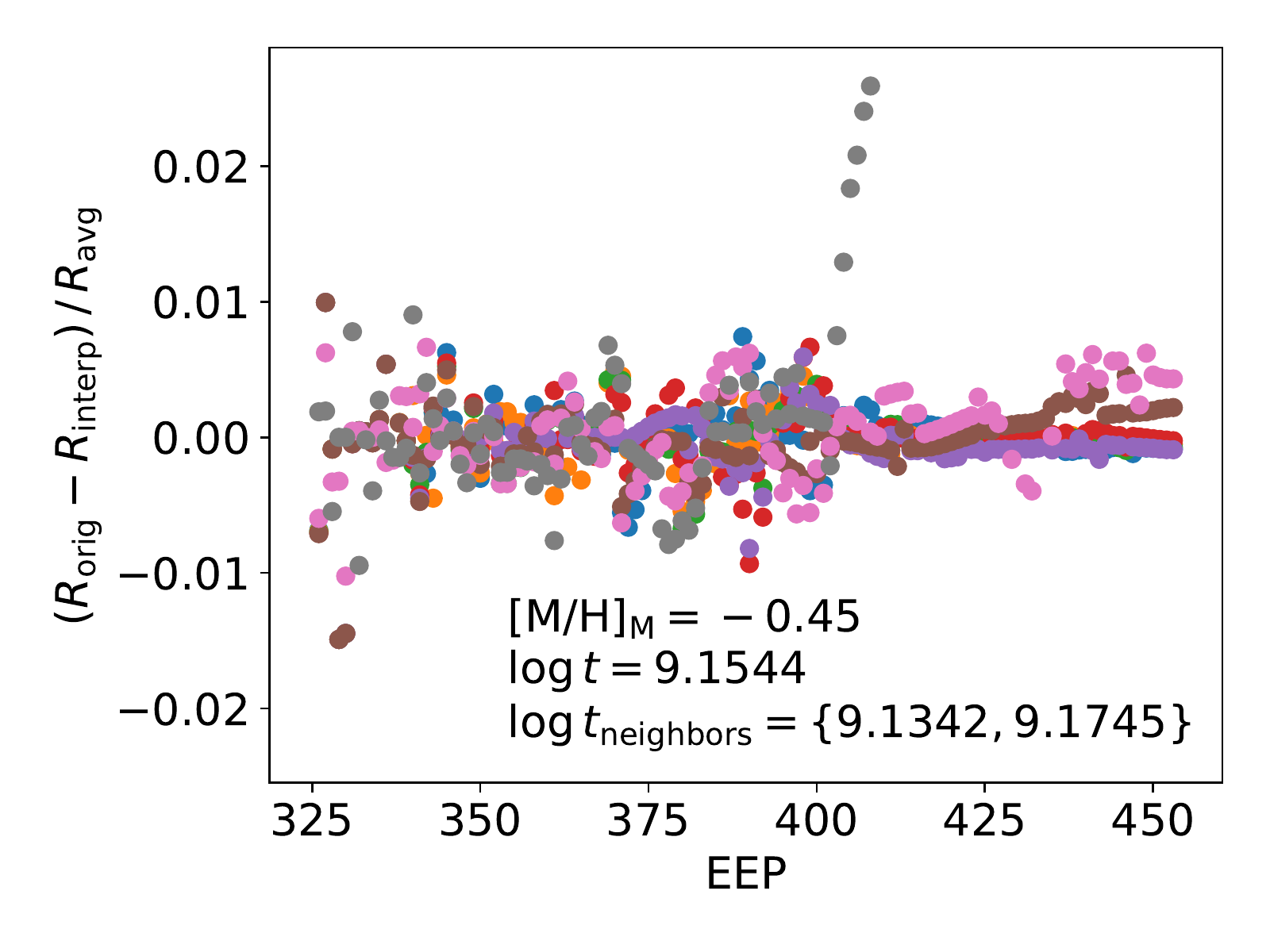}
\caption{Accuracy of interpolation between the isochrones. Left panel: proportional difference between the luminosity at the original \texttt{MIST} isochrone with $\log{t} = 9.1544$ and the luminosity that is interpolated between the isochrone's neighbors in age space. Right panel: same, for the specially averaged radius $R_{\rm M}$. Average luminosity is $L_{\rm avg} = (L_{\rm orig} + L_{\rm interp}) / 2$ and average radius is similar. Different colors indicate different initial rotational velocities in the same way they do in Figure \ref{fig:mist}. For example, gray markers correspond to $\omega_{\rm M} = 0.7$, pink -- to $\omega_{\rm M} = 0.6$, and so on. Average distance between known isochrones in the course of actual interpolation that produces our model grid is half the distance in this test case, so that we expect the former interpolation procedure to be significantly more accurate than that which is pictured here.}
\vspace{10pt}
\label{fig:age_interp_accuracy}
\end{figure*}

The main focus of this subsection has been to check and make sure that our interpolation between isochrones results in a model grid on which we accurately compute the probability densities, e.g., in Equations \ref{eq:rho_jkp_2} and \ref{eq:rho_jkp_t}. We also want to check that this interpolation is accurate, in itself. To this end, we compare models on the isochrone at $\log{t} = 9.1544$ from the original \texttt{MIST} grid with the models we obtain by interpolating between this isochrone's neighbors, at $\log{t} = 9.1342$ and $\log{t} = 9.1745$. Specifically we compare two model parameters that determine magnitude -- luminosity $L$ and specially averaged radius $R_{\rm M}$. Figure \ref{fig:age_interp_accuracy} shows that most differences in luminosity between the original isochrone and the interpolated version are $\sim1\%$ and most differences in radius are lower. In the course of actual interpolation that yields our model grid, the average age difference between known isochrones is half the difference in this test case. Thus, we expect interpolation errors to be even lower for the actual interpolation procedure, by a factor of $\sim 4$ if linear and quadratic terms dominate the local series expansions of luminosity and radius as functions of age. 

\section{Computation of Bayesian Probability Density} \label{bayesian_computation}

\subsection{Likelihood} \label{like_computation}

From the latter portions of Section \ref{cluster_like}, recall that we want to compute integrals of the form
\begin{linenomath*}\begin{equation} \label{eq:prob_t_2}
    P(\mu_t, \sigma_t) = \int {\rm d}w_0\,{\rm d}w_2\,P({\bm \phi'}) \propto \int {\rm d}w_0\,{\rm d}w_2\,{\cal L}({\bm \phi'}),
\end{equation}\end{linenomath*}
where the likelihood function ${\cal L}$ over a limited set of cluster parameters ${\bm \phi'}= \{\mu_t, \sigma_t, w_0, w_2\}$ is 
\begin{linenomath*}\begin{equation} \label{eq:like_phi_prime_2}
    {\cal L}({\bm \phi'}) = \int {\rm d}q\,{\rm d}b \,{\cal L}({\bm \phi})
\end{equation}\end{linenomath*}
and the full likelihood function over all cluster parameters ${\bm \phi} = \{q, b, \mu_t, \sigma_t, w_0, w_2\}$ is the product of data point likelihood factors $\varrho_p$:
\begin{linenomath*}\begin{equation} \label{eq:like_2}
    {\cal L}({\bm \phi}) = \prod^n_p \varrho_p.
\end{equation}\end{linenomath*}

We also recall that Bayesian probability density is $P({\bm \phi}) = {\cal L}({\bm \phi}) Z_{\cal L}^{-1}$, where $Z_{\cal L}$ is a normalization constant, equal to
\begin{linenomath*}\begin{equation} \label{eq:z_l_2}
    Z_{\cal L} = \int {\rm d}{\bm \phi} \,{\cal L}({\bm \phi}).
\end{equation}\end{linenomath*}

In order to evaluate the rightmost integral in Equation \eqref{eq:prob_t_2}, we first wish to determine ${\cal L}({\bm \phi})$ up to a multiplicative constant for all ${\bm \phi}$ where the function is appreciable. The naive approach to this task, suggested by Equation \eqref{eq:like_2}, can easily encounter numerical overflow and underflow, due to the fact that the number of factors $n = 2353$ in Equation \eqref{eq:like_2} is large and the fact that the differences between individual factors are also large.

In particular, standard, positive floating point values in the \texttt{Python} programming language cannot be closer than $f_{\rm min} = 2.2 \times 10^{-308}$ to zero or farther than $f_{\rm max} = 1.8 \times 10^{308}$ from zero. Thus, for example, if each factor is below $\sqrt[n]{f_{\rm min}} = 0.74$, Equation \eqref{eq:like_2} evaluates to zero. If each factor is above $\sqrt[n]{f_{\rm max}} = 1.35$, the equation evaluates to infinity. In practice, $\log{\varrho_p}$ are more or less randomly distributed throughout some range in our implementation of the analysis. When ${\bm \phi}$ is closer to its maximum-likelihood value, this range is higher, and products of the form $\prod^{k}_p \varrho_p$ with $k \in [1, \ldots, n]$ can evaluate to values greater than $f_{\rm max}$. 

A solution to this problem that is relatively slow but guaranteed to work involves taking the logarithm of every $\varrho_p$, at every ${\bm \phi}$. In this case, we can first evaluate
\begin{linenomath*}\begin{equation} \label{eq:like_log}
    \ln{{\cal L}({\bm \phi})} = \sum^n_p \ln{\varrho_p} \quad \forall{\bm \phi},
\end{equation}\end{linenomath*}
then compute likelihood as 
\begin{linenomath*}\begin{equation}
    \exp{\left[\ln{{\cal L}({\bm \phi})} - \max_{\bm \phi}{\ln{{\cal L}({\bm \phi})}}\right]},
\end{equation}\end{linenomath*}
which is Equation \eqref{eq:like_2} divided by $\max_{\bm \phi}{{\cal L}}$.

We offer an alternate solution, one that is faster by about a factor of two in our implementation. Towards this end, we define $\rho_{jp}(\mu_t, \sigma_t) = \sum_{i} w_{i}\, \rho_{ijp}({\bm x_{\bm p}}; \mu_t, \sigma_t)$ and use equations \eqref{eq:rho_p_phi} and \eqref{eq:varrho} to express likelihood factors $\varrho_p$ as
\begin{linenomath*}\begin{equation}\label{eq:varrho_AB}
\varrho_p \equiv \varrho_p({\bm \phi}) = 1 + q A_p({\bm \phi'}) + q b B_p({\bm \phi'}),
\end{equation}\end{linenomath*}
where ${\bm \phi'} \equiv \{\mu_t, \sigma_t, w_0, w_2\}$,
\begin{linenomath*}\begin{equation} \label{eq:A}
    A_p({\bm \phi'}) = \frac{\rho_{0p}(\mu_t, \sigma_t)}{\rho_{{\rm b}p}({\bm x_{\bm p}})} - 1,
\end{equation}\end{linenomath*}
and 
\begin{linenomath*}\begin{equation} \label{eq:B}
    B_p({\bm \phi'}) = \frac{\rho_{1p}(\mu_t, \sigma_t) - \rho_{0p}(\mu_t, \sigma_t)}{\rho_{{\rm b}p}({\bm x_{\bm p}})}.
\end{equation}\end{linenomath*}

For each ${\bm \phi'}$, we choose some constant $C$, divide every $\rho_p$ by this constant and multiply the resulting factors together. The constant should be large enough that there is no overflow, i.e.,
\begin{linenomath*}\begin{equation} \label{eq:c_req_1}
    \prod^{n}_p \frac{\varrho_p}{C} = \frac{1}{C^n} \prod^{n}_p \varrho_p < f_{\rm max} \quad \forall k,q,b,
\end{equation}\end{linenomath*}
yet small enough that the maximum of the product is much greater than $f_{\rm min}$:
\begin{linenomath*}\begin{equation} \label{eq:c_req_2}
    \max_{q, b}{\prod^{n}_p \frac{\varrho_p}{C}} = \frac{1}{C^n} \max_{q, b}{\prod^{n}_p \varrho_p} \gg f_{\rm min}.
\end{equation}\end{linenomath*}
In this case, the likelihood in Equation \eqref{eq:like_2} is divided by $C^n$. To obtain a value of $C$ that satisfies Equations \eqref{eq:c_req_1} and  \eqref{eq:c_req_2}, we aim to find the maximum of ${\cal L}$ across $b$ and $q$, for a given set of the remaining cluster parameters ${\bm \phi'}$. Towards this goal, we use Equations \eqref{eq:like_2}, \eqref{eq:varrho}, \eqref{eq:A}, and \eqref{eq:B} to write down
\begin{linenomath*}\begin{align} \label{eq:like_prime}
\begin{split}
    \frac{\partial\ln{\cal L}}{\partial q} &= \sum_p \frac{1}{q + q_p(b)} \quad {\rm and}\\
    \frac{\partial\ln{\cal L}}{\partial b} &= \sum_p \frac{1}{b + b_p(q)},
\end{split}
\end{align}\end{linenomath*}
where
\begin{linenomath*}\begin{align}
\begin{split}
    q_p(b) &= \frac{1}{A_p + B_p b}, \\
    b_p(q) &= \frac{1 / q + A_p}{B_p},
\end{split}
\end{align}\end{linenomath*}
$A_p$ and $B_p$ are given by Equations \eqref{eq:A} and \eqref{eq:B}, and we have suppressed arguments ${\bm \phi'}$. Further differentiating \eqref{eq:like_prime}, we get
\begin{linenomath*}\begin{align} \label{eq:like_double_prime}
\begin{split}
    \frac{\partial^2\ln{\cal L}}{\partial q^2} &= -\sum_p \frac{1}{\left[q + q_p(b)\right]^2} \quad {\rm and}\\
    \frac{\partial^2\ln{\cal L}}{\partial b^2} &= -\sum_p \frac{1}{\left[b + b_p(q)\right]^2}.
\end{split}
\end{align}\end{linenomath*}
One can show that neither $q_p(b)$ nor $b_p(q)$ can be on the open interval $(-1, 0)$. Thus, as long as $q \in (0, 1)$ and $b \in (0, 1)$, second derivatives in Equation \eqref{eq:like_double_prime} are always defined and negative. This suggests that the likelihood function has a single extremum on this domain -- a maximum. We assume that the latter assertion is true, solely for the purposes of finding a constant to divide $\varrho_p$. We set both derivatives in Equation \eqref{eq:like_prime} to zero and solve the system of equations using a modified Powell's method to find $\tilde{q}$ and $\tilde{b}$ -- the values of $q$ and $b$ where the likelihood reaches $\tilde{\cal L}$, which is probably its maximum. We set $C = \sqrt[n]{\tilde{\cal L}}$ and perform the procedure discussed around Equations \eqref{eq:c_req_1} and \eqref{eq:c_req_2}. Consider that, in the course of this procedure, we reduce the order of magnitude of every $\varrho_p$ by the average order of magnitude at maximum likelihood:
\begin{linenomath*}\begin{equation}
    \log{\frac{\varrho_p(q, b)}{C}} = \log{\varrho_p(q, b)} - \frac{1}{n}\sum_p\log{\varrho_p(\tilde{q}, \tilde{b})}. 
\end{equation}\end{linenomath*}
In other words, even at $(\tilde{q}, \tilde{b})$, the average order of magnitude is now zero. Thus, Equation \eqref{eq:c_req_2} is satisfied, and it is very likely that Equation \eqref{eq:c_req_1} is satisfied as well. On the other hand, the reduction in the magnitude of $\varrho_p$ has made underflow more probable. We can think of the multiplication on the left hand side expression of Equation \eqref{eq:c_req_1} as a biased pseudo-random walk in the magnitude of the running product, starting at zero. The longer the walk, the more likely it is to reach $\log{f_{\rm min}}$ and result in underflow. To minimize the probability of such an occurrence, we split the multiplication into 10 products, each composed of $\sim 235$ factors. Having completed the multiplication, we compute
\begin{linenomath*}\begin{equation} \label{eq:like_phi}
    \ln{{\cal L}({\bm \phi})} = \ln{\prod_p \frac{\varrho_p({\bm \phi})}{C({\bm \phi'})}} + \ln{\tilde{\cal L}({\bm \phi'})},
\end{equation}\end{linenomath*}
and determine $\hat{{\bm \phi}}$, where ${\cal L}({\bm \phi})$ has its maximum over ${\bm \phi}$. The additive term $\ln{\tilde{\cal L}}$ removes the effect of dividing $\varrho_p$ by ${\bm \phi'}$-dependent $C$ and ${\bm \phi}$ have been restored as arguments. We compute the logarithm of ${\cal L}({\bm \phi})$, since the function itself could be larger than $f_{\rm max}$. 

In addition to maximum log likelihood in Equation \eqref{eq:like_phi}, we now aim to calculate the logarithm of the likelihood marginalized in $q$ and $b$, on a grid of ${\bm \phi'}$: 
\begin{linenomath*}\begin{equation} \label{eq:like_qb}
    \ln{{\cal L}({\bm \phi'})} = \ln{\int {\rm d}q\,{\rm d}b\, \prod_p \frac{\varrho_p({\bm \phi})}{C({\bm \phi'})}} + \ln{\tilde{\cal L}({\bm \phi'})},
\end{equation}\end{linenomath*}
with ${\cal L}({\bm \phi})$ and ${\cal L}({\bm \phi'})$ as defined in Equations \eqref{eq:like_2} and \eqref{eq:like_phi_prime_2}, respectively. The integrand in Equation \eqref{eq:like_qb} is a negligible fraction of its maximum over most of its domain. When this kind of an integrand is approximated via Monte Carlo methodologies, frequent sampling of the domain where the integrand is large ensures accuracy. Although our integration method is deterministic, we will similarly sample the domain densely where the integrand is large by making the $(q, b)$ grid in such portions of the domain relatively fine. We apply the following procedure to find a grid that meets this requirement. First, we define a subset of the domain ${\cal R} \equiv [q_0, q_1]\cap[b_0, b_1]$ where the grid will be fine. We initialize $q_0 = b_0 = 0$, $q_0 = q_1 = 1$, an equally spaced grid of 11 values between $q_0$ and $q_1$, and similarly in the $b$ dimension. Then, at every other value of each component of ${\bm \phi'}$, we compute the integrand value $I$ versus the fraction of the total integral that accumulates at the locations where the integrand is above $I$, fit this dependence to a linear spline, and use the latter to compute $I$ corresponding to 99.9\% of the total integral. We next narrow down ${\cal R}$ as much as possible under the condition that the integrand is always below $I$ outside the new ${\cal R}$. In each dimension, we allot 11 points to ${\cal R}$ as before and 4 additional points to the complement of ${\cal R}$ and repeat computation of $I$. We repeat this procedure, starting with the narrowing down of ${\cal R}$, until the new ${\cal R}$ does not differ much from its value in the previous iteration. After three iterations, this procedure results in $[q_0, q_1] = [0.70, 0.84]$ and $[b_0, b_1] = [0.40, 0.76]$.

At this point, we perform the integral in Equation \eqref{eq:like_qb} for every ${\bm \phi'}$, on an equally spaced grid of 21 values in each dimension to cover ${\cal R}$ and 7 additional points allotted to the complement of ${\cal R}$ in each dimension. We check that $99.9\%$ of the integral still always falls within ${\cal R}$, so that most of the integration is over the fine grid. We then subtract from $\ln{{\cal L}({\bm \phi'})}$ its maximum and take the exponent of the result, which gives us a new version of ${\cal L}({\bm \phi'})$, one that is below $f_{\rm max}$ and with a maximum significantly above $f_{\rm min}$. It turns out that ${\cal L}({\bm \phi'})$ is unimodal.

\subsection{Bayesian Probability Density} \label{bayesian}

Computation of ${\cal L}({\bm \phi'})$ in Section \ref{like_computation} is on a grid of 21 equally spaced values in each component of ${\bm \phi'}$, centered around $\hat{{\bm \phi'}}$. This grid covers set ${\cal T} \cap {\cal W}$. Here, ${\cal T}$ is the intersection of $\mu_t \in [9.154, 9.165]$ and $\sigma_t \in [0.036, 0.047]$; ${\cal W}$ is the intersection of $w_0 \in [0.025, 0.225]$ and $w_2 \in [0.4, 0.9]$. However, the integral in Equation \eqref{eq:z_l_2}, for example, is over a much larger formal region of normalization in ${\bm \phi}$. To calculate such integrals, we make the following approximations with respect to ${\cal L}({\bm \phi'})$ outside ${\cal T} \cap {\cal W}$. 

We assume that when the likelihood integral in Equation \eqref{eq:prob_t_2} is limited to ${\cal W}$, it is multiplied by some value that is slightly less than 1 and doesn't depend on $(\mu_t, \sigma_t)$. This ensures that the limited integral, which we call ${\cal L}(\mu_t, \sigma_t)$, is still proportional to $P(\mu_t, \sigma_t)$. We now wish to compute the corresponding normalization constant,
\begin{linenomath*}\begin{equation} \label{eq:z_t}
    Z_{t} = \int {\rm d}\mu_t\, {\rm d}\sigma_t\, {\cal L}(\mu_t, \sigma_t).
\end{equation}\end{linenomath*}
On ${\cal T}$, we numerically approximate the integral in Equation \eqref{eq:z_t} in the usual fashion. To estimate the integral outside ${\cal T}$, we calculate $p = {\cal L}_{\rm p} / {\cal L}_{\rm max}$. Here, ${\cal L}_{\rm p}$ is the average ${\cal L}(\mu_t, \sigma_t)$ on the perimeter of ${\cal T}$ and ${\cal L}_{\rm max}$ is the maximum ${\cal L}(\mu_t, \sigma_t)$ over ${\cal T}$. We then approximate $P(\mu_t, \sigma_t)$ as a two-dimensional normal density distribution with zero covariance and a maximum at the location of ${\cal L}_{\rm max}$. In this case, cumulative density over the locations where density drops below fraction $p$ of its peak is simply equal to $p$. Thus, we assume that fraction $p$ of the integral in Equation \eqref{eq:z_t} is outside ${\cal T}$, so that
\begin{linenomath*}\begin{equation} \label{eq:z_t_roc}
    Z_{t} = \frac{1}{1 - p}\int_{\cal T} {\rm d}\mu_t\, {\rm d}\sigma_t\, {\cal L}(\mu_t, \sigma_t)
\end{equation}\end{linenomath*}
and $P(\mu_t, \sigma_t) = {\cal L}(\mu_t, \sigma_t) / Z_t$. We exchange the roles of ${\cal T}$ and ${\cal W}$ in the above procedure to calculate $P(w_0, w_2)$. Figure \ref{fig:prob} presents the 35\%, 65\% and 95\% confidence regions for $P(\mu_t, \sigma_t)$ and $P(w_0, w_2)$. 

\bibliography{refs.bib}
\bibliographystyle{aasjournal}
\end{document}